\documentclass[%
 reprint,
 amsmath,amssymb,
aps,
prx,
floatfix,
nofootinbib]{revtex4-2}

\usepackage[T1]{fontenc}
\usepackage{graphicx}
\usepackage{dcolumn}
\usepackage{bm}

\usepackage{amsmath,scalerel}	
\usepackage{amssymb}	
\usepackage{txfonts}
\usepackage[bb=libus]{mathalpha}
\usepackage{bbold}
\usepackage{relsize}
\usepackage{mathtools}
\usepackage{bigints}
\usepackage{sidecap}
\usepackage{caption,subcaption}
\usepackage{hyperref}

\usepackage{floatrow}
\usepackage{soul}
\usepackage{comment}
\usepackage[normalem]{ulem}

\usepackage{macros_ub}
\graphicspath{{./}{figures/}}

\begin{document}

\preprint{APS/123-QED}

\title{Wave interference as the origin of the cyclic magnetorotational dynamo in accretion disks: insights from weakly nonlinear theory and local shearing box simulations} 

\author{Uddipan Banik}
\email{uddipanbanik@ias.edu}
\affiliation{Institute for Advanced Study, Einstein Drive, Princeton, NJ 08540, USA\\Department of Astrophysical Sciences, Princeton University, 112 Nassau Street, Princeton, NJ 08540, USA\\Perimeter Institute for Theoretical Physics, 31 Caroline Street N., Waterloo, Ontario, N2L 2Y5, Canada}

\author{Amitava Bhattacharjee}
\email{amitava@princeton.edu}
\affiliation{Department of Astrophysical Sciences, Princeton University, 112 Nassau Street, Princeton, NJ 08540, USA}

\author{James M. Stone}
\email{jmstone@ias.edu}
\affiliation{Institute for Advanced Study, Einstein Drive, Princeton, NJ 08540, USA}

%




\date{\today}

\begin{abstract}

Long-period cyclic reversals of the large-scale magnetic field are a prominent feature of the dynamo associated with the magnetorotational instability (MRI) in accretion disks, but their physical origin remains unclear. We develop a quasilinear theory (QLT) of the MRI dynamo where the electromotive force (emf) is computed from the linear eigenfunctions under the WKB approximation. The emf $\pmb{\varepsilon}$ depends on the mean field $\bB$ more generally than standard mean-field closures allow. In the unstratified case, the leading order contribution to the large-scale dynamo is the shear-current effect: the emf is given in terms of the current $\bJ$ by $\pmb{\varepsilon} = \pmb{\beta}\cdot\bJ$, with a tensor $\pmb{\beta}(\bB,t)$ that oscillates with time $t$ and whose off-diagonal components generate the mean field. The oscillations arise from beats between the two branches of magnetorotational eigenfrequencies. Since the beat frequency varies only weakly with wavenumber, the beats remain coherent and drive the long-period butterfly cycle seen in local shearing box simulations. We predict a dominant cycle period $\sim 30\sqrt{1+a^2}t_{\rm orb}$, with $a$ the vertical-to-radial aspect ratio and $t_{\rm orb}$ the orbital period, and an amplitude scaling $\sim a^2$ before saturation at $a\gtrsim 5$. Both trends agree with zero-net-flux unstratified shearing box simulations with \texttt{Athena++}. A carrier-envelope analysis of the simulation spectra shows that the same interference mechanism extends beyond strict QLT, through higher-order linear combinations of the eigenfrequencies, with observed cycles arising from pairwise beats within this spectral network. These results identify coherent interference between nearly degenerate eigenfrequencies as a key mechanism behind large-scale cyclic dynamos, with implications for magnetic variability in protoplanetary disks, X-ray binaries, and AGNs.

\end{abstract}

\maketitle


\section{Introduction}\label{sec:intro}

Cyclic large-scale magnetic fields are a striking manifestation of self-organization in rotating conducting fluids and plasmas. They are observed or inferred across a wide range of systems. The 11-year sunspot cycle \citep[][]{Hathaway.15} traces a large-scale oscillating field in the solar interior. Geomagnetic reversals are recorded in the paleomagnetic record \citep[][]{Hagay.etal.10}, while magnetic variability and reversals have been inferred for giant planets \citep[][]{Hathaway.86}. Quasi-periodic modulations of X-ray binary and AGN light curves \citep[][]{Alston.etal.16} are often interpreted in terms of cyclic magnetic activity in accretion disks \citep[][]{ONeill.etal.11,Zhou.Lai.24}. A common ingredient in many of these systems is differential rotation. In weakly magnetized accretion flows, the magnetorotational instability (MRI) \citep[][]{Balbus.Hawley.91,Balbus.95} is the dominant mechanism for amplifying seed fields to dynamically significant strengths. MRI-driven turbulence rapidly develops in both local and global MHD simulations \citep{Hawley.etal.95,Hawley.etal.96,Stone.etal.96}. Once the linear instability saturates, a large-scale dynamo generically emerges. In particular, the mean toroidal field undergoes quasi-periodic polarity reversals on timescales of order $10$--$100$ orbital periods, producing the well-known butterfly diagram \citep{Brandenburg.etal.95,Stone.etal.96,Gressel.10,Dhang.etal.24}. Similar cyclic behavior appears not only in stratified and/or net-flux simulations \citep[][]{Stone.etal.96,Miller.Stone.00,Simon.etal.11,Lesur.etal.13,Bai.Stone.13} but also in unstratified zero-net-flux simulations \citep{Hawley.etal.96,Fromang.Papaloizou.07,Fromang.etal.07,Guan.etal.09,Simon.etal.09,Shi.etal.16}, showing that shear alone can drive the cycle. The strength of the dynamo depends sensitively on box geometry: tall boxes exhibit pronounced cycles and converged turbulent (Maxwell and Reynolds) stresses, whereas equal-aspect-ratio boxes show weak or no cycles and stresses that decrease with increasing resolution \citep{Fromang.Papaloizou.07,Fromang.etal.07,Shi.etal.16}. Despite this extensive numerical evidence, the physical origin of the long cycle period and its aspect-ratio dependence remains unclear. Prior work has shown that magnetized shear-flow dynamos can be subcritical, sustained by finite-amplitude perturbations and non-modal growth much like hydrodynamic shear-flow turbulence \citep[][]{Rincon.etal.07,Rincon.etal.08,Squire.Bhattacharjee.14a,Squire.Bhattacharjee.14b}. The dynamo can also sustain itself through a non-linear feedback loop consisting of shear-driven poloidal-to-toroidal conversion followed by nonlinear regeneration of poloidal field through non-axisymmetric MRI perturbations \citep[][]{Rincon.etal.07,Riols.etal.13}. However, a quantitative theory for the period of these feedback loops, and of MRI dynamo cycles more generally, is still lacking.

The standard framework for large-scale dynamos is mean-field theory \citep{Brandenburg.Subramanian.05,Radler.14,Brandenburg.18}, which expands the electromotive force (emf) in the mean magnetic field $\bB$ and its derivatives, e.g. $\pmb{\varepsilon}=\pmb{\alpha}\cdot\bB+\pmb{\beta}\cdot\bJ+\cdots$, where $\bJ=(c/4\pi)(\nabla\times\bB)$. In the classic $\alpha$--$\Omega$ dynamo \citep{Parker.55,Steenbeck.Krause.Radler.66}, $\alpha$ is proportional to the mean kinetic helicity. While this framework has been phenomenologically successful, it faces two major difficulties in a fully-developed MHD turbulence. First, the kinematic $\alpha$ is quenched by the current helicity generated as the large-scale field grows \citep{Pouquet.etal.76, Gruzinov.Diamond.94, Bhattacharjee.Yuan.95, Blackman.Field.99, Radler.Rheinhardt.07}, and at high magnetic Reynolds numbers relevant to astrophysical disks the $\alpha$ undergoes catastrophic quenching \citep{Vainshtein.Cattaneo.92}. Second, the expansion is not a controlled closure: its convergence is not guaranteed, and the tensors $\pmb{\alpha}$ and $\pmb{\beta}$ are often inferred from simulations rather than derived from first principles. In non-helical flows, where $\pmb{\alpha}$ is weak, the dominant contribution is instead the shear-current effect $\pmb{\beta}\cdot\bJ$, whose off-diagonal component can amplify the mean field \citep{Squire.Bhattacharjee.15a,Squire.Bhattacharjee.15b,Squire.Bhattacharjee.15c,Squire.Bhattacharjee.15d,Squire.Bhattacharjee.16}. Rotating systems can also admit an additional term $\pmb{\Upsilon}\cdot(\nabla\times\bU)$, with $\pmb{\Upsilon}$ proportional to the cross-helicity \citep{Yoshizawa.90,Tripathi.etal.26}. Even with these extensions, however, mean-field theory, in its standard form, has not helped to explain the long, geometry-dependent cycles seen in unstratified shearing boxes, where the dynamo operates primarily through an off-diagonal component of $\pmb{\beta}$ \citep[][]{Squire.Bhattacharjee.16,Shi.etal.16}.

These limitations motivate a more systematic route: quasilinear theory (QLT), in which the emf is computed directly from linear eigenmodes rather than postulated through a low-order closure. QLT is most appropriate when fluctuations remain weak relative to the mean field.  It has been used in laboratory plasmas to determine the self-consistent dynamo in reversed-field pinches unstable to tearing modes \citep[][]{Bhattacharjee.Hameiri.86,Strauss.86}. The onset of the dynamo cycles before the MRI turbulence reaches a statistical steady state \citep{Bhat.etal.16} indicates that QLT should be able to capture at least the initial dynamo phase. Earlier work on accretion disks established important aspects of the quasilinear program. \citet{Lesur.Ogilvie.07} used a perturbative expansion for non-axisymmetric shearing waves and identified a non-linear feedback loop for dynamo cycles \citep[c.f.][]{Rincon.etal.07}, but did not derive the cycle period. \citet{Ebrahimi.Blackman.16} computed the quasilinear emf from linear MRI eigenfunctions, but did not evolve the time-dependent dynamo equations. Thus, a dynamical explanation for the butterfly period and its geometry dependence remains missing. A complementary line of work has approached the zero-net-flux MRI dynamo from a nonlinear dynamical systems perspective, identifying subcritical periodic orbits (identified as fixed points of the stroboscopic/Poincar\'e map constructed from the simulation data) and their bifurcations (with varying magnetic Reynolds number) as the organizing structures of MRI dynamo action \citep{Riols.etal.13}. Our goal here is different: rather than characterizing the phase-space skeleton of the fully nonlinear dynamo, we seek a first-principles physical explanation for the cycle period from a weakly nonlinear perspective. We show that toroidal cycles may arise from the interference of axisymmetric oscillatory modes, provided turbulence mediated by non-axisymmetric fluctuations supplies weak poloidal fields.

In this paper we focus on the weakly nonlinear theory for axisymmetric MRI modes. We compute the full quasilinear emf from the linear MRI eigenfunctions in the WKB approximation, without assuming a mean-field closure, and then integrate the resulting dynamo equations for the large-scale toroidal field. The central result is that the MRI dispersion relation contains two branches of shear Alfv\'en waves whose frequency difference $\Delta\omega_{\bk'}=\omega_{1\bk'}-\omega_{2\bk'}$
is weakly dependent on wavenumber for $k'\gtrsim k'_\rmc$, where $k'_\rmc$ is the marginally unstable wavenumber. Since the emf is an integral over contributions from all small-scale modes, the key factor that determines the level of coherence is the degree of phase-mixing across the spectrum. Because $\Delta\omega_{\bk'}$ varies weakly over the dominant range of $\bk'$, these contributions add nearly coherently instead of phase-mixing away. The large-scale induction therefore develops a slow envelope modulation whose characteristic period is set by the narrow spread of $\Delta\omega_{\bk'}$ across the contributing modes. In this sense, the cyclic MRI dynamo is a large-scale manifestation of wave-interference.

This mechanism leads to concrete, testable predictions. We show analytically in Section~\ref{sec:cycle_period_analytic_QLT} that the dominant cycle period scales as
\begin{align}
T_{\rm cycle}\sim 30\sqrt{1+a^2}\;t_{\rm orb},
\end{align}
where $a=L_z/L_R$ is the vertical-to-radial aspect ratio of the shearing box and $t_{\rm orb}=2\pi/\Omega$ is the orbital period. We also find that the cycle amplitude scales as $\sim a^2$ before saturating at $a\gtrsim 5$. Both trends arise directly from the anisotropic character of the MRI eigenmodes. We test these predictions against unstratified, zero-net-flux \texttt{Athena++} shearing box simulations spanning $a=1$--$8$ and extending to $500$ orbital periods. The QLT-predicted dominant periods and their aspect-ratio dependence agree well with the simulations. We further show, through a carrier-envelope analysis of the simulation power spectra in Section~\ref{sec:cycle_period_analytic_sim}, that the interference mechanism is not merely an artifact of the quasilinear approximation. The simulations contain higher-order linear combinations of the MRI eigenfrequencies, and the observed cycle periods arise from pairwise beats within this broader spectral network. The basic organizing principle of the MRI dynamo cycles is therefore wave interference.

The paper is organized as follows. Section~\ref{sec:linear} reviews the linear MRI and the eigenmode structure underlying the dynamo. Section~\ref{sec:quasilinear} develops the quasilinear theory, derives the emf, integrates the dynamo equations numerically, and obtains analytic estimates for the cycle period. Section~\ref{sec:sim} presents the \texttt{Athena++} simulations, and compares them with the quasilinear predictions. Section~\ref{sec:discussion} discusses the physical interpretation, the limitations of QLT, and the astrophysical implications of the results.

\section{Linear theory}\label{sec:linear}
The physical system under consideration is a small annulus of the accretion disk, modeled as a vertically thick ring of radius $R$. The evolution of the hydrodynamic quantities, density $\rho$, velocity $\bv$, pressure $P$ and magnetic field $\bB$, in this system can be described by the ideal MHD equations:

\begin{align}
&\frac{\partial \rho}{\partial t} + \nabla \cdot \left(\rho \bv\right) = 0,\nonumber\\
&\frac{\partial \bv}{\partial t}  + \bv \cdot \nabla\bv = -\frac{1}{\rho}\nabla\left(P+\frac{B^2}{8\pi}\right) + \frac{1}{4\pi \rho}\bB \cdot \nabla \bB - \nabla \Phi,\nonumber\\
&\frac{\partial \bB}{\partial t} = \nabla\times \left(\bv\times \bB\right),\nonumber\\
&\nabla\cdot \bB = 0,\nonumber\\
&\frac{\partial}{\partial t}\ln{\left(\frac{P}{\rho^\gamma}\right)} + \bv \cdot \nabla\ln{\left(\frac{P}{\rho^\gamma}\right)} = 0,
\label{MHD_eqs}
\end{align}
where $\gamma$ is the adiabatic coefficient and the specific entropy is conserved along a streamline \citep[][]{Balbus.Hawley.91}.

\begin{figure*}
\centering
\includegraphics[width=1\textwidth]{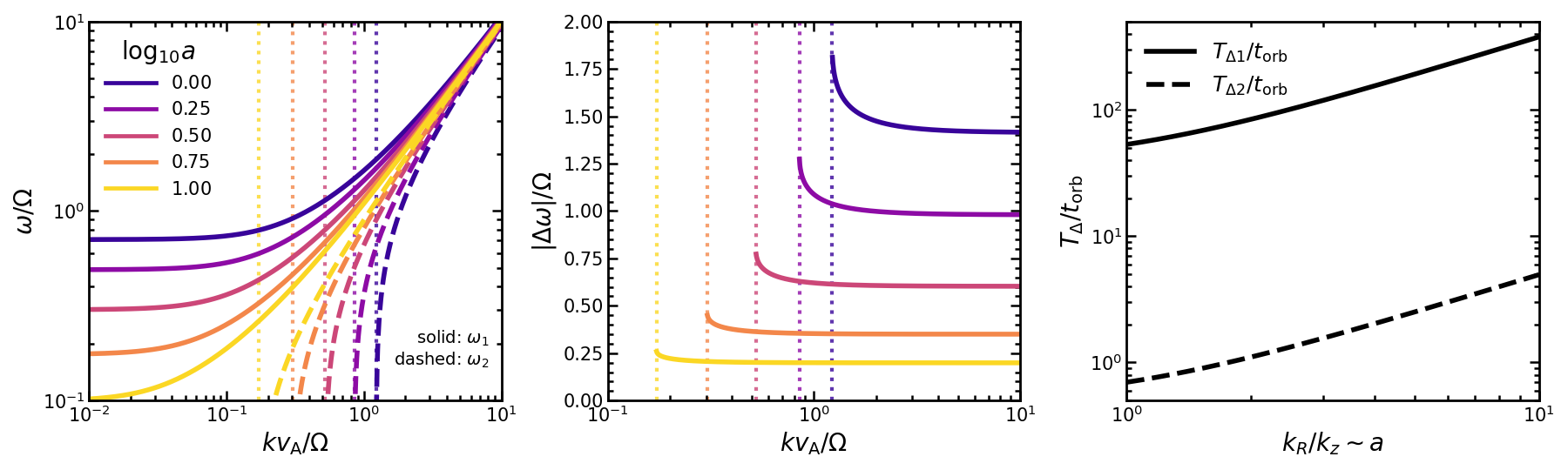}
\caption{The linear eigenfrequencies, beat frequencies and beat periods as functions of $kv_\rmA/\Omega$ and $a\equiv k_R/k_z$. The left panel shows the eigenfrequencies $\omega_1$ (solid) and $\omega_2$ (dashed) as functions of $kv_\rmA/\Omega$, for five values of $a$ equally spaced in $\log_{10}a$ between $1$ and $10$. The middle panel shows the beat frequency $|\Delta\omega| = |\omega_1-\omega_2|$ for the same values of $a$. The right panel shows the corresponding slow and fast beat periods, $T_{\Delta1}/t_{\rm orb}$ (solid) and $T_{\Delta2}/t_{\rm orb}$ (dashed), as functions of $a$, where $T_{\Delta1}=4\pi/|\Delta\omega(2k_\rmc)-\Delta\omega_\infty|$ and $T_{\Delta2}=4\pi/|\Delta\omega(2k_\rmc)+\Delta\omega_\infty|$, with $\Delta\omega_\infty=(2k_z/k)\Omega$ (see Section~\ref{sec:cycle_period_analytic_QLT}). In the left and middle panels, the faint vertical dotted lines indicate the marginally stable value $k_\rmc v_\rmA/\Omega=\sqrt{3}\,k_z/k$, corresponding to $\omega_2=0$ for Keplerian rotation with $\kappa=\Omega$. The frequencies are normalized by $\Omega$, and the periods by the orbital time $t_{\rm orb}=2\pi/\Omega$.}
\label{fig:lin_MRI}
\end{figure*}

We linearize the above equations about a background shear flow with $v_{0\phi}=R\,\Omega(R)$ ($\Omega$ is the angular frequency) and $v_{0R}=v_{0z}=0$, assuming that the mean-field components $B_{0R}$, $B_{0z}$, and $B_{0\phi}$ evolve on space- and time-scales much longer than those of the linear perturbations. As shown by \citet[][]{Balbus.Hawley.91}, in the WKB limit, $k_RR\gtrsim 1$ and $k_zR\gtrsim 1$, the linear problem can be written as a generalized eigenvalue problem, with the response at each $\bk$ expressed as a superposition of normal modes. Because shear makes the linear MHD operator non-self-adjoint, the eigenvalues can in general be complex and the eigenfunctions non-orthogonal. For axisymmetric modes $(m=0)$, however, the WKB limit simplifies the problem: the eigenvalues are real or purely imaginary, and the eigenfunctions form a complete bi-orthogonal set, such that the inner product of the left and right eigenvectors vanishes for distinct eigenvalues and is non-zero otherwise. In this paper we restrict attention to axisymmetric modes, which already capture the essential dynamo physics; extending the analysis to non-axisymmetric modes is straightforward. To isolate the basic mechanism, we further specialize to the unstratified case, taking $P$ and $\rho$ to be spatially uniform, and assume scale separation so that the large-scale field $\bB$ varies on spatial and temporal scales much larger than those of the fluctuations and may therefore be treated as uniform in the perturbative analysis. The generalization  to include spatial inhomogneities or stratification in the linear and quasilinear treatments is left for future work.

In the limit of $\gamma \gg 1$ or $v_\rmA = B / \sqrt{4\pi\rho} \ll c_\rms = \sqrt{\gamma P/\rho}$, where $v_\rmA$ and $c_\rms$ are the Alfv\'en and sound speeds respectively, which holds for high $\beta = 8\pi P / B^2$ plasmas, we are in the incompressible limit, $\nabla \cdot \bv = 0$. We perform our perturbative analysis in the limit of unstratified incompressible flow under the WKB approximation. Let us linearize the ideal MHD equations, i.e., expand each quantity $\bQ(\bx,t)$ ($\bv$ or $\bB$) as a power series, $\bQ(\bx,t) = \bQ_0 + \epsilon \bQ_1 + \epsilon^2 \bQ_2 + ...$, where $\epsilon = \left|\delta\bB\right|/\left|\bB\right|$ is a small parameter. This yields the following linearized equations for the evolution of the fluctuations:

\begin{align}
&\nabla \cdot \bv_1 = 0,\nonumber\\
&\frac{\partial \bv_1}{\partial t}  + \bv_0 \cdot \nabla \bv_1 + \bv_1 \cdot \nabla\bv_0 = -\frac{1}{\rho_0}\nabla\left(\frac{\bB_0\cdot\bB_1}{4\pi}\right) + \frac{1}{4\pi \rho_0}\bB_0 \cdot \nabla \bB_1,\nonumber\\
&\frac{\partial \bB_1}{\partial t} = \nabla\times \left(\bv_1\times \bB_0\right) + \nabla\times \left(\bv_0\times \bB_1\right),\nonumber\\
&\nabla\cdot \bB_1 = 0.
\end{align}

We perform the Fourier transform of the linear perturbations $\bv_1$ and $\bB_1$ in space and time, i.e., express these as a Fourier series, $\bQ_1(\bx,t) = \int_{-\infty}^\infty \rmd \omega \int \rmd k_R \int \rmd k_z \, \bQ_{1\omega \bk} \exp{\left[i\left(\bk\cdot\bx - \omega t\right)\right]}$, where $\bk = \left(k_R,k_z\right)$ is the wavenumber and $\omega$ is the frequency\footnote{Since we are interested in the long time evolution of the system rather than the initial transients, a temporal Fourier transform yields the same results as that of the Laplace transform.}. We adopt the WKB approximation, $k_R R \gtrsim 1, k_z R \gtrsim 1$, i.e., study the evolution of perturbations on scales smaller than the extent of the ring. Solving the linearized equations in the WKB limit in cylindrical coordinates $(R,\phi,z)$ following \citet[][]{Balbus.Hawley.91} yields the following dispersion relation for the normal modes:

\begin{align}
\Tilde{\omega}^4 - \frac{k^2_z}{k^2}\kappa^2\Tilde{\omega}^2 - \frac{4 k^2_z}{k^2}\Omega^2 {\left(\bk \cdot \bv_\rmA \right)}^2 = 0,
\label{lin_disp_rel}
\end{align}
where the radial epicyclic frequency $\kappa$ is given by  $\kappa^2 = 4\Omega^2 + \rmd \Omega^2/\rmd \ln{R}$, ${\Tilde{\omega}}^2 = \omega^2 - {\left(\bk\cdot \bv_\rmA\right)}^2$, and $\bv_\rmA = \bB_0/\sqrt{4\pi \rho_0}$ is the Alfv\'en velocity. Including density stratification adds ${\left((k_R/k_z)N_z-N_r\right)}^2$ to $\kappa^2$, where $N_R$ and $N_z$ are the radial and vertical pieces of the square of the Brunt-V$\ddot{\rma}$is$\ddot{\rma}$l$\ddot{\rma}$ frequency $N$ for buoyancy waves (akin to gravity waves), i.e., $N^2 = -(\nabla P/\gamma \rho) \cdot \nabla(\ln{P\rho^{-\gamma}}) = N^2_R + N^2_z$ \citep[][]{Balbus.Hawley.91}. In a stably stratified medium (with $((k_R/k_z)N_z-N_R)^2>-\kappa^2$ in unmagnetized shear-flows), these are often much smaller than the epicyclic frequency.

Solving equation~(\ref{lin_disp_rel}) yields the following solution for the eigenmode frequencies for unstratified incompressible flow:

\begin{align}
\omega^2_{1 \bk} = {\left(\bk\cdot \bv_\rmA\right)}^2 + \frac{k^2_z \kappa^{2}}{2 k^2}\left(1+\sqrt{1+\frac{16\, k^2}{k^2_z}\frac{\Omega^2{\left(\bk\cdot\bv_\rmA\right)}^2}{\kappa^{4}}}\right),\nonumber\\
\omega^2_{2 \bk} = {\left(\bk\cdot \bv_\rmA\right)}^2 + \frac{k^2_z \kappa^{2}}{2 k^2}\left(1-\sqrt{1+\frac{16\, k^2}{k^2_z}\frac{\Omega^2{\left(\bk\cdot\bv_\rmA\right)}^2}{\kappa^{4}}}\right).
\label{MRI_freqs}
\end{align}
These correspond to waves that behave like Alfv\'en waves on small scales and inertial waves on large scales. It is not hard to see that $\omega^2_{1 \bk}$ is positive definite and corresponds to an oscillatory mode: the Alfv\'en wave modified by shear. This mode has a faster phase-velocity, since magnetic tension and shear (radial epicycles) reinforce each other as restoring forces. On the other hand, $\omega_{2 \bk}$ is the frequency of the slower mode (also shear-modulated Alfv\'en wave), where tension and shear oppose each other. Shear wins over tension and $\omega^2_{2 \bk}$ changes sign at a particular combination of $\bk$ and $\bv_\rmA$, leading to an instability, famously known as the magnetorotational instability (MRI) \citep[][]{Velikhov.59,Chandrasekhar.61,Balbus.Hawley.91}. The occurrence of this instability requires the following criterion \citep[][]{Balbus.Hawley.91,Balbus.95}:

\begin{align}
&{\left(\bk\cdot\bv_\rmA\right)}^2 < \left(4\Omega^2 - \kappa^2\right) \frac{k^2_z}{k^2} = -\frac{\rmd\Omega^2}{\rmd\ln R} \frac{k^2_z}{k^2},\nonumber\\
&\frac{\rmd\Omega^2}{\rmd\ln R} < 0.
\label{MRI_criterion}
\end{align}
The instability criterion is satisfied on large enough scales in trailing shear flows with ${\rmd\Omega^2}/{\rmd\ln R}<0$. Therefore, MRI is active in the outer parts of accretion disks around compact objects, where the flow is Keplerian, i.e., $\Omega$ radially decreases outwards as $R^{-3/2}$. It is also evident from equation~(\ref{MRI_criterion}) that MRI is dominant for weaker fields and for perturbations more perpendicularly aligned with $\bB$ (smaller $\bk\cdot\bv_\rmA$). In all these cases, shear wins over magnetic tension. Weak field lines act as springs connecting patches of the disk, draining angular momentum from the inner patch and giving it to the outer patch through shear, which drives accretion \citep[][]{Balbus.Hawley.91}. When the above inequality becomes an equality, the unstable mode becomes marginally stable with $\omega^2_{2 \bk} = 0$. In an unstably stratified medium, where the Brunt-V$\ddot{\rma}$is$\ddot{\rma}$l$\ddot{\rma}$ frequencies are sufficiently negative, i.e., below a critical threshold determined by shear, the Taylor/buoyancy/interchange instability becomes active instead of the MRI. This is particularly relevant in the sub-Keplerian inner regions of accretion disks that are MRI stable but show interesting dynamo action (e.g., quasi-periodic flux-eruptions in magnetically arrested disks around black holes). We will discuss this buoyancy-driven dynamo in future work.


Let us briefly discuss the asymptotic behavior of the eigenfrequencies given by equations~(\ref{MRI_freqs}) (for unstratified, incompressible, magnetized shear-flows). We denote the critical $\bk$ for which the system is marginally stable as $\bk_\rmc$. For ${\bk\cdot\bv_\rmA \lesssim \bk_\rmc\cdot\bv_\rmA}$, the frequency of the oscillatory mode is given by $\omega^2_{1\bk} = (4 k^2_z/k^2)\,\Omega^2 + {\left(\bk\cdot\bv_\rmA\right)}^2$. In the limit of $\bk\cdot\bv_\rmA \ll \bk_\rmc\cdot\bv_\rmA$, we have $\omega^2_{1\bk} \approx (k^2_z/2k^2)\,\kappa^2 + (1 + 4\Omega^2/\kappa^2){\left(\bk\cdot\bv_\rmA\right)}^2 \xrightarrow[]{\bk\cdot\bv_\rmA/\bk_\rmc\cdot\bv_\rmA \to 0} (k^2_z/2k^2)\,\kappa^2$, and $\omega^2_{2\bk} \approx (1 - 4\Omega^2/\kappa^2){\left(\bk\cdot\bv_\rmA\right)}^2 \xrightarrow[]{\bk\cdot\bv_\rmA/\bk_\rmc\cdot\bv_\rmA \to 0} 0^{-}$. In the opposite limit, $\bk\cdot\bv_\rmA \gg \bk_\rmc\cdot\bv_\rmA$, we have $\omega^2_{1\bk} \approx (k^2_z/2k^2)\,\kappa^2 + \,{\left(\bk\cdot\bv_\rmA\right)}^2\, + 2\left|k_z/k\right|\bk\cdot\bv_\rmA \Omega$ and $\omega^2_{2\bk} \approx (k^2_z/2k^2)\,\kappa^2 + {\left(\bk\cdot\bv_\rmA\right)}^2 - 2\left|k_z/k\right|\bk\cdot\bv_\rmA \Omega$. These limiting cases are going to come in handy in analyzing the dynamo. A key point to note here is the fact that $\omega_{1\bk} - \omega_{2\bk} \sim ({2 k_z}/{k})\,\Omega\left[1 + ({k^2_z}/{4 k^2}){\left({\bk\cdot\bv_\rmA}/{\Omega}\right)}^{-2}\calF(\kappa/\Omega)\right]$ ($\calF$ is an $\calO(1)$ function of $\kappa/\Omega$) for $\bk\cdot\bv_\rmA > \bk_\rmc\cdot\bv_\rmA$. Thus, on small scales, \(\Delta\omega_{\bk}\) becomes only weakly dependent on \(k\), and is especially small when \(|k_z|/k \ll 1\), i.e. for \(k_R/k_z \simeq a \gg 1\). The eigenfrequencies \(\omega_{1\bk}\) and \(\omega_{2\bk}\), as well as their difference \(|\Delta\omega_{\bk}|\), are plotted as functions of \(k v_{\rmA}/\Omega\) in the left and middle panels of Fig.~\ref{fig:lin_MRI}, for different values of \(a = k_R/k_z\). As \(a\) increases, the marginal scale shifts to smaller \(k v_{\rmA}/\Omega\), the two stable branches become nearly parallel at large \(k v_{\rmA}/\Omega\), and \(|\Delta\omega_{\bk}|/\Omega\) approaches a lower asymptotic value. The corresponding beat periods therefore increase with aspect ratio, with the longer one, $T_{\Delta1}=4\pi/|\Delta\omega(2k_\rmc)-\Delta\omega_\infty|$, reaching hundreds of orbital times for \(a \sim 10\), as shown in the right panel of Fig.~\ref{fig:lin_MRI}. This slow, weakly varying beat frequency is the key linear ingredient behind the long-period large-scale dynamo cycles discussed in Section~\ref{sec:cycle_period_analytic_QLT}.

The linear perturbations can be expressed as a linear combination of the (right) eigenfunctions:

\begin{align}
\begin{bmatrix}
v_{1\bk R}\left(\omega\right)\\ \\ v_{1\bk \phi}\left(\omega\right) \\ \\ B_{1\bk R}\left(\omega\right) \\ \\ B_{1\bk \phi}\left(\omega\right)
\end{bmatrix} &=
\begin{bmatrix}
1\\ \\ \dfrac{i\omega}{\omega^2 - {\left(\bk\cdot\bv_\rmA\right)}^2} \left(-\dfrac{\kappa^2}{2\Omega} + \dfrac{{\left(\bk\cdot\bv_\rmA\right)}^2}{\omega^2}\dfrac{\rmd\Omega}{\rmd\ln R}\right) \\ \\ -\dfrac{\bk\cdot \bB_0}{\omega} \\ \\ \dfrac{2 i\, \Omega\, \bk\cdot\bB_0}{\omega^2 - {\left(\bk\cdot\bv_\rmA\right)}^2}
\end{bmatrix}\nonumber\\
&\times\delta v_{\bk R}\left(\omega\right),
\label{lin_eig_fourier}
\end{align}
where $\delta v_{\bk R}\left(\omega\right) = \delta v_{\bk R}\sum_{i} c_{i\bk}\,\delta\left(\omega - \omega_{i\bk}\right)$ is the Fourier transform of the radial velocity perturbation, with amplitude $\delta v_{\bk R}$. The dispersion relation is quadratic in $\omega^2_\bk$, so it has roots $\pm \omega_{1\bk}$ and $\pm \omega_{2\bk}$. The $+$ and $-$ signs correspond to the right and left circular polarizations of the wave, i.e., positive and negative helicities. Taking the inverse Fourier transform of equation~(\ref{lin_eig_fourier}) yields the following solution for the linear perturbations in the time domain:

\begin{align}
v_{1\bk R}\left(t\right)
&=
\delta v_{\bk R}\nonumber\\
&\times\sum_{i=1,2}
\Big[
\left(c_{i\bk}^+ + c_{i\bk}^- \right)\cos{\left(\omega_{i\bk}t\right)}
-i\left(c_{i\bk}^+ - c_{i\bk}^- \right)\sin{\left(\omega_{i\bk}t\right)}
\Big],
\nonumber\\
v_{1\bk\phi}\left(t\right)
&=
\delta v_{\bk R} \sum_{i=1,2}
\frac{\omega_{i\bk}}{\omega^2_{i\bk} - {\left(\bk\cdot\bv_\rmA\right)}^2}
\nonumber\\
&\times
\left(
-\frac{\kappa^2}{2\Omega}
+ \frac{{\left(\bk\cdot\bv_\rmA\right)}^2}{\omega^2_{i\bk}}
\frac{\rmd\Omega}{\rmd\ln R}
\right)
\nonumber\\
&\times
\Big[
\left(c_{i\bk}^+ + c_{i\bk}^- \right)\sin{\left(\omega_{i\bk} t\right)}
+i\left(c_{i\bk}^+ - c_{i\bk}^- \right)\cos{\left(\omega_{i\bk} t\right)}
\Big],
\nonumber\\
B_{1\bk R}\left(t\right)
&=
i\, \bk\cdot\bB_0\, \delta v_{\bk R} \sum_{i=1,2}
\frac{1}{\omega_{i\bk}}
\nonumber\\
&\times
\Big[
\left(c_{i\bk}^+ + c_{i\bk}^- \right)\sin{\left(\omega_{i\bk} t\right)}
+i\left(c_{i\bk}^+ - c_{i\bk}^- \right)\cos{\left(\omega_{i\bk} t\right)}
\Big],
\nonumber\\
B_{1\bk \phi}\left(t\right)
&=
2 i\, \Omega\, \bk\cdot\bB_0\, \delta v_{\bk R} \sum_{i=1,2}
\frac{1}{\omega^2_{i\bk} - {\left(\bk\cdot\bv_\rmA\right)}^2}
\nonumber\\
&\times
\Big[
\left(c_{i\bk}^+ + c_{i\bk}^- \right)\cos{\left(\omega_{i\bk}t\right)}
-i\left(c_{i\bk}^+ - c_{i\bk}^- \right)\sin{\left(\omega_{i\bk}t\right)}
\Big].
\label{lin_eig}
\end{align}
where we ensure the reality condition, $c^{+}_{i-\bk} = c^{-\ast}_{i\bk}$, $c^{-}_{i-\bk} = c^{+\ast}_{i\bk}$, and $\omega_{i-\bk} = \omega_{i\bk}$ (the imaginary $i$ is not to be confused with the subscript $i$ for the eigenmode index). There are three important points to note here. First, $\left(c_{i\bk}^+ + c_{i\bk}^- \right)$ denotes the total strength of fluctuations at each $\bk$ and $\left(c_{i\bk}^+ - c_{i\bk}^- \right)$ the net helicity of those fluctuations, i.e., the difference between the amplitudes of right and left circularly polarized waves. Second, the azimuthal perturbations are proportional to $\Omega$ and the shear, $\rmd\Omega/\rmd\ln R$, which represents the shear-driven tangential stretching of the field lines and the associated development of azimuthal velocity and magnetic perturbations. Third, $v_{1\bk\phi}$ and $B_{1\bk R}$ get stronger near marginal stability $\left(\omega_{2\bk}\to 0\right)$, implying that the strength of the emf and the large-scale field would be dominated by modes closer to marginal stability.

\begin{figure*}
\centering
\includegraphics[width=0.98\textwidth]{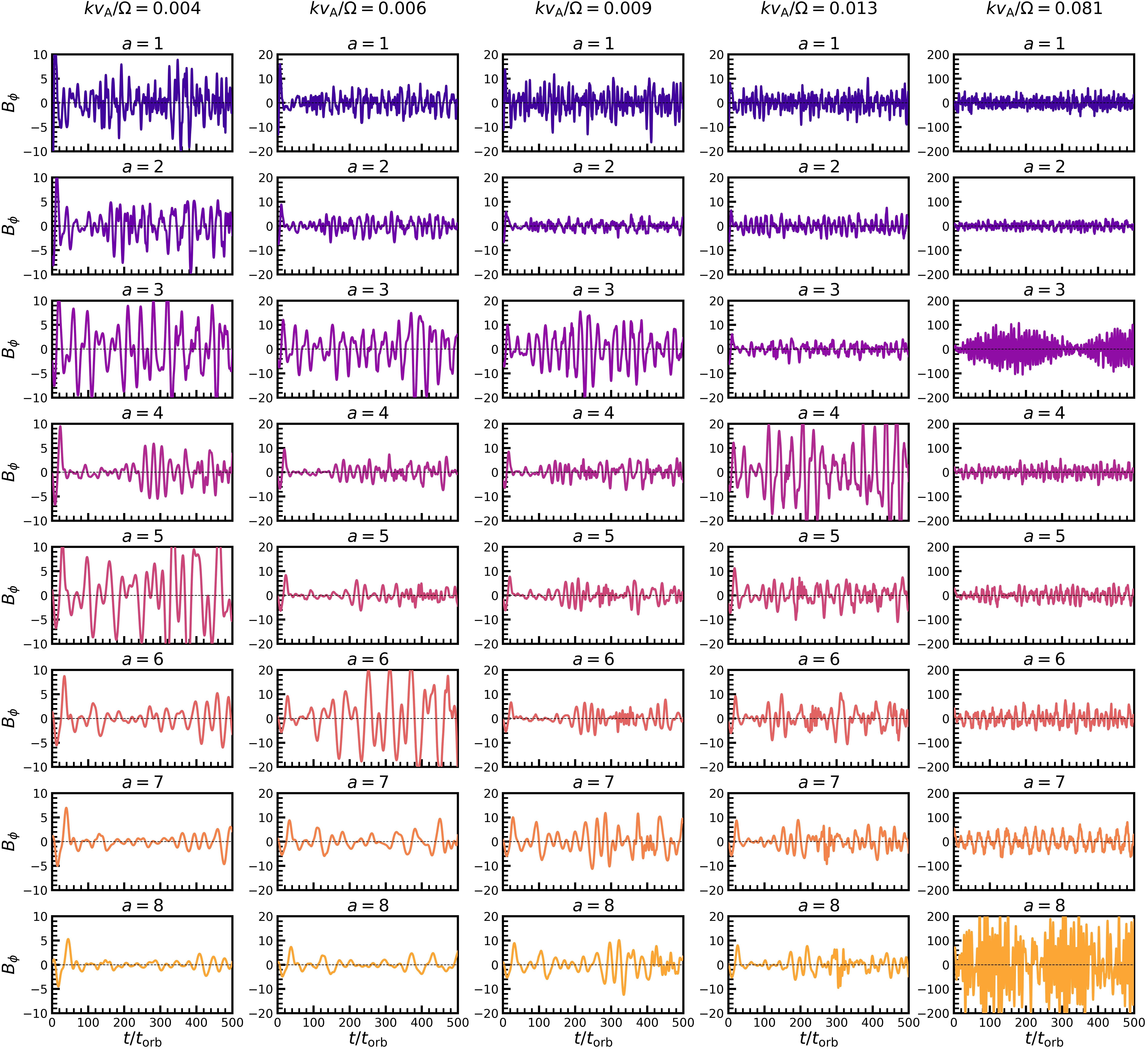}
\caption{Quasilinear theory: Large scale $B_\phi$ in the shearing frame vs $t$, for different aspect ratios $a = 1-8$ (rows), and different values of $k v_\rmA/\Omega$ as shown (columns), computed by numerically integrating the quasilinear dynamo equation (second of equations~[\ref{dynamo_eqs_concise}]) at a constant $B_R$ and $B_z$ that satisfy $k_R B_R + k_z B_z = 0$. The shearing frame $B_\phi$ is obtained by subtracting $(\rmd\Omega/\rmd\ln R)\, B_R\, t$ from the inertial frame $B_\phi$. The cyclic behavior is more pronounced at larger $a$ and smaller $k v_\rmA/\Omega$.}
\label{fig:Bphi_vs_t_a}
\end{figure*}

\section{Weakly nonlinear theory}\label{sec:quasilinear}

So far we have discussed the spatiotemporal behavior of linear perturbations on a fixed background shear flow with a large-scale magnetic field. The large-scale field itself, however, evolves through the back reaction of the fluctuations, on timescales longer than the linear wave period: of order the orbital time $2\pi/\Omega$ on large scales and the Alfv\'en crossing time $2\pi/\bk\cdot\bv_\rmA$ on small scales. The large-scale field is $\left<\bB\right> = \left<\bB_2\right> = \bB_0$, where the brackets denote the limit $k_i\to k_{i,\rm min}=2\pi/L_i$. Here $i$ labels the radial or vertical component of $\bk$, and $L_i$ is the corresponding box size. In this limit, $\left<\bB_1\right>=0$, since $\bB_{1\bk}\propto \bk\cdot\bB_0=0$ by $\nabla\cdot\bB_0=0$.

More generally, we write $\bB=\left<\bB\right>+\delta\bB$ and $\bv=\left<\bv\right>+\delta\bv$, where $\delta\bB$ and $\delta\bv$ are nonlinear fluctuations and $\left<\bB\right>$ and $\left<\bv\right>$ are the mean fields. In the present case, $\left<\bv\right>=R\Omega\hat{\phi}$. Substituting into the induction equation yields
\begin{align}
&\frac{\partial \left<\bB\right>}{\partial t} = \nabla\times \left(\left<\bv\right>\times \left<\bB\right>\right) + \nabla\times \left<\delta\bv\times \delta\bB\right>,
\label{induction_eq_mean_plus_fluc}
\end{align}
where we used $\left<\delta\bB\right>=\left<\delta\bv\right>=0$. The first term is the mean-flow contribution, i.e. the $\Omega$ effect that stretches radial field into toroidal field. The second is the dynamo term, driven by the inductive electric field from correlated fluctuations. This field, $\pmb{\varepsilon}=\left<\delta\bv\times\delta\bB\right>$, is the electromotive force (emf). In fully nonlinear MHD, evaluating the emf is difficult. Mean-field theory aims to express it in terms of mean quantities, e.g. $\pmb{\varepsilon}=\pmb{\alpha}\cdot\left<\bB\right>+\pmb{\beta}\cdot\left<\bJ\right>+\cdots$, where $\pmb{\alpha}$ and $\pmb{\beta}$ depend on properties of the fluctuations such as helicity. The truncation of this series is generally uncertain and often imposed through an ad hoc closure. Here, however, we show that both the full emf and the dynamo coefficients can be evaluated systematically in perturbation theory.

We now compute the emf at leading order using second-order perturbation theory, or quasilinear theory. In this approximation, the fluctuations are taken to be linear, and the large-scale field is generated by their correlations. The second-order magnetic perturbation then obeys
\begin{align}
&\frac{\partial \bB_2}{\partial t} = \nabla\times \left(\bv_0\times \bB_2\right) + \nabla\times \left(\bv_2\times \bB_0\right) + \nabla\times \left(\bv_1\times \bB_1\right),
\label{induction_eq_order_2}
\end{align}
where the subscripts $0$, $1$, and $2$ denote the mean field, the linear perturbation, and the second-order perturbation. As noted above, the large-scale field $\left<\bB_2\right>$ is just $\bB_0$, and the background flow is $\bv_0=R\Omega(R)\hat{\phi}$. The second-order velocity perturbation $\bv_2$ would follow from the quasilinear force equation, but $\nabla\times\left(\bv_2\times\bB_0\right)$ is $\calO\!\left((kR)^{-1}\right)$ and can be neglected in the WKB limit. The term $\nabla\times\left(\bv_0\times\bB_2\right)$ reduces to $\left({\rmd {\Omega}}/{\rmd \ln{R}}\right) B_{2R}\hat{\phi}$, the usual $\Omega$ effect. The nonlinear coupling of the linear perturbations, through $\left<\bv_1\times\bB_1\right>$, produces the emf that drives the quasilinear evolution of the magnetic field on timescales longer than the orbital time. In reality, the perturbations $\bv_1$ and $\bB_1$ are themselves modified by nonlinear mode coupling. Here we neglect those nonlinear terms and use their linear forms in the quasilinear induction equation. As shown below, this is sufficient to capture the dynamo periodicity, though not its amplitude. We briefly discuss this in section~\ref{sec:beyond_QLT} but leave a detailed analysis for future work.

\begin{figure*}
\centering
\includegraphics[width=0.9\textwidth]{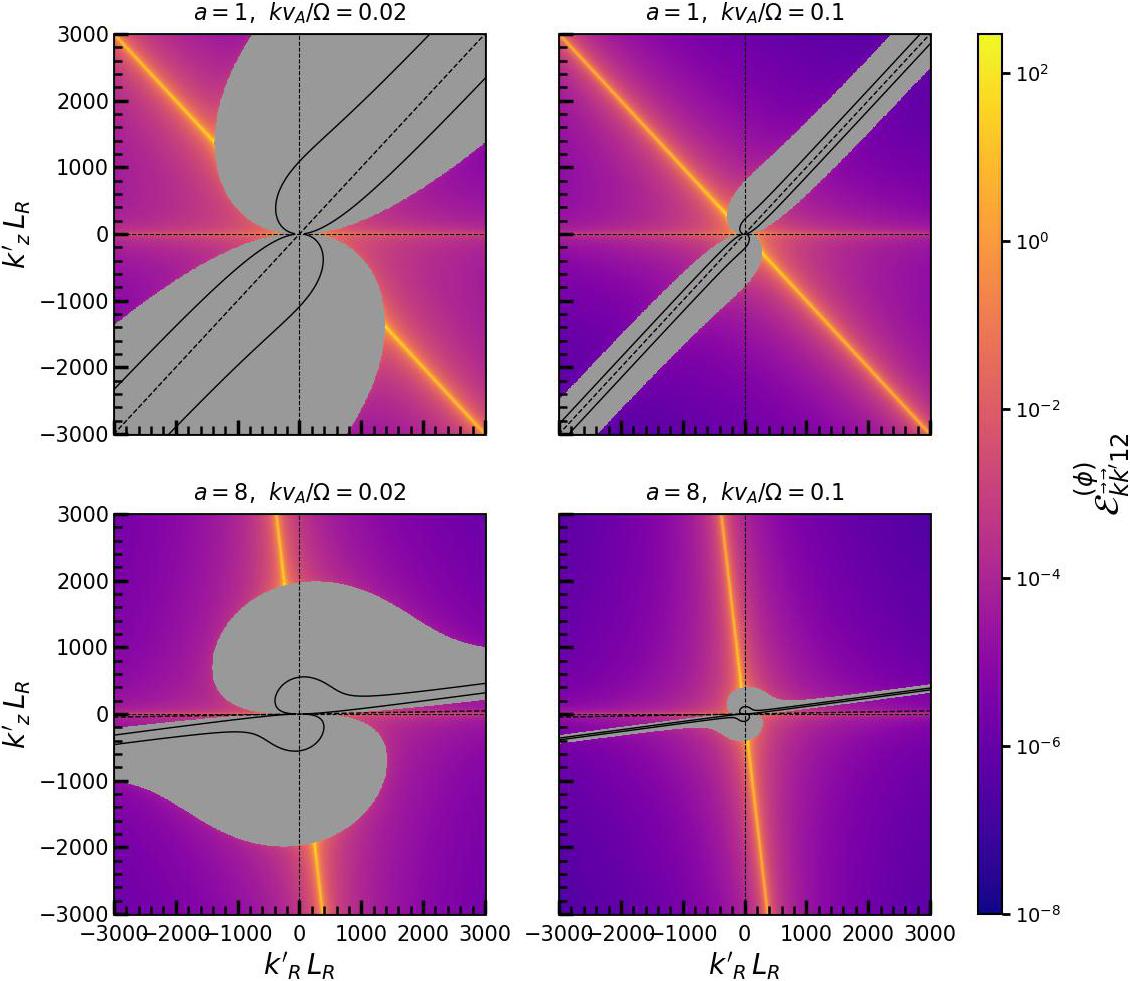}
\caption{Contribution to the toroidal induction as a function of $k'_R$ and $k'_z$ for two different values of $k v_\rmA/\Omega$ and two different aspect ratios (columns). Colors denote the part of the integrand proportional to $\cos{\left(\left(\omega_{1\bk'} - \omega_{2\bk-\bk'}\right)t\right)}$ in equation~(\ref{emf_toroidal_final}). Gray shaded region shows the unstable range of $\bk'$, where either $\omega^2_{2\bk'}$ or $\omega^2_{2\bk-\bk'}$ is negative. The black curve shows the linear instability zone, for which $\omega^2_{2\bk'}<0$, obtained from equation~(\ref{MRI_criterion}). The cyclic dynamo action in our theory arises from the stable region. Note that the primary contribution is from a narrow strip around $\bk'\perp\bk$, i.e., $\bk'\parallel\bB$. A smaller portion of this strip fits in a box of larger $a$, which is why the waves get less phase-mixed, giving rise to longer period cycles at larger $a$. The unstable (stable) region shrinks (expands) more at larger $k v_\rmA/\Omega$, which implies that phase-mixing washes away cyclic behavior at stronger $B_R$ and/or smaller (larger $\bk$) boxes.}
\label{fig:Ik}
\end{figure*}

\subsection{Dynamo equations under the quasilinear approximation}\label{sec:dynamo_eqs}

The analysis of the large-scale dynamo is simplified in the Fourier space of $\bx$. Substituting the Fourier modes of the linear perturbations from equations~(\ref{lin_eig}) in the induction equation~(\ref{induction_eq_order_2}), we obtain the following equation for the evolution of the fields at second order (see Appendix~\ref{App:emf} for a detailed derivation):

\begin{align}
\frac{\partial B_{\bk R}}{\partial t} &= {\bf \calE}_{\bk R}\left(B_R,B_z,t\right),\nonumber\\
\frac{\partial B_{\bk \phi}}{\partial t} &= {\bf \calE}_{\bk \phi}\left(B_R,B_z,t\right) + \frac{\rmd {\Omega}}{\rmd \ln{R}} B_{\bk R},\nonumber\\
\frac{\partial B_{\bk z}}{\partial t} &= {\bf \calE}_{\bk z}\left(B_R,B_z,t\right).
\label{dynamo_eqs_concise}
\end{align}
Here, the wavenumber $\bk$ corresponds to the largest scale field that resides in the system, and the induction $\pmb{\calE}_\bk$ is equal to $i\bk\times \pmb{\varepsilon}_\bk$, with $\pmb{\varepsilon}_\bk = \int \rmd \bk' \bv_{1\bk - \bk'}(t)\times \bB_{1\bk'}(t)$ the Fourier transform of the emf, $\pmb{\varepsilon} = \bv_1 \times \bB_1$. It is evident from equations~(\ref{lin_eig}) that $\bB_{1\bk}$ is proportional to the large-scale field $\bB$. Hereon we drop the subscript $0$ from $\bB_0$ and denote the large-scale/mean field as $\bB$ instead of $\langle\bB\rangle$ or $\bB_0$. We substitute the linear perturbations from equations~(\ref{lin_eig}). For compactness, define $s_{i\bk'} = (c^+_{i\bk'} + c^-_{i\bk'})/2$ and $d_{i\bk'} = (c^+_{i\bk'} - c^-_{i\bk'})/2$ and
\begin{align}
\Psi_{ij}^{\pm}
&\equiv
s_{i\bk'}s_{j,\bk-\bk'}
\pm d_{i\bk'}d_{j,\bk-\bk'},
\nonumber\\
\Theta_{ij}^{\pm}
&\equiv
s_{i\bk'}d_{j,\bk-\bk'}
\pm d_{i\bk'}s_{j,\bk-\bk'},
\nonumber\\
\Delta_{ij}^{\pm}
&\equiv
\frac{1}{\omega_{i\bk'}}
\pm \frac{1}{\omega_{j,\bk-\bk'}},
\nonumber\\
\Lambda_j
&\equiv
\frac{\kappa^2}{2\Omega^2}
-\frac{{\left(\bk'\cdot\bv_\rmA\right)}^2}{\omega_{j,\bk-\bk'}^2}
\frac{\rmd\ln\Omega}{\rmd \ln R},
\nonumber\\
\Gamma_{ij}^{\pm}
&\equiv
1 \pm \frac{\omega_{j,\bk-\bk'}}{2\omega_{i\bk'}}\Lambda_j.
\label{eq:emf_shorthand}
\end{align}

We find the following expression for the toroidal induction:
\begin{align}
{\bf\calE}_{\bk\phi}\left(\bB,t\right)
&=
\Omega \int \rmd \bk'\,
\left(k_z \frac{k'_R}{k'_z} - k_R\right)\,
\bk'\cdot\bB\;
\delta v_{\bk'R}\,\delta v_{\bk-\bk'R}
\nonumber\\
&\times
\sum_{i,j}
\frac{
\calR^{(\phi)}_{\bk\bk'ij}\left(\bB,t\right)
+i\calI^{(\phi)}_{\bk\bk'ij}\left(\bB,t\right)
}{
\omega_{j,\bk-\bk'}^2 - {\left(\bk'\cdot\bv_\rmA\right)}^2
},
\nonumber\\
\calR^{(\phi)}_{\bk\bk'ij}\left(\bB,t\right)
&=
\Psi_{ij}^{-}\,\Gamma_{ij}^{-}
\cos\!\left[\left(\omega_{i\bk'}-\omega_{j,\bk-\bk'}\right)t\right]
\nonumber\\
&+
\Psi_{ij}^{+}\,\Gamma_{ij}^{+}
\cos\!\left[\left(\omega_{i\bk'}+\omega_{j,\bk-\bk'}\right)t\right],
\nonumber\\
\calI^{(\phi)}_{\bk\bk'ij}\left(\bB,t\right)
&=
\Theta_{ij}^{-}\,\Gamma_{ij}^{-}
\sin\!\left[\left(\omega_{i\bk'}-\omega_{j,\bk-\bk'}\right)t\right]
\nonumber\\
&-
\Theta_{ij}^{+}\,\Gamma_{ij}^{+}
\sin\!\left[\left(\omega_{i\bk'}+\omega_{j,\bk-\bk'}\right)t\right].
\label{emf_toroidal_final}
\end{align}

Similarly, the radial induction can be expressed as
\begin{align}
{\bf\calE}_{\bk R}\left(\bB,t\right)
&=
\frac{k_z}{2}
\int \rmd \bk'\,\frac{k'_R}{k'_z}\,\bk'\cdot\bB\;
\delta v_{\bk'R}\,\delta v_{\bk-\bk'R}
\nonumber\\
&\times
\sum_{i,j}
\left[
\calR^{(R)}_{\bk\bk'ij}\left(\bB,t\right)
+i\calI^{(R)}_{\bk\bk'ij}\left(\bB,t\right)
\right],
\nonumber\\
\calR^{(R)}_{\bk\bk'ij}\left(\bB,t\right)
&=
\Psi_{ij}^{-}\,\Delta_{ij}^{-}
\sin\!\left[\left(\omega_{i\bk'}-\omega_{j,\bk-\bk'}\right)t\right]
\nonumber\\
&+
\Psi_{ij}^{+}\,\Delta_{ij}^{+}
\sin\!\left[\left(\omega_{i\bk'}+\omega_{j,\bk-\bk'}\right)t\right],
\nonumber\\
\calI^{(R)}_{\bk\bk'ij}\left(\bB,t\right)
&=
-\Theta_{ij}^{-}\,\Delta_{ij}^{-}
\cos\!\left[\left(\omega_{i\bk'}-\omega_{j,\bk-\bk'}\right)t\right]
\nonumber\\
&+
\Theta_{ij}^{+}\,\Delta_{ij}^{+}
\cos\!\left[\left(\omega_{i\bk'}+\omega_{j,\bk-\bk'}\right)t\right].
\label{emf_radial_final}
\end{align}
The vertical induction $\calE_{\bk z}$ is equal to $-k_R \calE_{\bk R}/k_z$, which ensures that $k_R B_{\bk R} + k_z B_{\bk z} = 0$ at all times.

The $(\rmd\Omega/\rmd\ln R) B_{\bk R}$ term in the toroidal induction represents the $\Omega$ effect, the shear-driven azimuthal stretching of the field lines. $\pmb{\calE}_\bk$ denotes the evolution of the mean field due to the inductive electric field generated by small-scale $\bv$ and $\bB$ fluctuations. Since $\pmb{\calE}_\bk = i\bk\times \pmb{\varepsilon}_\bk$, we have $\calE_{\bk R} = -i k_z \varepsilon_{\bk \phi}$, $\calE_{\bk\phi} = i\left(k_z\varepsilon_{\bk R} - k_R\varepsilon_{\bk z}\right)$, and $\calE_{\bk z} = i k_R\varepsilon_{\bk \phi}$, and the emf $\pmb{\varepsilon}$ can be easily obtained from the above expression for the induction $\pmb{\calE}$. Note that the imaginary part is zero if $d_{i\bk'} = (c^+_{i\bk'} - c^-_{i\bk'})/2$, which is proportional to the net helicity, is zero. The real part, on the other hand, is non-zero even if the helicity $(\propto d_{i\bk'})$ is zero. As we show in Appendix~\ref{App:dynamo_coeff}, the leading order (in $\bk$) term in the imaginary part corresponds to an emf of the form $\pmb{\alpha}\cdot\bB$ with $\pmb{\alpha}$ explicitly depending on the mean helicity of the flow, while the real part results in a $\pmb{\beta}\cdot\bJ$ form where $\pmb{\beta}$ mainly depends on the Maxwell ($\langle \delta \bB \otimes \delta \bB \rangle$) and Faraday ($\langle \delta \bv \otimes \delta \bB \rangle$) stresses.

The induction and the emf are aggregate responses from fluctuations on all scales $k'^{-1}$ smaller than the scale $k^{-1}$ under consideration. A key ingredient of the emf is $i\bk' \delta v_{\bk'R}\, \delta v_{\bk-\bk'R}$, which is the Fourier transform of $\delta v_R \nabla \delta v_R$, a quantity that is related to the derivative of the Reynolds stress responsible for turbulent transport. The coefficient $\delta v_{\bk'R}\, \delta v_{\bk-\bk'R}$ is related to the power-spectrum ${\left|\delta v_{\bk' R}\right|}^2$ as follows:

\begin{align}
\delta v_{\bk'R}\, \delta v_{\bk-\bk'R} = {\left|\delta v_{\bk' R}\right|}^2 - \frac{k_l}{2} \frac{\partial}{\partial k'_l}\left({\left|\delta v_{\bk' R}\right|}^2\right) + ...\,.
\label{conv_PS}
\end{align}
It is well known that the power spectrum for MRI turbulence in an unstratified box is typically of the form 

\begin{align}
{\left|\delta v_{\bk' R}\right|}^2 \approx \calN\, {\left|k'_\perp\right|}^{-\delta},
\label{PS}
\end{align}
with a positive exponent $\delta$ and $\bk'_\perp = \bk'\times \hat{\bB}$ the component of $\bk'$ perpendicular to the large-scale field $\bB$.


\subsection{The quasilinear emf}\label{sec:emf}

Evidently, the emf has a non-trivial dependence on both $\bB$ and $t$. Beyond the overall proportionality to $\bB$, the field dependence enters through the linear eigenfunctions and eigenfrequencies, while the time dependence is either exponential or oscillatory depending on whether the linear instability has saturated. Some insight may nevertheless be gained by expanding the emf order by order in $\bk$. At zeroth order, the emf can be recast in an effective $\alpha$-dynamo form, $\pmb{\varepsilon} = \pmb{\alpha}\left(\bB,t\right) \cdot \bB$, with $\alpha_{ij}$ roughly proportional to $\lim_{\bk\to 0} i\int \rmd\bk'\, (v_{1\bk-\bk'l}\,\varepsilon_{lin}\, k'_j\, v_{1\bk'n})/\omega_{\bk'}$ ($\varepsilon_{lin}$ is the Levi-Civita symbol and should not be confused with the emf). Here, $\omega_{\bk'}$ is of order $\Omega$ on large scales and $\bk'\cdot\bv_\rmA$ on small scales. Since the integral is dominated by large scales, where $\omega_{\bk'}$ varies slowly, we may pull $\omega^{-1}_{\bk'}$ outside the integral, identify it with the turbulent correlation time $t_{\rm corr}$, and write $\pmb{\alpha}$ approximately as $\alpha_{ij} = \alpha\,\delta_{ij}$ with $\alpha \approx t_{\rm corr} \left<\bv_1\cdot\pmb{\omega_1}\right>$. Here we used the identity that $\lim_{\bk\to 0}i\int \rmd\bk'\, v_{1\bk-\bk'l}\,\varepsilon_{lin}\, k'_i\, v_{1\bk'n}$ is equal to $\lim_{\bk\to 0} i\int \rmd\bk'\,\bv_{1\bk-\bk'}\cdot\left(\bk'\times\bv_{1\bk'}\right) = i\int \rmd\bk'\,\bv^\ast_{1\bk'}\cdot\left(\bk'\times\bv_{1\bk'}\right)$, namely the net kinetic helicity $\left<\bv_1\cdot\pmb{\omega_1}\right>$, with $\pmb{\omega_1} = \nabla\times\bv_1$. This argument is heuristic, however, and applies only to a kinematic dynamo, since it ignores self-consistency through the force equation. As shown in Appendix~\ref{App:alpha}, a more rigorous analysis of the toroidal induction using the linear eigenfunctions indeed yields the $\pmb{\alpha}\cdot\bB$ form at zeroth order in $\bk$, with $\alpha_{RR}=\alpha_{zR}$, $\alpha_{Rz}=\alpha_{zz}$, and $\alpha_{RR}+\alpha_{Rz} =\alpha_{RR}+\alpha_{zz} =\alpha_{zR}+\alpha_{Rz} =\alpha_{zR}+\alpha_{zz}\sim \int \rmd t\, \left(\calH_{\rm cur}-\calH_{\rm kin}\right)$, where $\calH_{\rm kin} = \left<\bv_1\cdot\pmb{\omega}_1\right>$ is the net kinetic helicity and $\calH_{\rm cur} = \left<\bJ_1\cdot\bB_1\right>/4\pi\rho_0$ is the net current helicity. Hence, the flow must either be initialized with, or self-consistently develop, non-zero helicity for the $\alpha$ dynamo to operate.

Magnetic-helicity conservation constrains the component of the emf parallel to the mean field. In a statistically steady (weakly resistive) state, small-scale helicity conservation in the inertial range of the MRI turbulence gives
$\pmb{\varepsilon}\cdot\bB=-\nabla\cdot\pmb{\calF}_{\rm SS}/2$
\citep{Vishniac.Cho.01,Ebrahimi.Bhattacharjee.14}, where $\pmb{\calF}_{\rm SS}$ is a small-scale helicity flux. This implies that the zeroth-order $\pmb{\alpha}$ emf must satisfy (see Appendix~\ref{App:helicity_constraint} for a detailed discussion)
\[
\bB\cdot\pmb{\alpha}\cdot\bB=0.
\]
Using the tensor structure derived in Appendix~\ref{App:alpha}, this gives
\begin{align}
\frac{B_\phi}{B_R+B_z}
=
-
\frac{\alpha_{RR}B_R+\alpha_{Rz}B_z}
{\alpha_{\phi R}B_R+\alpha_{\phi z}B_z},
\end{align}
if $\pmb{\alpha}$ is non-zero. Estimating the ratio near marginal stability, where the integrals are dominated by $\bk'\sim\bk'_\rmc$, using
$\bk'_\rmc\cdot\bv_{\rm A}/\Omega=\sqrt{4-\kappa^2/\Omega^2} k'_z/k'$ with
$k'/k'_z=\sqrt{1+a^2}$, and taking $B_R\approx B_z$ yields
\begin{align}
\left|\frac{B_\phi}{B_R}\right|
\sim \sqrt{1+a^2},
\end{align}
which is found to approximately hold true, modulo order unity factors, in the \texttt{Athena++} simulations discussed later. Thus the helicity constraint naturally predicts a toroidal-to-poloidal ratio that increases with the aspect-ratio parameter $a=k'_R/k'_z$, consistent with the dominance of the toroidal field in the MRI dynamo.

The first-order contribution to the emf takes the form of the shear-current effect \citep[][]{Squire.Bhattacharjee.16}. At this order in $\bk$, the emf may be written as $\pmb{\varepsilon} = \pmb{\beta}\left(\bB,t\right)\cdot\bJ$, where $\bJ = (c/4\pi)\,\nabla\times\bB$ is the mean current and $\pmb{\beta}$ is a tensor with non-trivial dependence on $\bB$ and $t$ (see Appendix~\ref{App:beta}). Unlike $\pmb{\alpha}$ which depends on the helicity (difference between the kinetic and current helicities), $\pmb{\beta}$ depends on the Faraday and Maxwell stresses, e.g., $\beta_{R\phi}$ depends on $\langle v_{1\phi} B_{1R} \rangle$ (see Appendix~\ref{App:beta}) and $\beta_{\phi R}$ mainly on $\langle B_{1z} B_{1z}\rangle$ \citep[][]{Mondal.etal.26}. The off-diagonal component $\beta_{R\phi}$ produces a radial emf and hence toroidal induction, while the diagonal component $\beta_{\phi\phi}$ generates a toroidal emf and radial induction. Past MRI saturation, both $\pmb{\alpha}$ and $\pmb{\beta}$ oscillate in time, with periods much longer than the orbital time (see Section~\ref{sec:cycle_period_analytic_QLT}). The mean part of $\beta_{R\phi}$ is positive, whereas that of $\beta_{\phi\phi}$ is negative, implying shear-driven growth of the toroidal field and turbulent diffusion of the radial field (see Appendix~\ref{App:beta}). Note that the axisymmetric modes yield $\varepsilon_R = \beta_{R\phi} J_\phi$, which generates $B_\phi$ in addition to the $\Omega$ effect. The inclusion of non-axisymmetric modes, on the other hand, would yield $\varepsilon_\phi = \beta_{\phi R} J_R$ (in addition to the axisymmetric $\beta_{\phi\phi} J_\phi$) with $\beta_{\phi R}$ depending on the Maxwell and Reynolds stresses \citep[][]{Mondal.etal.26}. This would grow $B_R$ and defeat the turbulent diffusion due to $\beta_{\phi\phi}$ sourced by the axisymmetric modes; in addition, it would contribute a $B_\phi$-draining term, $\varepsilon_R = \beta_{RR} J_R$, to the emf. The growth of the poloidal field due to the emf driven by a radial shear current through the non-axisymmetric modes $(\varepsilon_\phi = \beta_{\phi R} J_R)$ is the original shear-current effect proposed by \citet[][]{Squire.Bhattacharjee.16}. Here we show instead that the toroidal field is driven by the azimuthal shear current through the axisymmetric modes, which gives rise to $\varepsilon_R = \beta_{R\phi} J_\phi$.

Since the helicities are small in an unstratified shearing box, $\pmb{\alpha}$ is small \citep[][]{Shi.etal.16}; the unstratified MRI dynamo is therefore, to leading order, a $\beta$ dynamo rather than a conventional $\alpha$ dynamo. Still, as shown above, a small but non-zero $\pmb{\alpha}$ is required to yield a saturated toroidal-to-poloidal ratio through small-scale helicity conservation. The $\beta$ dynamo satisfies the helicity constraint, $\pmb{\varepsilon}\cdot\bB=-\nabla\cdot\pmb{\calF}_{\rm SS}/2$ (see Appendix~\ref{App:helicity_constraint}). Note that $\pmb{\alpha}$ and $\pmb{\beta}$ are only the lowest-order terms in a $\bk$ expansion of the emf. For practical calculations, we caution against truncating this expansion, since it may converge slowly when $\bk'\sim\bk$ or at large $t$. It is therefore preferable to evaluate the full emf directly from equations~(\ref{emf_toroidal_final}) and (\ref{emf_radial_final}), which can be done at little additional computational cost.

There are several key features of the emf:

\begin{itemize}
    \item Under the quasilinear approximation, the emf is generated by interference between the linear eigenmodes. In the unstable regime, it grows exponentially at roughly twice the linear MRI growth rate. In the stable regime, after linear saturation, it oscillates due to interference between the oscillatory modes.
    
    \item Averaged over the oscillations, the toroidal induction is generally positive, while the radial induction is negative for the axisymmetric modes (see Appendix~\ref{App:beta}). Thus the radial (or vertical) field decays after the linear growth phase, whereas the toroidal field can grow to large amplitudes. This reflects the action of rotational shear through the explicit dependence of the toroidal induction on $\Omega$ and $\rmd\ln\Omega/\rmd\ln R$ (see equation~[\ref{emf_toroidal_final}]), in addition to the usual $\Omega$-effect term $\left(\rmd\ln\Omega/\rmd\ln R\right) B_{\bk R}$ in $\partial B_{\bk\phi}/\partial t$ (see equation~[\ref{dynamo_eqs_concise}]). Non-axisymmetric modes would grow $B_R$ as well \citep[][]{Squire.Bhattacharjee.16}.
    
    \item In the toroidal induction, the coefficients of $\cos{\left(\left(\omega_{i\bk'} - \omega_{j\bk-\bk'}\right)t\right)}$ and $\cos{\left(\left(\omega_{i\bk'} + \omega_{j\bk-\bk'}\right)t\right)}$ differ; likewise for the oscillatory sine terms in the radial induction. This enables long-period cyclic behavior, which, as shown in Section~\ref{sec:cycle_period_analytic_QLT}, is driven by the $\cos{\left(\left(\omega_{i\bk'} - \omega_{j\bk-\bk'}\right)t\right)}$ term. The key reason is that $\Delta\omega_{\bk'}=\omega_{1\bk'} - \omega_{2\bk-\bk'}$ varies only weakly with $\bk'$, leading to inefficient phase mixing.
    
    \item The prefactor of the $\cos{\left(\left(\omega_{1\bk'} - \omega_{2\bk-\bk'}\right)t\right)}$ term in the toroidal induction tends to $1/2$ for $\bk'\gg \bk'_\rmc$, whereas the prefactor of the corresponding $\sin{\left(\left(\omega_{1\bk'} - \omega_{2\bk-\bk'}\right)t\right)}$ term in the radial induction tends to zero. Since the quasilinear cycles are driven by beats at large $\bk'$ (see Section~\ref{sec:cycle_period_analytic_QLT}), the toroidal field displays a much stronger cyclic signal than the radial field.
    
    \item The toroidal field amplitude exceeds the radial one by a factor $\sim k'/k'_z = \sqrt{1+a^2}$. Hence the cyclic behavior is more pronounced in $B_\phi$ than in $B_R$, for $a=L_z/L_R>1$.
\end{itemize}

For dynamo cycles to emerge, the linear MRI must first saturate. Although quasilinear evolution can in principle amplify the large-scale field until the MRI criterion in equation~(\ref{MRI_criterion}) is no longer satisfied, zero-net-flux simulations indicate that QLT substantially overestimates the saturated field strength. In practice, saturation likely occurs earlier through nonlinear mode coupling: the perturbations $\bv_{1\bk-\bk'}$ and $\bB_{1\bk'}$ that generate the emf can couple to neighboring $\bk'$ modes, producing a cascade, or get modified through the coupling of the eigenmodes at the same $\bk'$. These interactions modify the perturbation amplitudes through convolutions of the linear eigenfunctions evaluated at the resonances, $\sum_i \bk'_i = 0$ and $\sum_i \omega_{j\bk'_i} = 0$. Turbulent cascades as well as the non-linear coupling between stable (exponentially decaying) and unstable (exponentially growing) modes can shut off the linear instability and give rise to a saturated Maxwell stress. Moreover, parasitic instabilities like the Kelvin-Helmholtz and tearing mode instabilities may also saturate the MRI \citep[][]{Pessah.Goodman.09}. The overall topic of non-linear saturation of the linear MRI lies beyond the current quasilinear framework and will be addressed in a follow-up paper. 

In this paper, we therefore assume that such beyond-quasilinear effects saturate the linear MRI by driving the growth rate of the unstable mode, $\gamma_\bk = \sqrt{-\omega^2_{2\bk}}$, to zero, so that the unstable region becomes marginally stable. Equivalently, we set $\omega_{2\bk}=0$ in the linearly unstable region of $\bk'$. This is a crude approximation invoked to follow through with the current quasilinear calculation; a more general non-linear treatment of the MRI unstable region is deferred to future work. Since, within QLT, the long-period cycles of the large-scale field arise from the beat term $\sim \exp{\left[-i\left(\omega_{1\bk'} - \omega_{2\bk-\bk'}\right)t\right]}$ (see Section~\ref{sec:cycle_period_analytic_QLT}), they are sourced only by the stable region, where the eigenmodes are oscillatory, i.e., $\omega^2_{2\bk'}\geq 0$.

\subsection{Quasilinear evolution of the large-scale field}\label{sec:B_evol}

\begin{figure*}[t!]
\centering
\includegraphics[width=1\textwidth]{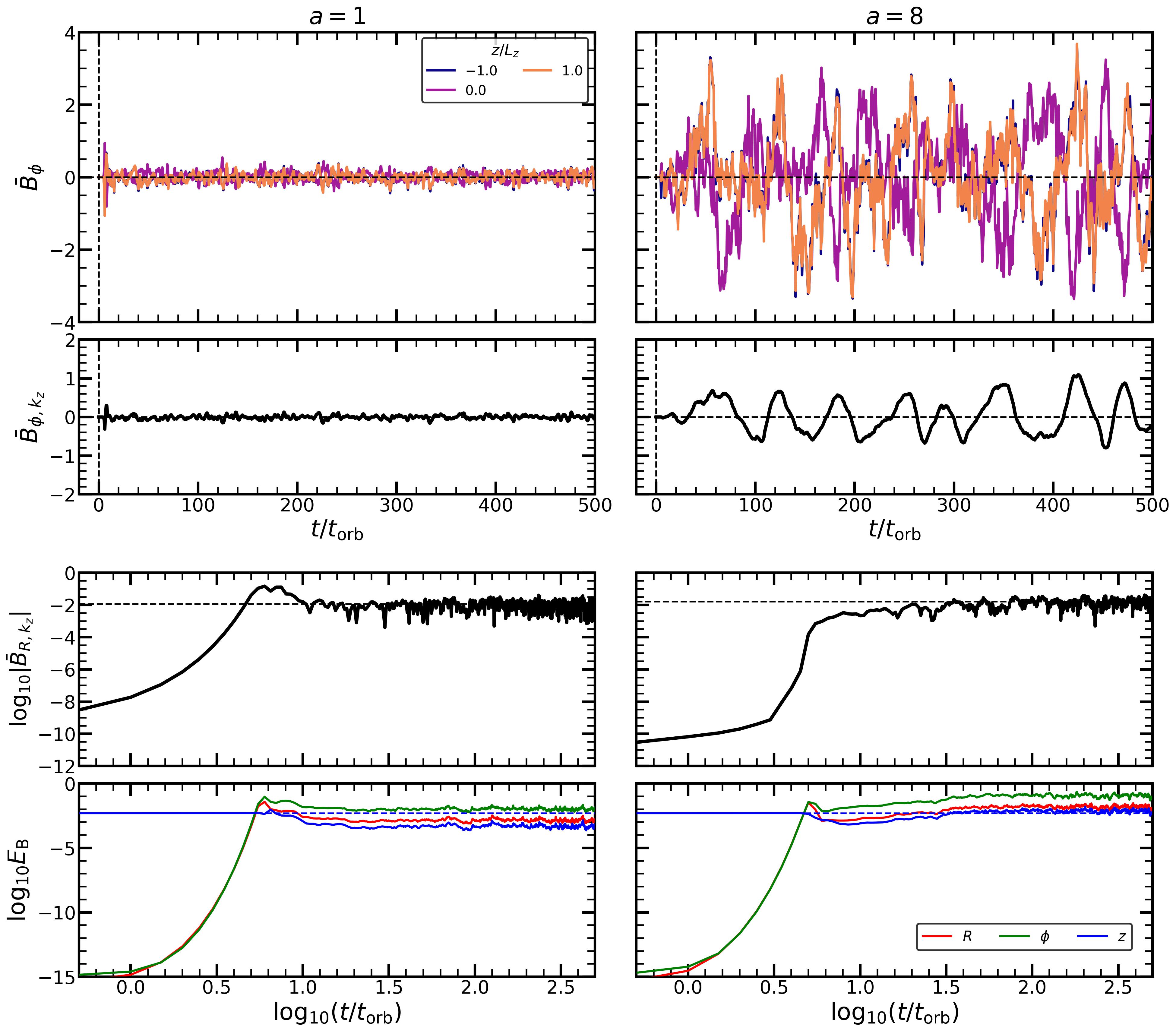}
\caption{Athena++ simulations: Mean toroidal and radial fields and magnetic energies for aspect ratios $a = L_z/L_R = 1$ and $8$. The first row depicts the mean toroidal field $\bar{B}_\phi$, averaged over $R$ and $\phi$, for different $z/L_z$, the second row shows the smallest $k_z$ (largest scale) mode of $\bar{B}_\phi$, the third row shows $\bar{B}_R$ (black horizontal dashed line indicates the long-time mean), and the fourth row plots the magnetic energies $E_\rmB$ along the radial, azimuthal and vertical directions (blue horizontal dashed line denotes the initial magnetic energy) as a function of $t/t_{\rm orb}$. Note that the saturated magnetic energies are an order of magnitude larger and the toroidal field cycles are stronger for $a=8$ than for $a=1$, although the large-scale radial field is of similar strength.}
\label{fig:Bphi_Emag_vs_t_sim}
\end{figure*}

To evolve the large-scale $\bB$, we evaluate the induction terms in equations~(\ref{dynamo_eqs_concise}) using equations~(\ref{emf_toroidal_final}) and (\ref{emf_radial_final}), and then numerically integrate equations~(\ref{dynamo_eqs_concise}). In the axisymmetric WKB limit, all three induction terms depend on $B_R$ and $B_z$ but not on $B_\phi$, so the toroidal equation decouples from the radial and vertical ones. The poloidal equations yield a monotonic decay of $B_R$ and $B_z$ to very small values, because the mean part of the corresponding emf is negative (Appendix~\ref{App:beta}). This is, however, purely an outcome of the turbulent diffusion caused by the axisymmetric modes. \texttt{Athena++} simulations show an initial poloidal decay (after linear MRI) followed by saturation at small but non-zero values, which shows that the emf driven by the non-axisymmetric modes (primarily through shear-current effect \citep[][]{Squire.Bhattacharjee.16,Mondal.Bhat.23,Mondal.etal.26}) can balance out the axisymmetric turbulent diffusion. Since the degree of variation of $B_R$ and $B_z$ is much less than that of $B_\phi$ in the simulations (past MRI saturation), in this quasilinear treatment, we evolve $B_\phi$, holding $B_R$ and $B_z$ fixed at small constant values satisfying $k_R B_R + k_z B_z = 0$. We explicitly check the validity of this long-term near-constancy of the large-scale poloidal field in the simulations, as shown in Fig.~\ref{fig:Bphi_Emag_vs_t_sim}.

We then integrate the azimuthal mean-field equation (the second of equations~[\ref{dynamo_eqs_concise}]) using the toroidal induction in equation~(\ref{emf_toroidal_final}), with initial condition $B_\phi=0$. The velocity power spectrum is taken to be ${\left|\delta v_{\bk' R}\right|}^2 = \calN {\left(k^{'2}_{\perp} + k^2\right)}^{-\delta/2}$, where $k = \left|\bk\right| = 2\pi\sqrt{1/L^2_R + 1/L^2_z}$ is of order the smallest wavenumber in the system. Using Parseval's theorem, $\int\rmd\bk\,\left|\bv_{\bk}\right|^2 = 2\pi\int\rmd\bx\,\left|\bv\right|^2 = 2\pi L_R L_z \langle\bv^2\rangle$, together with $\bB_\bk \approx \bB/(k_R k_z)$ ($k_R = 2\pi/L_R$, $k_z = 2\pi/L_z$), we estimate the normalization $\calN$. The variance $\langle\bv^2\rangle$ is measured from the simulation. We adopt $k_{R,\rm max} = 64\,k_R$ and $k_{z,\rm max} = 64\,a\,k_z$, with $k_z = k_R/a$ ($L_z = aL_R$), and verify that doubling the resolution changes the result negligibly. We assume that the helicity of the perturbations is zero, i.e., $c^+_{i\bk'}=c^{-}_{i\bk'}$, in accordance with the non-helical nature of the flow in zero-net-flux simulations \citep[][]{Shi.etal.16}. The field obtained this way is in the inertial frame; the shearing-frame $B_\phi$ is then found by subtracting the $(\rmd\Omega/\rmd\ln R)\,B_R\,t$ term (see the second of equations~[\ref{dynamo_eqs_concise}]).

Fig.~\ref{fig:Bphi_vs_t_a} shows the large-scale toroidal field in the shearing frame as a function of $t/t_{\rm orb}$, where $t_{\rm orb}=2\pi/\Omega$, for different $a$ and $k v_\rmA/\Omega$. Here $v_\rmA=\sqrt{(B_R^2+B_z^2)/4\pi\rho_0}$ is the Alfv\'en speed. The toroidal field exhibits beats, with characteristic periods ranging from $\sim 10\,t_{\rm orb}$ to a few $100\,t_{\rm orb}$. At fixed $a$, the beats are most pronounced for particular values of $k v_\rmA/\Omega$. These cycles arise from the $\cos{\left[\left(\omega_{1\bk'}-\omega_{2\bk-\bk'}\right)t\right]}$ term in the toroidal induction (equation~[\ref{emf_toroidal_final}]); as shown in Section~\ref{sec:cycle_period_analytic_QLT}, the long periods result from the slow variation of $\Delta\omega_{\bk'}=\omega_{1\bk'}-\omega_{2\bk-\bk'}$ with $\bk'$ at large $\bk'$.

The oscillation amplitude generally increases with $k v_\rmA/\Omega$, consistent with the overall linear dependence of the emf on $\bB$. For $a\lesssim 2$, the dynamo shows several comparable periods and is therefore more stochastic, whereas at larger $a$ the cycles become cleaner and stronger. The longer periods at larger $a$ reflect the approximate scaling of the cycle period as $\sqrt{1+a^2}$. Since the eigenfrequencies scale as $k'_z/k'$, they are more sparsely distributed in $\bk'$-space at larger $a$, which reduces phase mixing and enhances the cyclic behavior. The larger amplitudes at larger $a$ also partly follow from the factor $\omega^2_{j\bk-\bk'}-(\bk'\cdot\bv_\rmA)^2$ in the denominator, that scales as $\sim (k_z'^2/2k'^2)\,\kappa^2\sim a^{-2}\kappa^2$ at $k'\gg k$ (see equations~[\ref{MRI_freqs}] and [\ref{emf_toroidal_final}]). By contrast, increasing $k v_\rmA/\Omega$ shortens the overall cycle period and makes the evolution more stochastic. Overall, these trends reflect more efficient phase mixing at smaller $a$ and larger $B_R$.

The trends can be understood from Fig.~\ref{fig:Ik}, which shows $\calE^{(\phi)}_{\bk\bk'12}$, the part of the integrand proportional to $\cos{\left[\left(\omega_{1\bk'}-\omega_{2\bk-\bk'}\right)t\right]}$, as a function of $k_R'$ and $k_z'$ for $a=1$ and $8$ and $k v_\rmA/\Omega=0.02$ and $0.1$; the gray region marks the unstable modes. Since the quasilinear emf that yields the cycles is dominated by the stable (oscillatory) region, three features matter. First, the dominant contribution comes from a narrow strip around $\bk'\perp\bk$, i.e. $\bk'\parallel\bv_\rmA$, because the MRI turbulent power spectrum peaks for field-aligned modes (equation~[\ref{PS}]). Second, a smaller fraction of this strip lies inside the box when the box is taller, i.e. at larger $a$. Third, the stable region expands as $k v_\rmA/\Omega$ increases. Hence more modes contribute in smaller boxes and/or for stronger fields, which enhances phase mixing and makes the dynamo less cyclic and more stochastic. Overall, the long cycles originate from the MRI dispersion relation, namely the weak $\bk$ dependence of the difference between the fast and slow shear-Alfv\'en frequencies, while the trends with aspect ratio and field strength reflect the combined effects of this dispersion relation and the anisotropic, field-aligned MHD turbulence.

\subsection{Origin of the dynamo cycles in QLT}\label{sec:cycle_period_analytic_QLT}

In QLT, oscillations of the mean magnetic field arise from the interference between two normal modes with frequencies $\omega_{1\bk'}$ and $\omega_{2\bk'}$. Each $\bk'$ mode contributes at the sum and difference frequencies, $\Sigma\omega_{\bk'}=\omega_{1\bk'}+\omega_{2\bk'}$ and $\Delta\omega_{\bk'}=\omega_{1\bk'}-\omega_{2\bk'}$. The long cycle is set by the low-frequency branch $\Delta\omega_{\bk'}$. Since $\Delta\omega_{\bk'}$ varies only weakly with $\bk'$ for $k'\gtrsim k'_\rmc$, the spread $\Delta\omega_\infty-\Delta\omega_\rmc$ is small (see Fig.~\ref{fig:lin_MRI}), producing slow beats. Here $\Delta \omega_{\rmc} = \Delta\omega_{k'\approx 2 k'_\rmc}$ and $\Delta \omega_{\infty} = \Delta\omega_{\bk'\to\infty}$ approximately denote the minimum and maximum values of $\Delta \omega_{\bk'}$.

The dominant contribution to the radial emf, which drives the toroidal field, is

\begin{align}
\varepsilon_{\bk R}(t) &\sim \int \rmd^2 \bk'\, \left[f_{1\bk'}\cos{\left(\Delta \omega_{\bk'} t\right)} + f_{2\bk'}\cos{\left(\Sigma \omega_{\bk'} t\right)}\right]\,,
\end{align}
where $f_{1\bk'}$ and $f_{2\bk'}$ are functions of $\bk'$ involving the linear eigenfunctions and eigenfrequencies (see the first of equations~[\ref{emf_toroidal_final}]). Assuming $k<k'$, we transform from $\bk'$ to $\left(s=k'_z/k',{\bk}'\cdot\hat{\bv}_\rmA\right)$ and then from ${\bk}'\cdot\hat{\bv}_\rmA$ to $\Delta\omega_{\bk'}$ at fixed $s$, so that

\begin{align}
\varepsilon_{\bk R}(t) &\sim \int \rmd s\, \left[\int_{\Delta \omega_\rmc}^{\Delta \omega_\infty} \rmd \Delta \omega\, g_1\left(\Delta \omega\right)\, \cos{\left(\Delta \omega t\right)}\right.\nonumber\\
&\left.+ \int_{\Sigma \omega_\rmc}^{\Sigma \omega_\infty} \rmd \Sigma \omega\, g_2\left(\Sigma \omega\right)\, \cos{\left(\Sigma\omega t\right)}\right],
\label{emf_analytic}
\end{align}
where $g_1\left(\Delta \omega\right) = f_{1\bk'}/J_1$ and $g_2\left(\Sigma \omega\right) = f_{2\bk'}/J_2$, with $J_1 = \left|\frac{s,\;{\bk}'\cdot\hat{\bv}_\rmA}{k'_R,\;k'_z}\right|\left(\partial\Delta\omega_{\bk'}/\partial{\bk}'\cdot\hat{\bv}_\rmA\right)$ and $J_2 = \left|\frac{s,\;{\bk}'\cdot\hat{\bv}_\rmA}{k'_R,\;k'_z}\right|\left(\partial\Sigma\omega_{\bk'}/\partial{\bk}'\cdot\hat{\bv}_\rmA\right)$ the Jacobians of the transformation. Since $f_{i\bk'}$ contains terms scaling as $\omega^{-1}_{i \bk'}$ and ${\left(\omega_{i\bk'}\omega_{j\bk-\bk'}\right)}^{-1}$, it is largest near the marginally stable wavenumber $\bk'_\rmc$, where $\omega_{2 \bk'} = 0$ (see equation~[\ref{MRI_criterion}]). However, the $\sin{\left(\omega_{2\bk'}t\right)}/\omega_{2\bk'}$ term in the emf, arising through $v_{1\bk'\phi}$ and $B_{1\bk'R}$, can produce a large secular growth of the toroidal field at late times, which is not seen in the simulations. We avoid this by noting that nonlinear wave-wave interactions should broaden the marginal mode to $\omega'^2_{2\bk'_\rmc} = \omega^2_{2\bk'_\rmc-\bk''} \approx -\left(\partial\omega'^2_{2\bk'}/\partial k'\right)\Delta k'$, where $\Delta k' = k'-k'_\rmc$ is defined by $\left(\partial\omega'^2_{2\bk'}/\partial k'\right)\Delta k' \approx \left(\partial^2\omega'^2_{2\bk'}/\partial k'^2\right){\left(\Delta k'\right)^2}/2$ (with derivatives evaluated at fixed $s$ and at $k'_\rmc$). Assuming that three-wave interactions dominate this broadening, the linear dispersion relation yields $\Delta k' \approx k'_\rmc$. Thus, after accounting for nonlinear broadening, the minimum wavenumber that contributes to the quasilinear emf is approximately $2 k'_\rmc$. Although, to evolve $B_\phi$, we consider both MRI stable and unstable regions to compute the emf (assuming $\omega_{2\bk'}=0$ in the unstable region), here we focus on the stable region, since this is where the cycles arise in QLT.

The long-period cycle comes from the first term in equation~(\ref{emf_analytic}), which we denote by $\varepsilon_{1\bk R}$. Integrating by parts yields
\begin{align}
\varepsilon_{1 \bk R}(t) &= \int_{\Delta \omega_\rmc}^{\Delta \omega_\infty} \rmd \Delta \omega\, g_1\left(\Delta \omega\right)\, \cos{\left(\Delta \omega t\right)} \nonumber\\
&= \frac{1}{t}\int\rmd s\, \Bigg[{g_1\left(\Delta\omega_{\infty}\right) \sin{\left(\Delta\omega_{\infty} t\right)} - g_1\left(\Delta\omega_{\rmc}\right) \sin{\left(\Delta\omega_{\rmc} t\right)}}\nonumber\\
&\quad- \int_{\Delta \omega_{\rmc}}^{\Delta \omega_{\infty}}\rmd\Delta\omega\,\frac{\rmd g_1(\Delta\omega)}{\rmd \Delta \omega} \sin{\left(\Delta \omega t \right)}\Bigg],
\end{align}
and repeated integration by parts yields the asymptotic series
\begin{align}
&\varepsilon_{1 \bk R}(t) = \frac{1}{t}\nonumber\\
&\times\int\rmd s\,\left[\sum_{n=0}^{\infty} \frac{g^{(2n)}_1\left(\Delta\omega_{\infty}\right) \sin{\left(\Delta\omega_{\infty} t\right)} - g^{(2n)}_1\left(\Delta\omega_{\rmc}\right) \sin{\left(\Delta\omega_{\rmc} t\right)}}{t^{2n}}\right.\nonumber\\
&\left.+ \sum_{n=0}^{\infty} \frac{g^{(2n+1)}_1\left(\Delta\omega_{\infty}\right) \cos{\left(\Delta\omega_{\infty} t\right)} - g^{(2n+1)}_1\left(\Delta\omega_{\rmc}\right) \cos{\left(\Delta\omega_{\rmc} t\right)}}{t^{2n+1}}\right],
\end{align}
where $g^{(m)}_1(\Delta \omega)$ is the $m^{\rm th}$ derivative of $g_1$ with respect to $\Delta \omega$. At long times,
\begin{align}
&\varepsilon_{1\bk R}(t)
\approx \frac{1}{t}\int \rmd s \,
\Big[
g_1(\Delta\omega_{\infty}) \sin(\Delta\omega_{\infty} t)
- g_1(\Delta\omega_{\rm c}) \sin(\Delta\omega_{\rm c} t)
\Big]
\nonumber\\
&\approx \frac{1}{t}\int \rmd s \,
\Bigg[
\big(g_1(\Delta\omega_{\infty})-g_1(\Delta\omega_{\rm c})\big)
\nonumber\\
&\quad\times
\sin\!\left(\frac{\Delta\omega_{\infty}+\Delta\omega_{\rm c}}{2}t\right)
\cos\!\left(\frac{\Delta\omega_{\infty}-\Delta\omega_{\rm c}}{2}t\right)
\nonumber\\
&\quad+
\big(g_1(\Delta\omega_{\infty})+g_1(\Delta\omega_{\rm c})\big)
\nonumber\\
&\quad\times
\cos\!\left(\frac{\Delta\omega_{\infty}+\Delta\omega_{\rm c}}{2}t\right)
\sin\!\left(\frac{\Delta\omega_{\infty}-\Delta\omega_{\rm c}}{2}t\right)
\Bigg].
\label{emf_analytic}
\end{align}
with the second term dominant because $\Delta\omega_\infty$ and $\Delta\omega_\rmc$ are close. The same analysis applies to $\varepsilon_{2\bk R}$ (second term of equation~[\ref{emf_analytic}]) after replacing $g_1\to g_2$ and $\Delta\to\Sigma$.

Thus $\varepsilon_{1\bk R}$ exhibits fast oscillations modulated by slow
beats, with characteristic periods
\begin{equation}
T_{\Delta 1}
=
\frac{4\pi}
{\left|\Delta\omega_{\infty}-\Delta\omega_{\rm c}\right|},
\quad
T_{\Delta 2}
=
\frac{4\pi}
{\Delta\omega_{\infty}+\Delta\omega_{\rm c}}.
\end{equation}
As shown in Appendix~\ref{App:Deltaomega_large_k}, expanding the eigenfrequencies in the limit $\bk'\cdot\bv_{\rm A}/\Omega\gg1$ gives
\begin{align}
\Delta\omega_{\bk'}
&\equiv
\omega_{1\bk'}-\omega_{2\bk'}
\nonumber\\
&\approx
\frac{2k'_z}{k'}\Omega
\left[
1+
\frac{k_z'^2}{4k'^2}
{\left(
\frac{\bk'\cdot\bv_{\rm A}}{\Omega}
\right)}^{-2}
\left(
2-\frac{\kappa^2}{\Omega^2}
+\frac{\kappa^4}{8\Omega^4}
\right)
+\cdots
\right].
\end{align}
Thus $\Delta\omega_{\bk'}$ approaches
$\Delta\omega_\infty=2k'_z\,\Omega/k'$ at large $k'$, so small-scale modes
oscillate at nearly the same beat frequency. This weak dispersion is
the key ingredient behind a long period cycle.

\begin{figure}
\centering
\includegraphics[width=0.95\textwidth]{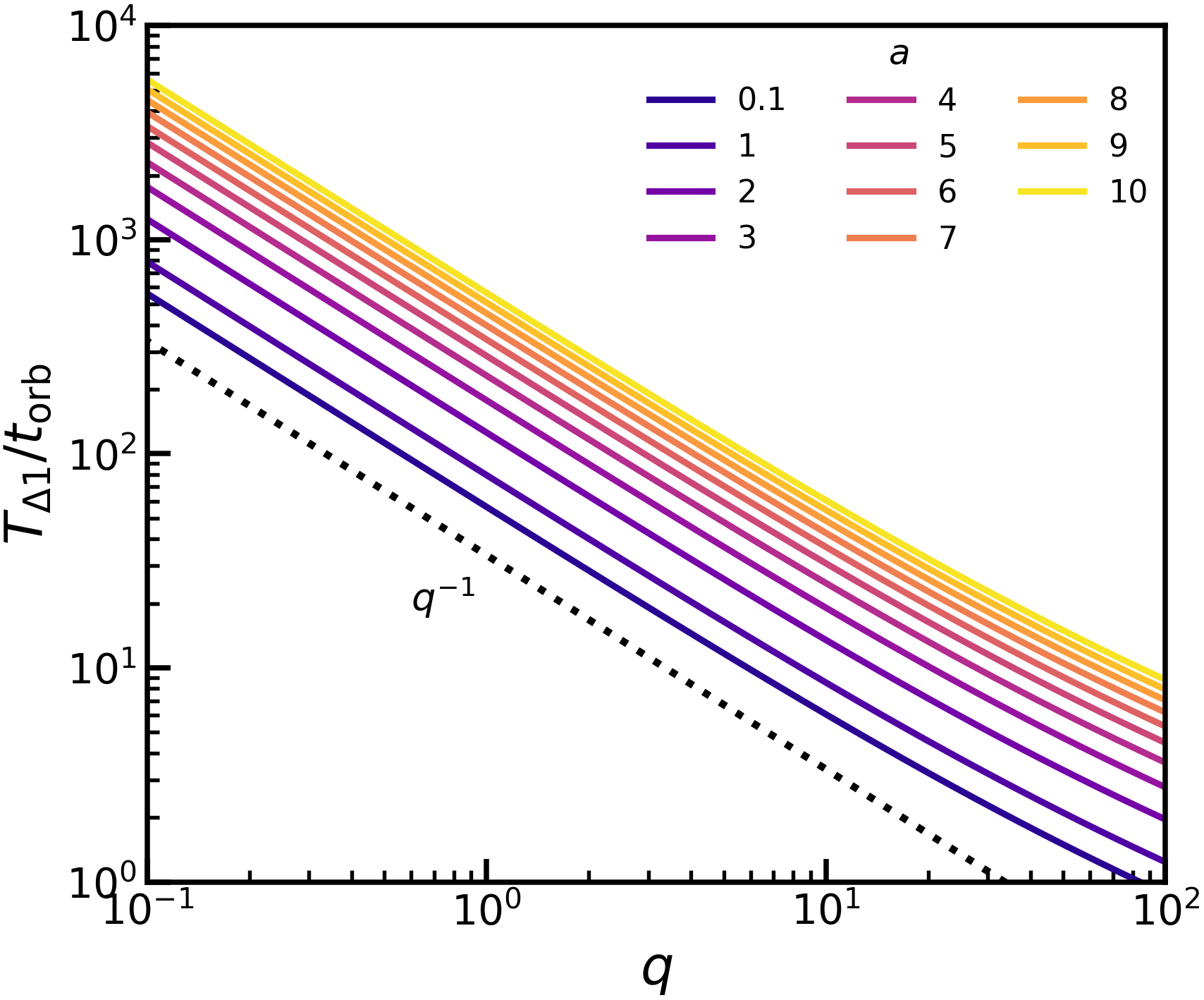}
\caption{Cycle period $T_{\Delta 1}$ (computed using equation~[\ref{T_cycle_exact}]) vs the shear, $q = -\rmd\ln\Omega/\rmd\ln R$, for modes with $\bk'\parallel \bk$, and for different aspect ratios $a = L_z/L_R$ as shown. The period scales as $\sim \sqrt{1+a^2}$ and decreases with increasing shear as $\sim q^{-1}$, but is still quite long at large $q$, provided $a$ is large.}
\label{fig:Tbeat_vs_q}
\end{figure}

Writing $\kappa^2=(4-2q)\Omega^2$, with
$q=-\rmd\ln\Omega/\rmd\ln R$, and evaluating the expansion through the next
order at $k'=2k'_{\rm c}$ using
${(\bk'_{\rm c}\cdot\bv_{\rm A}/\Omega)}^2=2q\,k_z'^2/k'^2$, we find the approximate expressions for $T_{\Delta 1}$ and $T_{\Delta 2}$ (see Appendix~\ref{App:Deltaomega_large_k}):
\begin{align}
T_{\Delta 1}^{\rm app}
&\approx
\frac{8192}{q(144-q)}
\frac{k'}{k'_z}\,
t_{\rm orb},
\nonumber\\
T_{\Delta 2}^{\rm app}
&\approx
\frac{8192}{16384+q(144-q)}
\frac{k'}{k'_z}\,
t_{\rm orb},
\label{T_cycle_analytic}
\end{align}
where $t_{\rm orb}=2\pi/\Omega$. These compact expressions agree well with
the exact periods obtained from the unexpanded frequencies, up to
$q\sim30$. To compute the exact values, write
$\Delta\omega_{\rm c}^{\rm ex}=(k'_z/k')\,\Omega F(q)$, with
$s=4-2q$, $\chi=[1+128q/s^2]^{1/2}$, and
\begin{equation}
F(q)=
\left|
\left[
8q+\frac{s}{2}(1+\chi)
\right]^{1/2}
-
\left[
8q+\frac{s}{2}(1-\chi)
\right]^{1/2}
\right|.
\end{equation}
Then the exact expressions are given by
\begin{align}
T_{\Delta 1}^{\rm ex}
&=
\frac{2}{|F(q)-2|}
\frac{k'}{k'_z}\,
t_{\rm orb},
\nonumber\\
T_{\Delta 2}^{\rm ex}
&=
\frac{2}{F(q)+2}
\frac{k'}{k'_z}\,
t_{\rm orb}.
\label{T_cycle_exact}
\end{align}
For Keplerian shear, $q=3/2$, the exact dispersion relation gives
\begin{align}
T_{\Delta 1}^{\rm ex}
&\approx
37.6\,t_{\rm orb}
\times
\begin{cases}
\sqrt{1+a^2}, & \bk'\parallel\bk,\\
\sqrt{1+a^{-2}}, & \bk'\perp\bk,
\end{cases}
\nonumber\\
T_{\Delta 2}^{\rm ex}
&\approx
0.493\,t_{\rm orb}
\times
\begin{cases}
\sqrt{1+a^2}, & \bk'\parallel\bk,\\
\sqrt{1+a^{-2}}, & \bk'\perp\bk.
\end{cases}
\end{align}
Here we have used the fact that $k'/k'_z = \sqrt{1+k_R'^2/k_z'^2}$. The longer period,
$T_{\Delta1}$, sets the upper limit to the quasilinear beat periods, while
$T_{\Delta2}$ sets the lower limit. In Fig.~\ref{fig:Tbeat_vs_q}, we plot the long period $T_{\Delta1}$ (for modes with $\bk'\parallel \bk$) as a function of $q$ for different $a$ from $0.1$ to $10$. Evidently, the period decreases with increasing shear as $q^{-1}$ (see equation~\ref{T_cycle_analytic}) but is still substantially long ($\sim 10\,t_{\rm orb}$) even for $q$ as large as $10$, provided $a>1$. The presence of nearly degenerate eigenfrequencies is all that is required for the existence of long period cycles in our quasilinear framework. Sheared rotational flow is one such case, but persistent cyclic dynamos may also appear when the flow has non-zero shear but zero net rotation in a stratified background \citep[][]{Squire.etal.25}. Kelvin-Helmholtz dynamos occurring in shear-layers \citep[][]{Tripathi.etal.26} are known to exhibit cyclic behavior as well. This suggests that the MRI dynamo is not the only channel for cycles \citep[][]{Squire.etal.25}. 

\begin{figure*}[t!]
\centering
\begin{subfigure}{0.29\textwidth}
    \centering
    \includegraphics[width=\linewidth]{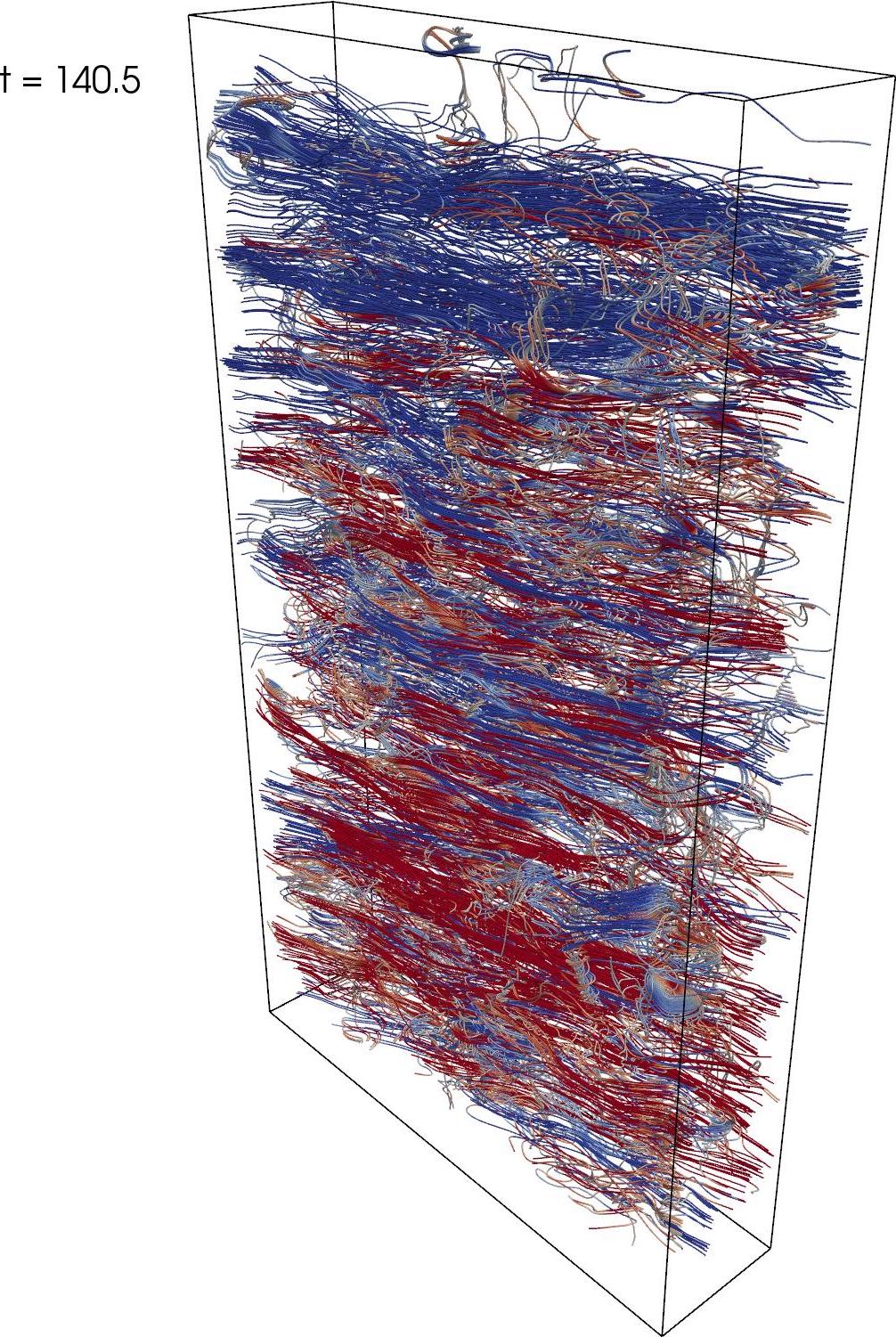}
\end{subfigure}
\hfill
\begin{subfigure}{0.29\textwidth}
    \centering
    \includegraphics[width=\linewidth]{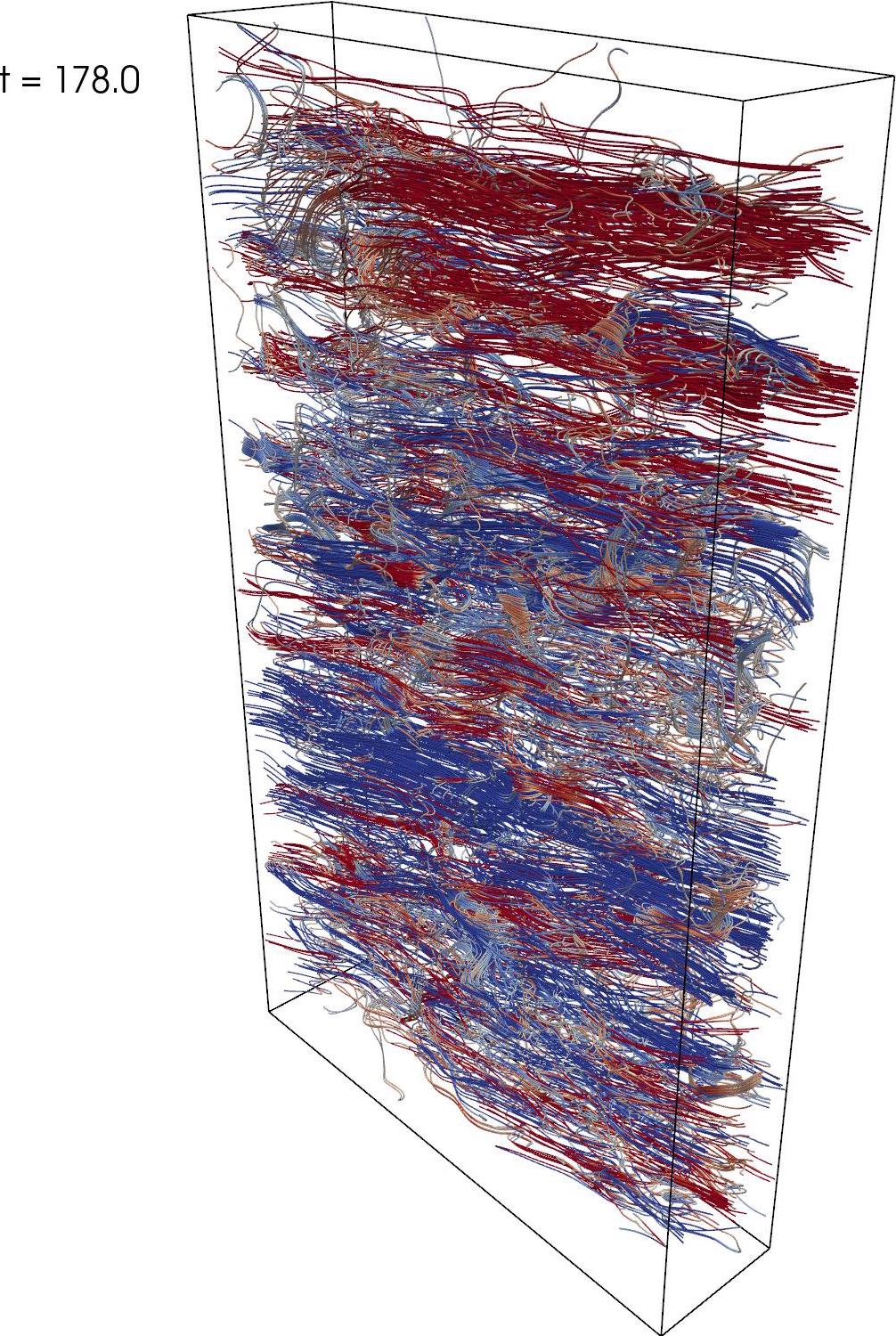}
\end{subfigure}
\hfill
\begin{subfigure}{0.325\textwidth}
    \centering
    \includegraphics[width=\linewidth]{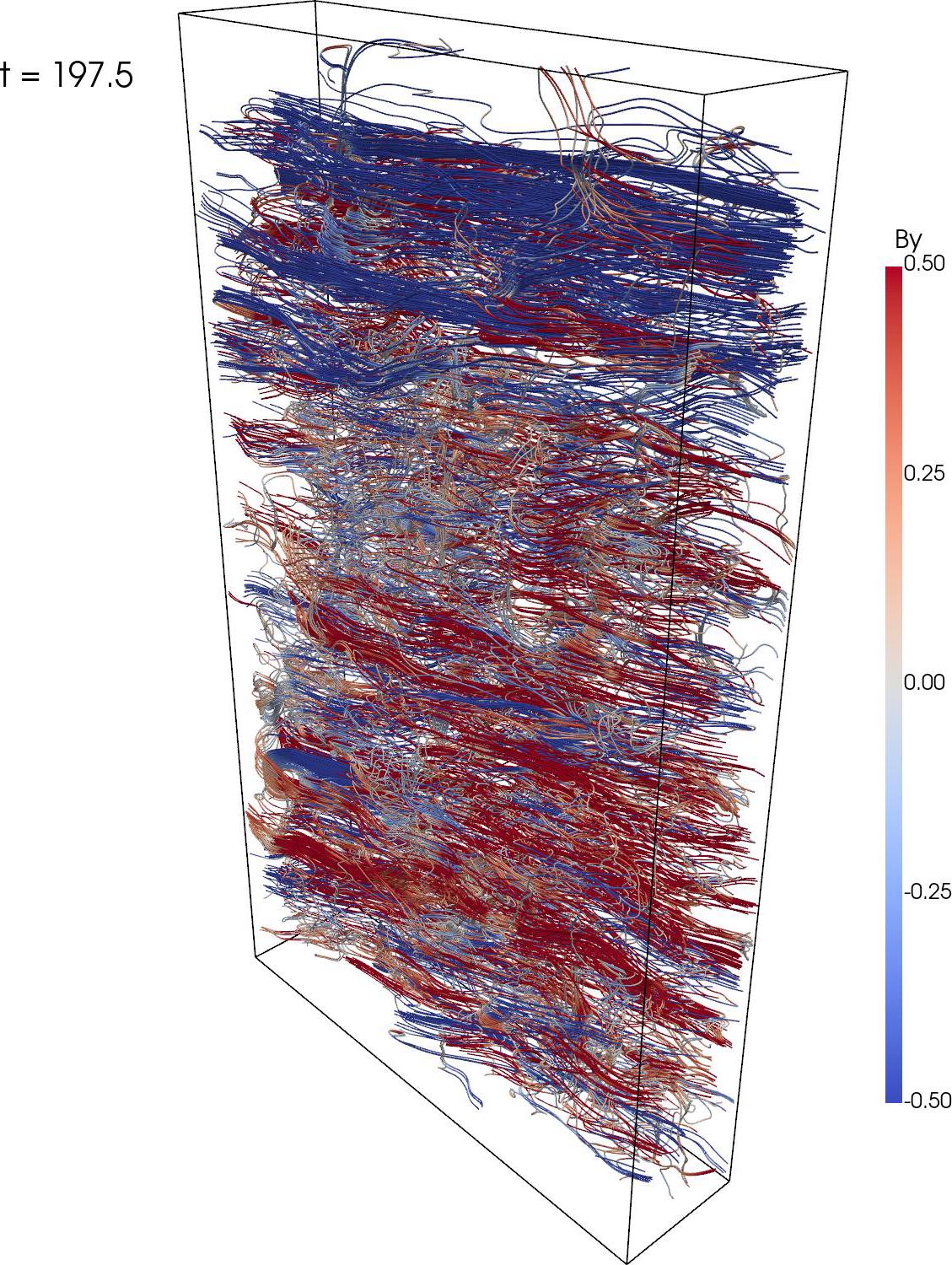}
\end{subfigure}
\caption{Athena++ simulations: streamlines of $\bB$, colored by the mean toroidal field $\bar{B}_y = \bar{B}_\phi = \langle B_\phi\rangle_{R,\phi}$, at three different times in a tall $(a = 7)$ box. Note the periodic large-scale field reversals. The cycle period is estimated to be roughly $60\,t_{\rm orb}$.}
\label{fig:B_streamlines}
\end{figure*}

\begin{figure*}
\centering
\includegraphics[width=0.95\textwidth]{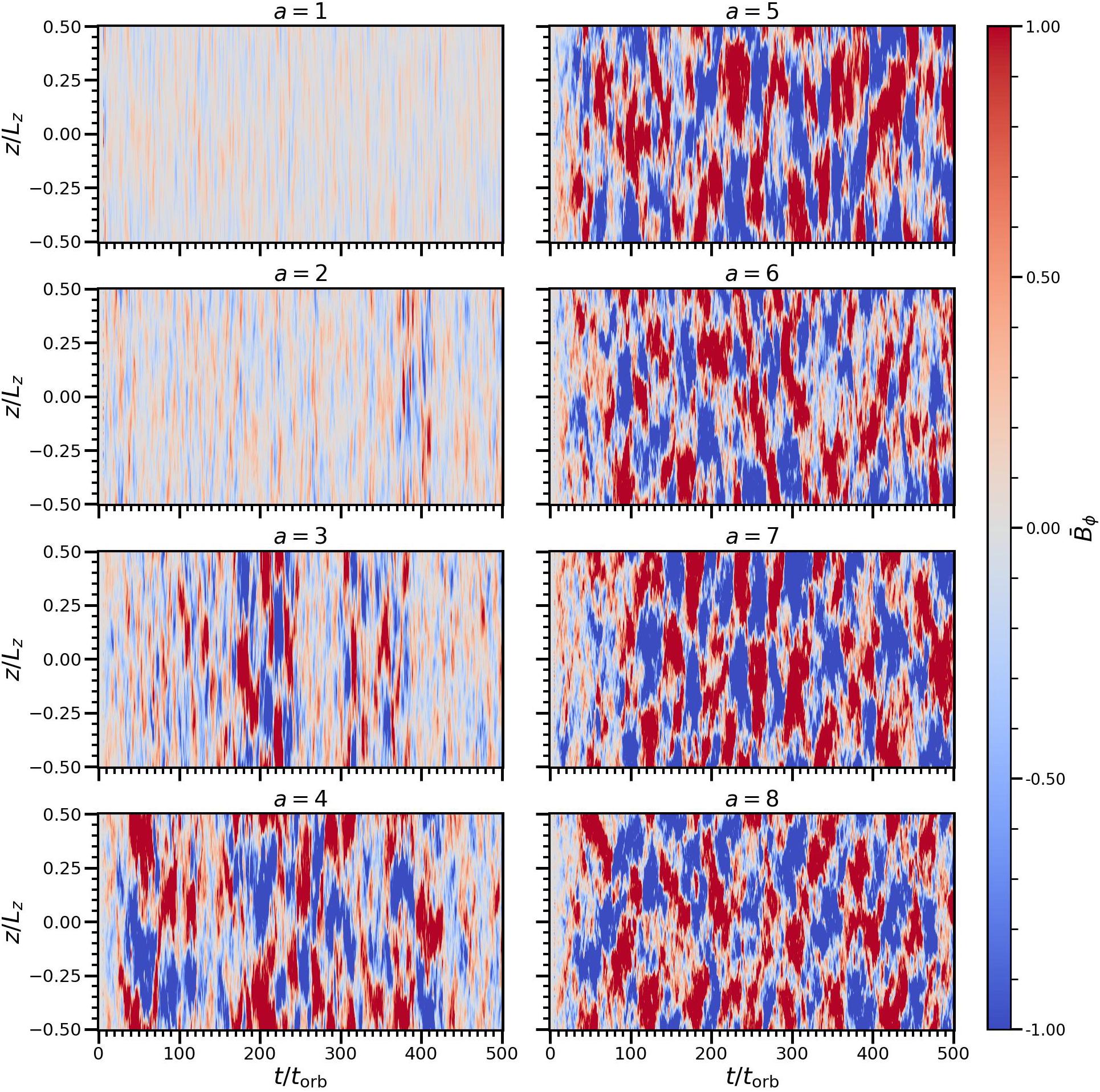}
\caption{Butterfly diagram from Athena++ simulations ($N_R = 64$, $N_\phi = 256$, $N_z=64 \,(a=1), 128\,(1<a<4), 256\,(4\leq a \leq 8)$): the mean toroidal field $\bar{B}_\phi$, averaged over $R$ and $\phi$, as a function of $z/L_z$ and $t/t_{\rm orb}$, for different aspect ratios $a = L_z/L_R$ from $1$ to $8$. Dynamo cycles appear at $a\gtrsim 3$.}
\label{fig:butterfly_diag_sim}
\end{figure*}




\begin{figure*}[t]
\centering
\begin{subfigure}[t]{0.85\textwidth}
\centering
\includegraphics[width=\textwidth]{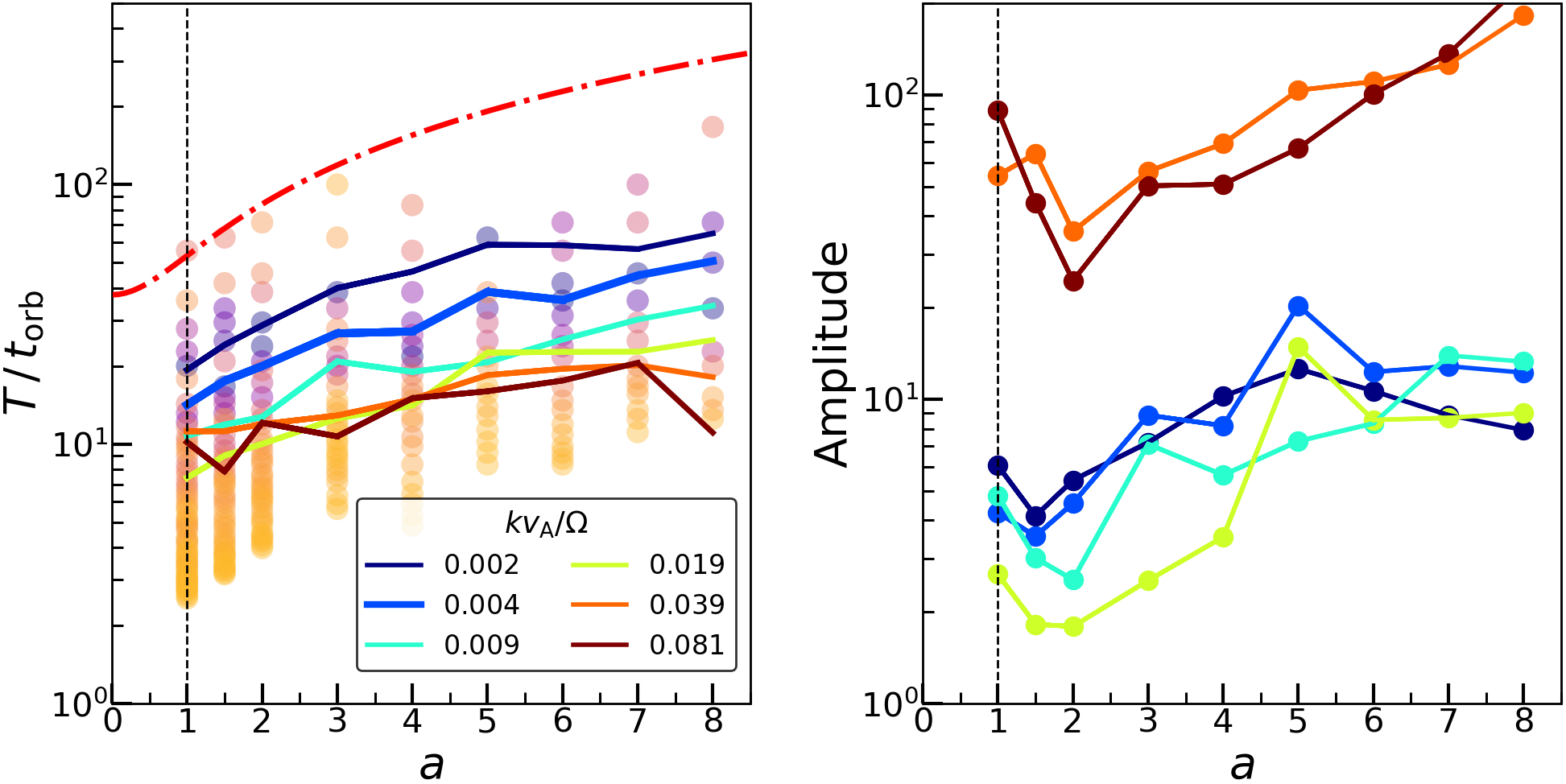}
\caption{Quasilinear theory.}
\label{fig:dom_per_amp_vs_a}
\end{subfigure}

\vspace{0.4cm}

\begin{subfigure}[t]{0.85\textwidth}
\centering
\includegraphics[width=\textwidth]{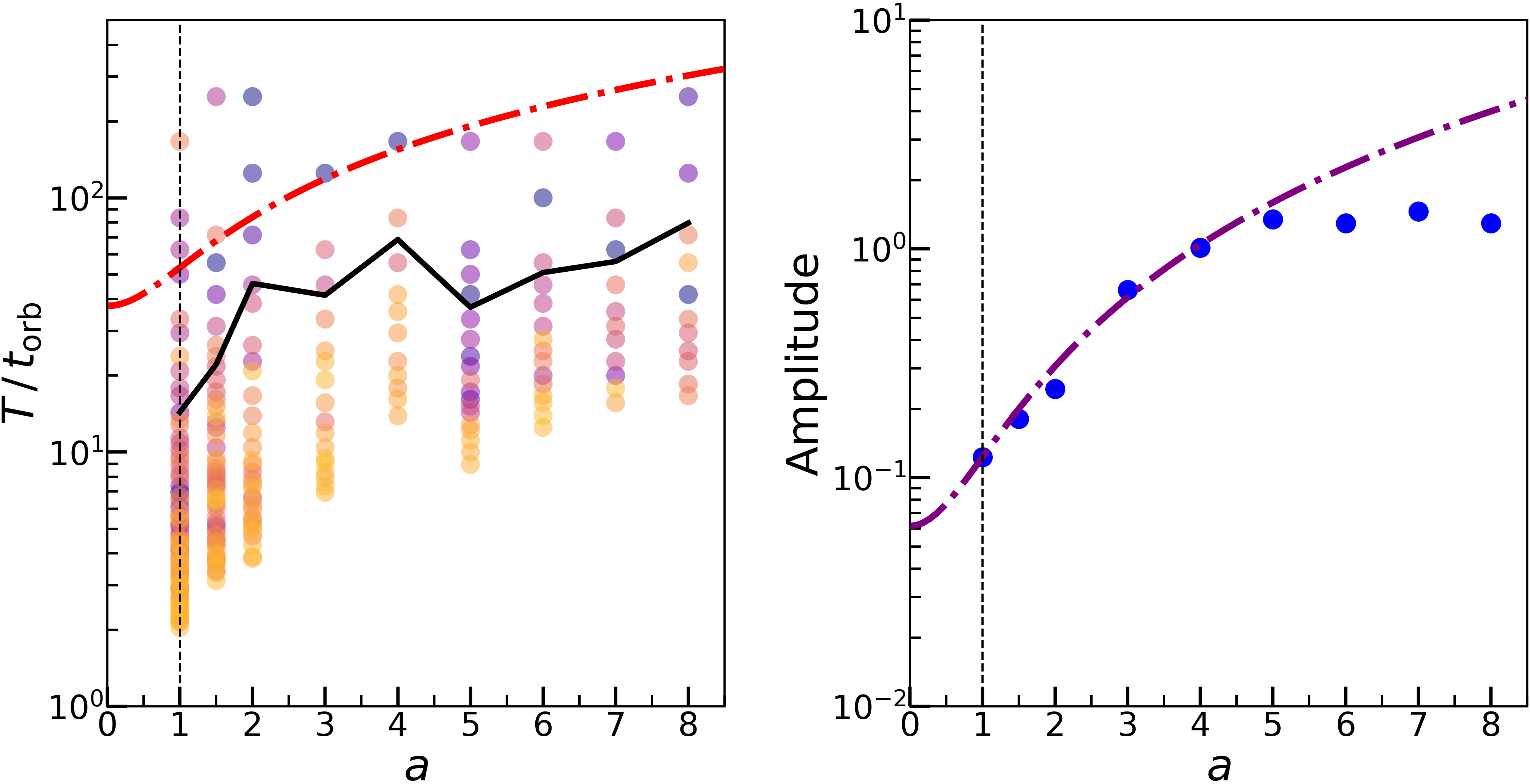}
\caption{\texttt{Athena++} simulations.}
\label{fig:per_amp_sim_vs_a}
\end{subfigure}
\caption{Fourier periods $T$ (left subpanels) and amplitudes (right subpanels) of the large-scale toroidal-field oscillations as functions of aspect ratio $a$. The amplitudes are defined as the dispersion $\sqrt{\langle B_\phi^2-{\langle B_\phi\rangle}^2\rangle}$, where the brackets denote the temporal average of $\bar{B}_{\phi k_z}(t)$ between $t=50$ and $500\,t_{\rm orb}$. Circles denote periods with amplitudes above $0.01$ times the peak amplitude, and the solid lines show the amplitude-weighted period, $T=\sum_i w_i T_i/\sum_i w_i$ ($T_i$ and $w_i$ are the period and amplitude of each Fourier mode). The dot-dashed red line shows the analytic estimate for $T_{\Delta 1}$ from equations~(\ref{T_cycle_exact}). Panel (a) shows the quasilinear theory results for different values of $k v_\rmA/\Omega$ ($kv_\rmA/\Omega=0.002$ closely corresponds to the simulation-based $B_R$ and $B_z$ values for $a = 8$, while $kv_\rmA/\Omega=0.019$ corresponds to that for $a = 1$). Panel (b) shows the corresponding \texttt{Athena++} results. In both cases, the dominant periods scale as $\sim\sqrt{1+a^2}$, while the amplitude grows approximately as $a^2$ at small $a$ before flattening at $a\gtrsim 5$. Smaller $k v_\rmA/\Omega$ in QLT shows longer cycles.}
\label{fig:per_amp_compare}
\end{figure*}

Getting back to the magnetorotational dynamo, the emf given in equation~(\ref{emf_analytic}) decays as $\sim \sin{\left(\Delta\omega_\infty t\right)}/t$ or $\sin{\left(\Delta\omega_\rmc t\right)}/t$, allowing the toroidal field to persist for a long time. Since $\varepsilon_{\bk z}$ has the same form as $\varepsilon_{\bk R}$ with $g_1$ replaced by $g'_1=(k'_z/k'_R)g_1$, and since $\partial B_\phi/\partial t = \calE_{\bk \phi}$ with $\calE_{\bk \phi} = i\left(k_z \varepsilon_{\bk R} - k_R \varepsilon_{\bk z}\right)$, the slowly varying part of $B_\phi$ is
\begin{align}
B_\phi(t) &= \int_0^t \rmd t'\calE_{\bk \phi}(t')\nonumber\\
&\approx \int \rmd s\, \left[h_1\left(\Delta\omega_\infty\right){\rm Si}\left(\Delta\omega_\infty t\right) - h_2\left(\Delta\omega_\rmc\right){\rm Si}\left(\Delta\omega_\rmc t\right)\right],
\end{align}
where ${\rm Si}(t)=\int_0^t \rmd x\,\sin(x)/x$ is the Sine integral, and $h_1=i\left(k_z g_1-k_R g'_1\right)=i\left(k_z-k_Rk'_z/k'_R\right)g_1$. Neglecting the fast contribution from $\varepsilon_{2\bk R}$ and using ${\rm Si}(t\to\infty)=\pi/2-\cos(t)/t$, we obtain
\begin{align}
B_\phi(t)
&\approx \frac{1}{t}\int \rmd s \,
\Big[
h_1(\Delta\omega_{\rm c}) \cos(\Delta\omega_{\rm c} t)
- h_1(\Delta\omega_{\infty}) \cos(\Delta\omega_{\infty} t)
\Big]
\nonumber\\
&\approx \frac{1}{t}\int \rmd s \,
\Bigg[
\big(h_1(\Delta\omega_{\rm c})-h_1(\Delta\omega_{\infty})\big)
\nonumber\\
&\quad\times
\cos\!\left(\frac{\Delta\omega_{\infty}+\Delta\omega_{\rm c}}{2}t\right)
\cos\!\left(\frac{\Delta\omega_{\infty}-\Delta\omega_{\rm c}}{2}t\right)
\nonumber\\
&\quad-
\big(h_1(\Delta\omega_{\rm c})+h_1(\Delta\omega_{\infty})\big)
\nonumber\\
&\quad\times
\sin\!\left(\frac{\Delta\omega_{\infty}+\Delta\omega_{\rm c}}{2}t\right)
\sin\!\left(\frac{\Delta\omega_{\infty}-\Delta\omega_{\rm c}}{2}t\right)
\Bigg].
\end{align}
Again the second term dominates because $\Delta\omega_\infty\approx\Delta\omega_\rmc$. Thus the toroidal field shows the same beating behavior as the emf, with slow modulation at the period $T_{\Delta1}$ in equation~(\ref{T_cycle_analytic}), as seen in Fig.~\ref{fig:Bphi_vs_t_a}. Phase mixing remains weak because $\Delta\omega_{\bk'}$ varies only slowly with $k'$. Physically, one mode is slightly faster because rotation reinforces magnetic tension in one branch and opposes it in the other. Their small frequency difference then drives the long-period modulation of the mean field. This is the quasilinear mechanism for the dynamo cycles. The simulations, however, are not limited to this channel and also contain additional frequencies generated by higher-order mode coupling.

\subsection{Beyond quasilinear effects}\label{sec:beyond_QLT}

So far we have considered the perturbations to be linear, i.e., they scale as $\exp{\left[-i\omega_{j\bk} t\right]}$. However, in a saturated turbulent state, this is not true in general. The perturbations themselves couple to the other modes (at the same as well as other $\bk$), which causes them to oscillate at not only their own eigenfrequencies but also frequencies that are linear combinations of the eigenfrequencies of the different modes. A generic perturbation therefore scales as

\begin{align}
&\delta Q^i_{\bk} (t) = \sum_j c_{1j\bk} U^i_{1j\bk} \exp{\left[-i\omega_{j\bk} t\right]}\nonumber\\
&+ \sum_j\sum_l\sum_n \int \rmd \bk' c_{2j\bk\bk'}  U^i_{2jln\bk\bk'} \exp{\left[-i\left(\omega_{l\bk'} + \omega_{n\bk-\bk'}\right) t\right]}\nonumber\\
&+ ...,
\end{align}
where the superscript $i$ represents the index of the state vector, the subscript $j,l$ and $n$ represent the index of the eigenmode, and $U^i_{1j\bk}$, $U^i_{2j\bk\bk'}$, etc. are complicated convolutions involving the linear eigenfunction $U^i_{j\bk}$. In the linear regime, we have $U^i_{1j\bk} = U^i_{j\bk}$ and $U^i_{2j\bk\bk'}$ is comparatively small. Upon substituting these nonlinear perturbations in the induction equation, it is not hard to see that the emf gains terms that scale as $\exp{\left[-i\left(\omega_{i\bk'} + \omega_{l\bk-\bk''} + \omega_{n\bk-\bk'+\bk''}\right) t\right]}$, $\exp{\left[-i\left(\omega_{i\bk''} + \omega_{j\bk'-\bk''} + \omega_{l\bk-\bk'''} + \omega_{n\bk-\bk'+\bk'''}\right) t\right]}$ and so on, in addition to the quasilinear terms $\sim \exp{\left[-i\left(\omega_{i\bk'} + \omega_{j\bk-\bk'}\right) t\right]}$ ($\bk'$, $\bk''$, $\bk'''$, etc. are the dummy wavenumbers to be integrated over). In section~\ref{sec:cycle_period_analytic_sim}, we explicitly show the presence of these frequencies in the power-spectrum generated from the simulation data. These additional terms have two key effects: (1) they give rise to a non-zero contribution to the cyclic dynamo from the unstable region in $\bk'$, which does not arise under the quasilinear approximation, and (2) they make the overall dynamo more turbulent and stochastic.

\section{Numerical simulations}\label{sec:sim}

Let us now compare the theory to simulations. The simulation setup is a local shearing box, which is well-described by the WKB approximation. However, besides being fully nonlinear rather than quasilinear, the simulations are more general than the theory in the sense that they (i) include both axisymmetric and non-axisymmetric (shearing waves) modes, (ii) show a turbulent fluctuation of $B_R$ and $B_z$ which are not constant, as assumed in the analytical theory, and (iii) self-consistently generate the turbulent power-spectrum of fluctuations, something that is beyond the scope of QLT. Despite these differences, it is interesting to see if some of the basic properties of the mean (large-scale) field in the simulations are reproduced by QLT. 

\subsection{Governing equations and code description}\label{sec:sim_desc}

We solve the compressible MHD equations for an unstratified shearing box corotating with the disk at an orbital frequency $\Omega\hat{\bz}$. We concern ourselves with high $\beta$ plasmas where the flow is fairly incompressible $(v \sim v_\rmA < c_\rms)$. Unlike the linear and quasilinear calculations, which we perform in a global inertial frame in cylindrical coordinates, the simulations are performed in a local shearing (non-inertial) frame in Cartesian coordinates. The equations include the continuity and induction equations (see equations~[\ref{MHD_eqs}]) along with the isothermal equation of state, $P = \rho c^2_\rms$, and the force equation that includes the non-inertial Coriolis and centrifugal forces: 

\begin{align}
\frac{\partial \bv}{\partial t} + \bv \cdot \nabla\bv &= -\frac{1}{\rho}\nabla\left(P+\frac{B^2}{8\pi}\right) + \frac{1}{4\pi \rho}\bB \cdot \nabla \bB\nonumber\\
&- 2\Omega\,\hat{\bz}\times\bv + 2q\Omega^2 x\, \hat{\bx},
\end{align}
where $\hat{\bx}$ refers to the radial direction, $\hat{\by}$ the azimuthal direction parallel to disk rotation, and $\hat{\bz}$ the vertical direction, while $q = -\rmd\ln\Omega/\rmd\ln R$ is the shear parameter. Since we study an unstratified shearing box, we have omitted the vertical tidal acceleration $\Omega^2 z \hat{\bz}$. We also neglect the non-ideal viscous and resistive effects.

We use the \texttt{Athena++} MHD code to perform our numerical simulations in a Cartesian grid, adopting the CTU integrator, a second order piecewise linear flux reconstruction, and the HLLE Riemann solver as the basic algorithms. The orbital advection scheme is implemented to increase the computational efficiency and reduce the mismatch of flux integrals on the radial faces due to remap. The conservation of the vertical magnetic flux is ensured to machine precision by using a remapping method.

\begin{figure*}[t!]
\centering
\includegraphics[width=1\textwidth]{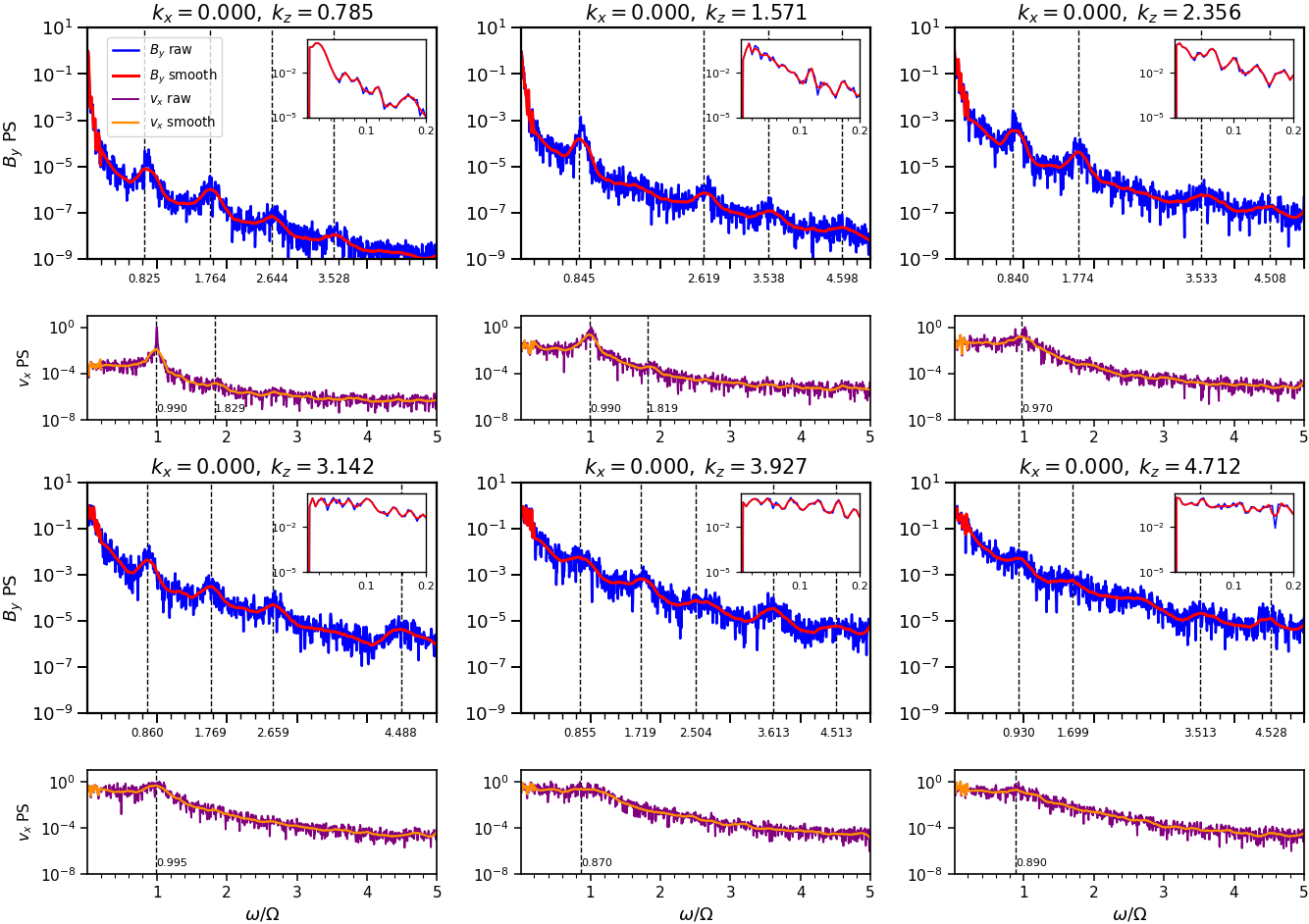}
\caption{Athena++ simulation $(a=8)$: Temporal power spectra $\left|B_y\right|^2 = \left|B_\phi\right|^2$ and $\left|v_x\right|^2 = \left|v_R\right|^2$ for the $a=8$ case for $k_x = 0$, $k_y = 0$ and six different values of $k_z$ $(k_z=2\pi n_z/L_z = \pi n_z/4,\; n_z = 1(1)6,\; L_z = 8)$. The vertical dashed lines indicate the peak frequencies of the smoothed power spectrum. The insets show the low frequency peaks corresponding to long period cycles (particularly in $B_y = B_\phi$ at small $k_z$). The peaks are more (less) pronounced at smaller (larger) $k_z$. These frequencies correspond to the linear eigenfrequencies and linear combinations thereof (see text in section~\ref{sec:cycle_period_analytic_sim}). The eigenfrequencies scale as $k_z/k\sim \sqrt{1+a^2}$, which is why they are well-spaced in $\bk$ and therefore protected from the stochastic zoo of turbulence at smaller $\bk$ and larger $a$.}
\label{fig:temporal_PS_sim}
\end{figure*}

\begin{figure*}[h!]
\centering
\includegraphics[width=1\textwidth]{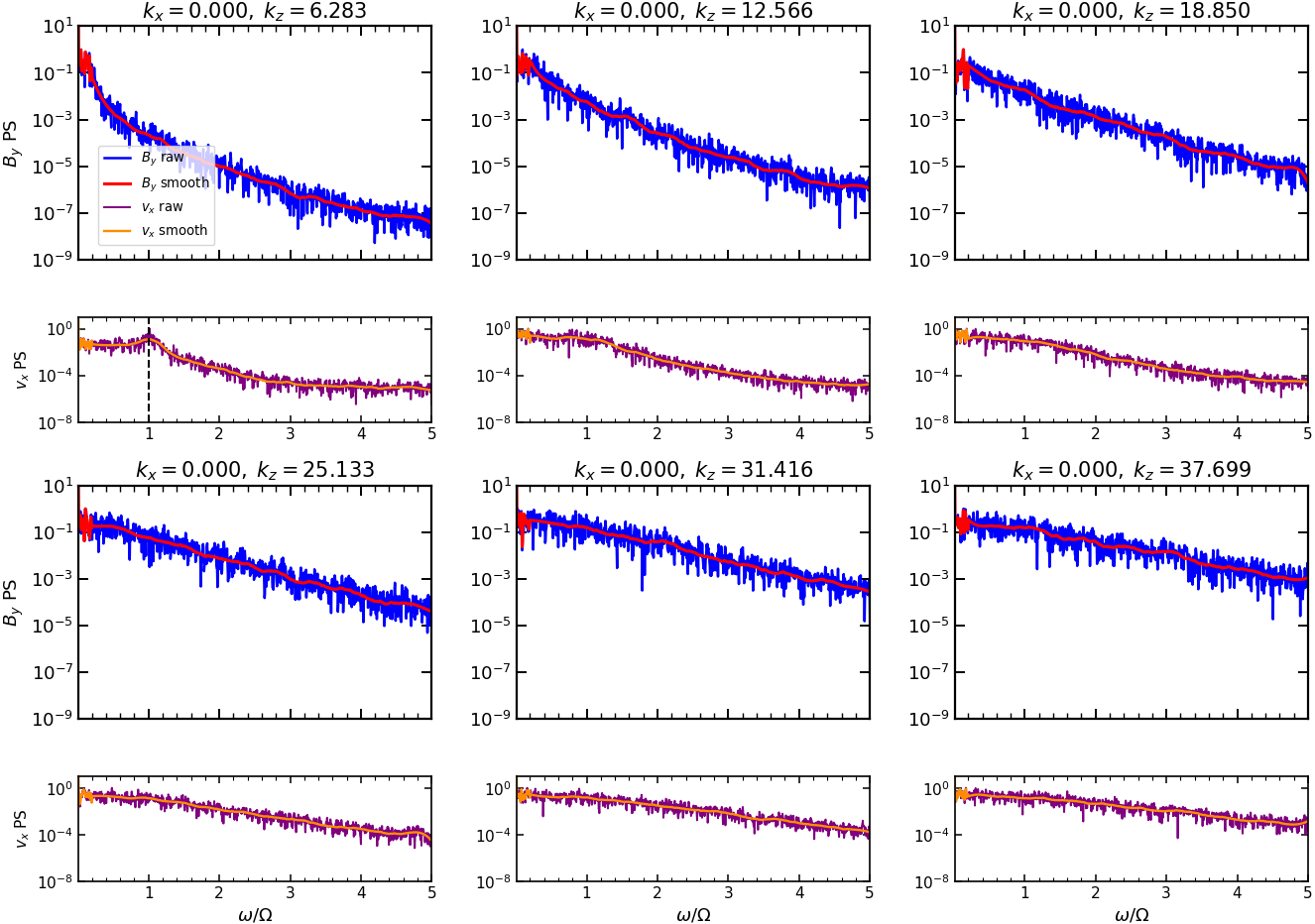}
\caption{Athena++ simulation $(a=1)$: Temporal power spectra $\left|B_y\right|^2 = \left|B_\phi\right|^2$ and $\left|v_x\right|^2 = \left|v_R\right|^2$ for the $a=1$ case, for $k_x = 0$, $k_y = 0$ and six different values of $k_z$. Only the lowest $k_z$ shows a discernible peak in the $v_x$ spectrum. Unlike larger $a$, the modes are closely spaced in $\bk$ and form a turbulent continuum.}
\label{fig:temporal_PS_sim_a1}
\end{figure*}

\begin{figure*}[t!]
\centering
\includegraphics[width=0.9\textwidth]{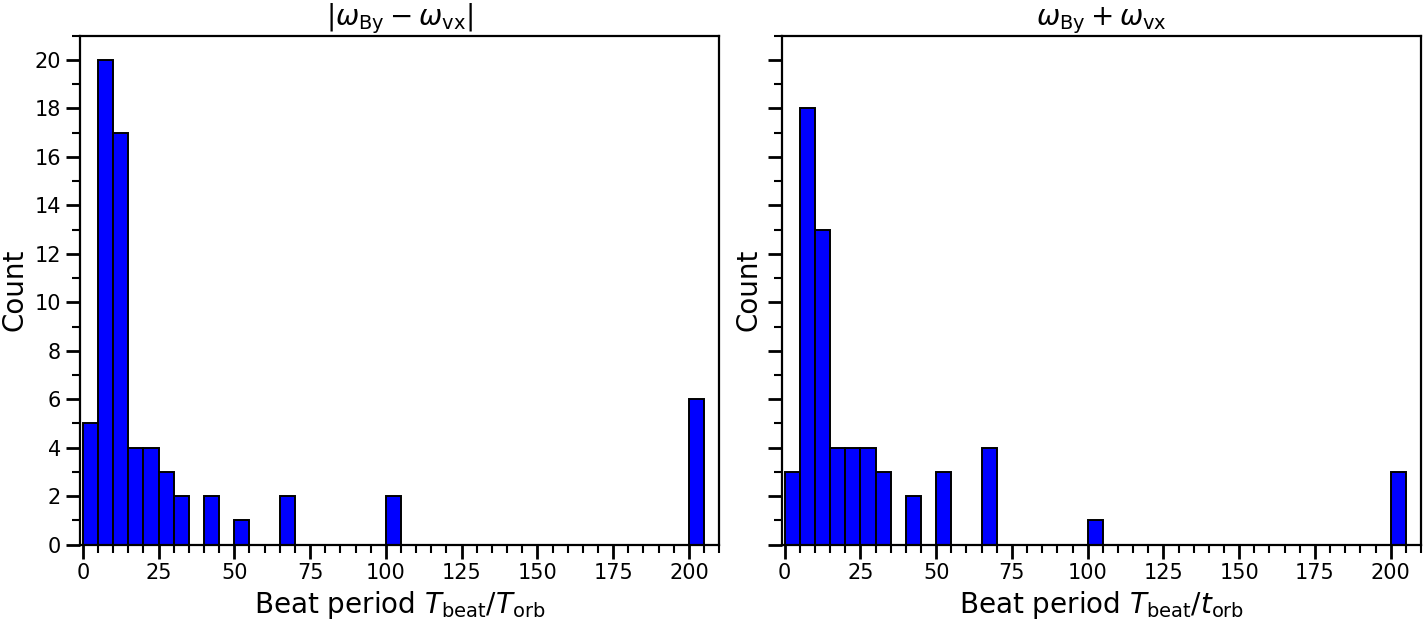}
\caption{Athena++ simulation $(a=8)$: Histogram of the dynamo cycle/beat period obtained from the differences between the peak frequencies shown in Fig.~\ref{fig:temporal_PS_sim} (see Section~\ref{sec:cycle_period_analytic_sim} for details). The peaks at $t\gtrsim 50\, t_{\rm orb}$ match well with the long period peaks of the toroidal field (see the $a=8$ case in the left panel of Fig.~\ref{fig:per_amp_sim_vs_a}).}
\label{fig:beat_period_sim_hist}
\end{figure*}

\subsection{Parameters and initial conditions}\label{sec:IC_params}

The initial density is taken to be a uniform distribution $\rho^{(0)} = 1$ throughout the box. The unit of time is taken to be $\Omega^{-1} = 1$ and that of length to be $H = c_\rms/\Omega = 1$. Therefore, the sound speed $c_\rms$ and the initial gas pressure $P^{(0)} = \rho^{(0)} c^2_\rms$ are unity. The magnetic field is initialized as $\bB = B^{(0)} \sin{\left(2\pi x / L_x\right)}\hat{\bz}$, where $B^{(0)}$ is such that the plasma $\beta^{(0)} = 8\pi P^{(0)}/B^{(0)\,2} = 100$ for all runs. This setup ensures that the net magnetic flux is zero in all three directions. The divergence-less nature of the magnetic field thus maintains the zero net flux condition at all times. We adopt a Keplerian shear flow, i.e., $q = 1.5$, and implement the periodic boundary conditions (BCs) in the azimuthal $(\hat{\by})$ and vertical $(\hat{\bz})$ directions and the shear-periodic BC in the radial $(\hat{\bx})$ direction.

We perform multiple sets of simulations with $L_x = H$, $L_y = 4 H$ and the vertical aspect ratio $a = L_z/L_x = \{1,1.5,2,3,4,5,6,7,8\}$. We run the simulations up to $t = 500\, t_{\rm orb}$. In all the runs, the number of grid cells along $x$, $N_x$, is taken to be $64$, while that along $y$, $N_y$, is taken to be $256$, such that the resolution is the same along $x$ and $y$ $\left(\Delta x = \Delta y\right)$. The number of cells along $z$, $N_z$, is taken as $64$ for $a = L_z/L_x = 1$, $128$ for $a = 1.5,2$ and $3$, and $256$ for $a = 4,5,6,7$ and $8$. The vertical cell size $\Delta z$ is therefore at maximum equal to $2\Delta x$ and at minimum $\Delta x$.

\subsection{Results}\label{sec:sim_results}

The simulations exhibit a large-scale dynamo: a coherent magnetic pattern generated spontaneously by the turbulence. Fig.~\ref{fig:B_streamlines} shows magnetic field lines for the $a=7$ run at three times, colored by the toroidal field averaged over the radial and azimuthal directions,
\[
\bar{B}_\phi=\bar{B}_y=\int \rmd x\,\rmd y\, B_y/L_xL_y.
\]
The polarity of this large-scale toroidal field reverses on a timescale of $\sim 60\,t_{\rm orb}$. This reversal is seen most clearly in the butterfly diagram, shown in Fig.~\ref{fig:butterfly_diag_sim}, where we plot $\bar{B}_\phi$ as a function of $z/L_z$ and $t/t_{\rm orb}=\Omega t/2\pi$ for $a=1,2,3,4,5,6,7$, and $8$. Fig.~\ref{fig:Bphi_Emag_vs_t_sim} shows (1) $\bar{B}_\phi$ at $z/L_z=-1,0,1$, (2) its largest-scale vertical Fourier mode,
\[
\bar{B}_{\phi k_z}(t)=\int \rmd z\, e^{-i k_z z}\bar{B}_\phi(z,t)/2\pi,
\]
with $k_z=2\pi/L_z$, (3) the same for $\bar{B}_R$, and (4) the magnetic energies in all three field components, all plotted against $t/t_{\rm orb}$. $\bar{B}_R$ shows a similar cyclic behavior with a similar period but much weaker amplitude and a relative phase-difference of $\pi/2$ with respect to $\bar{B}_\phi$.

Figs.~\ref{fig:dom_per_amp_vs_a} and \ref{fig:per_amp_sim_vs_a} compare the Fourier periods and amplitudes of the toroidal field in QLT and in the simulations. For QLT, we Fourier transform the $B_\phi$ obtained in Section~\ref{sec:B_evol} and define the amplitude as the dispersion $\sqrt{\langle B_\phi^2-{\langle B_\phi\rangle}^2\rangle}$, where the brackets denote the temporal average of $\bar{B}_{\phi k_z}(t)$ between $t=50$ and $500\,t_{\rm orb}$ (the initial transients are skipped while averaging). For the simulations, we Fourier transform $\bar{B}_\phi(z=0)=\langle B_\phi\rangle(z=0)$ and define the amplitudes in the same way. The left and right panels show the periods and amplitudes of the temporal Fourier modes of $\bar{B}_\phi$, respectively. In Fig.~\ref{fig:dom_per_amp_vs_a}, different colors correspond to different values of $k v_\rmA/\Omega$. The solid curves in the left panel show the amplitude-weighted periods, $T=\sum_i w_i T_i/\sum_i w_i$, where $T_i$ and $w_i$ are the period and relative amplitude of each mode. The circles mark all modes with amplitude exceeding $0.01$ times that of the strongest mode in the $kv_\rmA/\Omega=0.004$ case, and are color-coded by relative amplitude, from blue (strongest) to yellow (weakest). The solid dark blue curve for $kv_\rmA/\Omega=0.002$ closely corresponds to the simulation-based $B_R$ and $B_z$ values for $a = 8$, while the lime curve for $kv_\rmA/\Omega=0.019$ corresponds to that for $a = 1$. The red dot-dashed curve in the left panel of Fig.~\ref{fig:dom_per_amp_vs_a} shows the analytic estimate of the approximate upper limit $\sim 38\sqrt{1+a^2}t_{\rm orb}$ for the quasilinear periods (see Section~\ref{sec:cycle_period_analytic_QLT}), while the purple dot-dashed curve in the right panel of Fig.~\ref{fig:per_amp_sim_vs_a} shows the approximate amplitude scaling $\sim a^2$, obtained from QLT.

Several trends stand out. As seen from the energy and $\bar{B}_R$ evolution in Fig.~\ref{fig:Bphi_Emag_vs_t_sim}, the toroidal and radial fields first grow exponentially on a timescale $\sim \Omega^{-1}$, indicating linear MRI growth. At later times, $\bar{B}_\phi$ develops large-scale periodic oscillations whose amplitude increases with aspect ratio, as seen in both Figs.~\ref{fig:butterfly_diag_sim} and \ref{fig:Bphi_Emag_vs_t_sim}. No clear cyclic behavior is present for $a=1$, $1.5$ or $2$. Cycles begin to appear around $a=3$, though they remain stochastic for $a=3$ and $4$. For $a>4$, they persist throughout the simulation and typically show a few characteristic periods, the most prominent ones lying in the range $50$--$100\,t_{\rm orb}$ depending on $a$. The radial and vertical fields are much weaker than the toroidal field, by roughly an order of magnitude according to Fig.~\ref{fig:Bphi_Emag_vs_t_sim}, but they are nevertheless essential for producing large-amplitude modulations of $\bar{B}_\phi$. In the simulations, the absence (presence) of cycles for $a\lesssim 3$ ($a\gtrsim 3$) coincides with low (high) values of $B_R$ and $B_z$. Likewise, the quasilinear emf increases with $B_R$ and $B_z$ (see equations~[\ref{emf_toroidal_final}]), which explains why the amplitude of quasilinear $B_\phi$ increases with $k v_\rmA/\Omega$, as seen in the right panel of Fig.~\ref{fig:dom_per_amp_vs_a}.

The periodic behavior seen in the simulations agrees well with the quasilinear prediction, as is evident from comparing Figs.~\ref{fig:dom_per_amp_vs_a} and \ref{fig:per_amp_sim_vs_a}. At larger $a$, a few dominant periods stand out clearly above a broad background of weaker, stochastic modes. This is because the eigenfrequencies (and their difference) decrease at larger $a$ as $k_z/k\sim 1/\sqrt{1+a^2}$, which makes them more well-spaced, reduces the degree of phase-mixing, and gives rise to long period cycles (see Section~\ref{sec:cycle_period_analytic_QLT}). Both QLT and the simulations yield characteristic periods of $\sim 50$--$100\,t_{\rm orb}$. For $a\leq 2$, the Fourier spectrum is dense, reflecting the stochastic nature of $\bar{B}_\phi(t)$ and the lack of clear cyclic behavior (see Figs.~\ref{fig:Bphi_vs_t_a} and \ref{fig:Bphi_Emag_vs_t_sim}). As $a$ increases beyond $2$, the mode density decreases and a cyclic dynamo emerges. In both theory and simulations, the dominant period scales roughly as $\sqrt{1+a^2}$. The amplitude scales as $a^2$ before flattening for $a\gtrsim 5$; this scaling arises in QLT from the factor $\left[\omega^2_{j\bk-\bk'}-(\bk'\cdot\bv_\rmA)^2\right]^{-1}$ in the emf, which scales as $\sim k'^2/2k'^2_z \kappa^2 \sim a^{2}/\kappa^2$ at $k'\gg k$ (see equations~[\ref{emf_toroidal_final}]). However, QLT overestimates the amplitudes by a factor of a few. This is expected, since QLT constrains the temporal evolution of the mean field but not the turbulence level: the quasilinear emf depends on the linear eigenfunctions, but neglects their nonlinear convolutions (see Section~\ref{sec:beyond_QLT}).

\subsection{Origin of the dynamo cycles in simulations}\label{sec:cycle_period_analytic_sim}

Unlike QLT, the simulation spectra of the velocity and magnetic fluctuations contain not only the linear eigenmodes but also frequencies given by their linear combinations. This is a natural consequence of higher-order wave-interference effects, as briefly discussed in Section~\ref{sec:beyond_QLT}. As we show below, these additional frequencies produce a non-zero contribution to the dynamo cycles from the MRI unstable region, which is absent in QLT.

We compute the temporal power spectra of the $(k_x,k_y)=(0,0)$ modes of $B_y=B_\phi$ and $v_x=v_R$ for the $a=8$ and $a=1$ simulations, and plot them as functions of $\omega$ for different $k_z$ in Figs.~\ref{fig:temporal_PS_sim} and \ref{fig:temporal_PS_sim_a1}, respectively. The vertical dashed lines mark the dominant frequencies. The $a=8$ case shows clear spectral peaks, especially at large scales (small $k_z$), whereas the $a=1$ case shows no comparably distinct peaks and is much more continuum-like. This difference originates from the fact that the eigenfrequencies scale as $k_z/k$: for larger $a$, they are more widely separated in $\bk$-space, so phase mixing is weaker and discrete peaks remain visible, while for smaller $a$ they are more closely packed and blend into a stochastic continuum.

For $a=8$, the peaks roughly correspond to the linear eigenfrequency $\omega_{1,k_z=k_0}\approx \Omega$ and to combinations such as $\omega_{1,k_z=k_\rmc}\pm\omega_{2,k_z=k_\rmc}\approx \sqrt{7}\,\Omega$, $\omega_{1,k_z=k_\rmc}-\omega_{1,k_z=k_0}=(\sqrt{7}-1)\Omega$, $\omega_{1,k_z=k_\rmc}+\omega_{1,k_z=k_0}\approx (\sqrt{7}+1)\Omega$, and $\omega_{1,k_z=k_\rmc}+2\omega_{1,k_z=k_0}\approx (\sqrt{7}+2)\Omega$, where $k_0=2\pi/L_z=\pi/4h_z$ (since $L_z=8h_z$) is the smallest vertical wavenumber and $k_\rmc$ is the marginally unstable one. This follows from $\omega_{1,k_z=k_0}\approx k_z\kappa/k$ and $\omega_{1,k_z=k_\rmc}\approx \sqrt{7}\,k_z\kappa/k$ (see equations~[\ref{MRI_freqs}]), with $\kappa=\Omega$ for Keplerian flow. The critical frequency appears prominently because the fluctuations are strongest near marginal stability, as seen from the growth of $v_{1\bk\phi}$ and $B_{1\bk R}$ in the $\bk\to\bk_\rmc$ or $\omega_{2\bk}\to 0$ limit in equations~(\ref{lin_eig_fourier}). The exact peak positions vary somewhat with $k_z$ because of nonlinear coupling among modes near $k_0$ and $k_\rmc$. All of these modes lie in the MRI unstable region, since the large-scale $B_R$ and $B_z$ are quite small ($\sim 0.01$; see Fig.~\ref{fig:Bphi_Emag_vs_t_sim}).

We now show how interference between these rapidly oscillating waves can produce long-period modulations of the large-scale field. We extract the peak frequencies $\omega_{B_y}$ from the $B_y$ power spectrum at $k'_z = 2n\pi/L_z$ with $n=2(1)6$, and the frequencies $\omega_{v_x}$ from the $v_x$ spectrum at $k_0-k'_z$, where $k_0=2\pi/L_z$. The emf $\varepsilon_z$ at the large-scale mode $k_z=k_0$, which drives the toroidal field $B_y$, therefore oscillates at the frequency combinations $\omega_{B_y}\pm\omega_{v_x}$. Denoting this set of frequencies by $\omega_n$, we may write $\varepsilon_z$ as
\begin{align}
\varepsilon_z(t)=\sum_{n=1}^{N} a_n \cos(\omega_n t).
\end{align}
These frequencies cluster in groups, each of which has a narrow spread. Let us now study the emf in any one of these groups that has $N_\rmG$ such frequencies. Choose the median frequency $\omega_0$ as the representative carrier frequency for the group, and define $\delta_n \equiv \omega_n-\omega_0$.
Then
\begin{align}
\varepsilon_z(t)
&= \sum_{n=1}^{N_\rmG} a_n \cos\big[(\omega_0+\delta_n)t\big] \nonumber\\
&= \sum_{n=1}^{N_\rmG} a_n \left[\cos(\omega_0 t)\cos(\delta_n t)-\sin(\omega_0 t)\sin(\delta_n t)\right]\nonumber\\
&= A(t)\cos(\omega_0 t)-B(t)\sin(\omega_0 t),
\end{align}
where
\begin{align}
A(t) &= \sum_{n=1}^{N_\rmG} a_n \cos(\delta_n t), \quad
B(t) = \sum_{n=1}^{N_\rmG} a_n \sin(\delta_n t).
\end{align}
Hence the emf may be written in carrier--envelope form as
\begin{align}
\varepsilon_z(t)=E(t)\cos\!\big[\omega_0 t+\phi(t)\big],
\end{align}
with
\begin{align}
E(t) &= \sqrt{A^2(t)+B^2(t)}, \quad \phi(t) = \tan^{-1}\!\left(\frac{B(t)}{A(t)}\right).
\end{align}
The quantities $A(t)$ and $B(t)$ are sums over the individual offsets $\delta_n$, but the beating appears most directly in the envelope amplitude. Indeed,
\begin{align}
E^2(t)
&=A^2(t)+B^2(t) =\sum_{n=1}^{N_\rmG} a_n^2
 +2\sum_{m<n} a_m a_n \cos\big[(\delta_m-\delta_n)t\big].
\end{align}
Since $\delta_m-\delta_n=(\omega_m-\omega_0)-(\omega_n-\omega_0)=\omega_m-\omega_n$,
we obtain
\begin{align}
E^2(t)
=
\sum_{n=1}^{N_\rmG} a_n^2
+
2\sum_{m<n} a_m a_n \cos\big[(\omega_m-\omega_n)t\big].
\end{align}
Thus, the envelope is controlled by the pairwise frequency differences $\omega_m-\omega_n$, which are nothing but the beat frequencies. A similar exercise could also be performed for the radial emf $\varepsilon_x$. We club the frequencies $\omega_{B_y}\pm \omega_{v_x}$ in groups, compute the pairwise differences $\Delta\omega = \omega_m-\omega_n$ and the beating period $T_{\rm beat}$ in each group. We plot the histogram of these beating periods in Fig.~\ref{fig:beat_period_sim_hist}. These periods match well with the periods derived from the temporal power-spectrum of $B_y$ (see the $a = 8$ case in the left panel of Fig.~\ref{fig:per_amp_sim_vs_a}). Especially, the peaks at $\sim 50$, $100$ and $200\,t_{\rm orb}$ are correctly recovered. In this analysis, we have smoothed the $v_x$ and $B_y$ power-spectra for the identification of prominent peaks, which could introduce potential error in the frequency estimation. For consistency check, we varied the smoothing kernel and found that the histogram of beating periods appears similar with persistent peaks at the aforementioned periods in each case, which shows that the periods in Fig.~\ref{fig:beat_period_sim_hist} are not an artifact of mis-estimation of frequencies due to smoothing. All in all, this exercise shows that wave-interference is ultimately responsible for the emergence of long period dynamo cycles. However, the relevant beats are not restricted to the quasilinear $\Delta \omega_{\bk'} = \omega_{1\bk'}-\omega_{2\bk'}$ channel; they also involve higher order combinations of eigenfrequencies and therefore include beyond-quasilinear contributions.

\section{Discussion}\label{sec:discussion}

\subsection{Physical picture}

The central result of this paper is that the long-period MRI dynamo cycle can be understood as a wave-interference phenomenon. Starting from the linear MRI eigenfunctions in the WKB approximation, we computed the quasilinear emf, $\langle \bv_1\times\bB_1\rangle$, without assuming an \textit{a priori} mean-field closure, and used it to evolve the large-scale field. The resulting emf depends nontrivially on the mean field through the eigenfunctions and eigenfrequencies, and contains oscillatory contributions involving both sums and differences of the two eigenfrequency branches. The low-frequency part, associated with the difference,
\[
\Delta\omega_{\bk'}=\omega_{1\bk'}-\omega_{2\bk'},
\]
drives a slow modulation of the mean induction and hence of $B_\phi$. In this way, QLT provides a first-principles explanation for the origin of the dynamo cycles, i.e., the butterfly diagram.

The physical reason for the long period is that, for $k'v_{\rm A}>\Omega$, the two frequency branches become nearly Alfv\'enic and differ slightly by a shear-dependent term, and $\Delta\omega_{\bk'}$ becomes small and weakly dependent on $k'$. Physically, the fast and slow branches differ in how shear competes with magnetic tension: for the fast branch, tension is reinforced by shear, whereas for the slow branch they oppose each other. This small residual asymmetry leaves the two branches nearly degenerate, but not exactly so, producing a nonzero beat frequency. Since the emf is an integral over many $\bk'$ modes, a long-lived large-scale modulation requires weak phase mixing across the contributing spectrum. The MRI eigenstructure satisfies this requirement for $k'\gtrsim k'_{\rm c}\sim \Omega/v_{\rm A}$, so the low-frequency beat survives spectral averaging and the narrow spread in the beat frequency across the spectrum produces a long dynamo cycle. Within the present theory, the cycle period is therefore set by the weak dispersion of $\Delta\omega_{\bk'}$.

This picture also explains why the cycles depend strongly on box geometry. The aspect ratio $a=L_z/L_R$ sets $k'_z/k'$ and therefore the eigenfrequencies, which are roughly proportional to $k'_z/k'$. At larger $a$, the relevant frequencies are more sparsely distributed in $\bk'$-space, phase mixing is weaker, and the cycles become more pronounced and longer-lived. The same geometry also lowers $\Delta\omega_{\bk'}$, increasing the cycle period $T_{\rm cycle}$; within QLT this yields the approximate scaling: $T_{\rm cycle} = T_{\Delta 1} \sim 30\sqrt{1+a^2}t_{\rm orb}$. The linear response, and therefore the emf, also becomes stronger at larger $a$, increasing the cycle amplitude as $\sim a^2$ before flattening off beyond $a\sim 5$. These trends are broadly consistent with local shearing box simulations.

\subsection{Relation to mean-field theory}

In the large-scale limit, the full quasilinear emf may be expanded in a form analogous to the usual $\alpha$--$\beta$ mean-field expansion. In that sense, the quasilinear calculation contains the familiar mean-field terms, but is not limited to them. The lowest-order (in $\bk$) contribution has the structure of an (helicity-dependent) $\alpha$ effect, while the next-order term has the structure of a shear-current, or $\beta$, effect. For the unstratified zero-net-flux system studied here, the helicities are small, so the contribution $\pmb{\alpha}\cdot\bB$ to the emf, is small. A small but non-zero $\pmb{\alpha}$ is, however, required to yield a saturated ratio of the toroidal and poloidal components, $|B_\phi/B_R| \sim \sqrt{1+a^2}$, through the conservation of magnetic helicity on small scales. The dominant contribution to the emf is $\pmb{\varepsilon} = \pmb{\beta}\cdot\bJ$, which is the shear current effect. In the axisymmetric sector, the off-diagonal component $\beta_{R\phi}$ grows the toroidal field (along with the shear-driven $\Omega$ effect). If non-axisymmetric modes are included, $\beta_{\phi R}$ grows the poloidal field \citep[][]{Squire.Bhattacharjee.16,Rincon.etal.07,Riols.etal.13}. We intend to generalize the quasilinear framework of this paper to non-axisymmetric MRI in future work.

Let us briefly compare our perturbative treatment to direct statistical simulations \citep[][]{Mondal.Bhat.23,Mondal.etal.26}. \citet[][]{Mondal.etal.26} decompose each quantity $\bQ$ as $\bar{\bQ}+\delta\bQ$ in the shearing frame, with $\bar{\bQ}$ the horizontal mean, and show how a positive off-diagonal $\beta_{\phi R}$ (negative off-diagonal turbulent diffusivity) arises. The mean fields satisfy $\partial_t\bar B_R\simeq-\partial_z\varepsilon_\phi$ and $\partial_t\bar B_\phi\simeq-q\Omega\bar B_R + \partial_z \varepsilon_R - \partial_R \varepsilon_z$, where $q=-\rmd\ln\Omega/\rmd\ln R$: shear amplifies $\bar B_\phi$ from $\bar B_R$, while $\varepsilon_\phi\simeq\beta_{\phi R}\bar J_R$, with $\bar J_R=-\partial_z\bar B_\phi$ and $\beta_{\phi R}\simeq(\rho\Omega)^{-1}[(2-q)^{-1}\overline{\delta B_z\delta B_z}/4\pi+q^{-1}\rho\,\overline{\delta v_z\delta v_z}]>0$, regenerates $\bar B_R$. The magnetic tension force $\delta B_z\partial_z\bar B_\phi$ drives $\delta v_\phi$, and the Coriolis force converts this into $\delta v_R$, which yields $\varepsilon_\phi\simeq-\overline{\delta v_R\delta B_z}\propto\bar J_R$. Saturation occurs because nonlinear Lorentz-force triple correlations produce a term $\varepsilon_\phi^{(3)}$ such that $-\partial_z\varepsilon_\phi^{(3)}$ balances the current-driven source $-\partial_z(\beta_{\phi R}\bar J_R)$ in $\partial_t\bar B_R=-\partial_z\varepsilon_\phi$. The same $\bar{B}_\phi \to \delta v_\phi \to \delta v_R \to \bar{B}_R \to \bar{B}_\phi$ loop occurs in the quasilinear framework, when both axisymmetric and non-axisymmetric modes are taken into account. What our framework offers, though, but direct statistical simulations \citep[][]{Mondal.Bhat.23,Mondal.etal.26} do not, is the timescale over which this loop is closed: the period over which the small-scale velocity and magnetic fluctuations get back into phase with one another and give rise to a coherent large-scale dynamo.

Our perturbative analysis shows that the MRI dynamo should not be viewed as a conventional $\alpha$--$\beta$--$\Omega$ dynamo. Neither $\alpha$ nor $\beta$ should be regarded as a complete closure in this problem: they are only the first terms in the expansion of the full quasilinear emf. For practical purposes, it is more informative to evaluate the full emf directly (using equations~[\ref{emf_toroidal_final}] and [\ref{emf_radial_final}]) than to truncate the series at low order.

\subsection{What QLT captures, and what it misses}

The main success of QLT is that it explains the existence of long-period dynamo cycles and their aspect-ratio dependence, without imposing phenomenological time-dependent dynamo coefficients. It identifies the relevant slow timescale with the interference between the two eigenfrequency branches. At the same time, the simulations reveal that the full dynamo is spectrally richer than QLT alone. In the quasilinear picture, the dominant long-period cycle arises from beats between the two eigenfrequencies in the MRI stable region. In the simulations, by contrast, the spectrum contains not only the linear eigenfrequencies but also higher-order linear combinations, and the observed cycle periods arise from pairwise beats among a broader set of peaks, including a nonzero contribution from the MRI-unstable region.

QLT has well-defined limitations. It builds the mean field from the interference of linear perturbations while neglecting their nonlinear coupling with other modes. MRI saturation, however, is governed by such mode couplings---both between different $\bk$ and within a given $\bk$---which drain energy from the linearly unstable (exponentially growing) modes and redistribute it through the stable (exponentially decaying) modes. Preliminary calculations indicate that this redistribution of energy between the stable and unstable modes gives rise to dynamo cycles with periods similar to those arising from the interference of oscillatory modes or waves discussed in this paper. In a follow-up paper, we will present a more general theory based on a nonlinear dynamical systems analysis of the large-scale fields, complementary to the invariant solutions approach of \citet[][]{Riols.etal.13}.

\subsection{Astrophysical implications}

Although the present theory and simulations treat an idealized system---a local, unstratified, magnetized shear-flow---they suggest a broader organizing principle for large-scale cyclic dynamos: when the eigenspectrum contains nearly degenerate branches whose frequency differences vary only weakly across the modes that dominate the emf, the contributions to the emf oscillate with nearly the same beat frequency and therefore do not phase-mix rapidly.

How far this mechanism extends beyond the present setting remains to be established. In stratified disks, buoyancy \citep[][]{Skoutnev.etal.22} and an $\alpha$ effect emerging from non-helical flows \citep[][]{Dhang.etal.24} may open up additional dynamo channels. Explicit resistivity, finite magnetic Prandtl number ${\rm Pm}$ \citep[][]{Fromang.etal.07,Riols.etal.15}, non-axisymmetric modes \citep[][]{Lesur.Ogilvie.07}, and global curvature \citep[][]{Pessah.Psaltis.05,Das.etal.17} would modify the spectrum and the saturation pathway. It would be important to test whether the same interference picture continues to organize the large-scale dynamo in more realistic models, especially in the low-${\rm Pm}$ regime relevant to many astrophysical disks. A related problem is the dependence of the turbulent stress on the geometry and ${\rm Pm}$. The sharp decline of the stress with decreasing ${\rm Pm}$ and its turning off below a critical ${\rm Pm}>1$ in $a=1$ boxes with explicit dissipation \citep[][]{Fromang.etal.07} appears to be connected to the reduction in turbulent stress with increasing resolution in ideal MHD \citep[][]{Fromang.Papaloizou.07,Shi.etal.16}. However, this behavior is strongly geometry-dependent: \citet[][]{Shi.etal.16} found that vertically elongated boxes not only develop a coherent cyclic large-scale dynamo (which we have explained using QLT) but also resolution-independent stresses, while explicit-dissipation studies of elongated boxes show that sustained turbulence can persist to lower ${\rm Pm}$ when the vertical domain is extended \citep[][]{Nauman.Pessah.16,Nauman.Pessah.18}. \citet[][]{Held.Mamatsashvili.22} found a ${\rm Pm}^\delta$ (${\rm Rm}$ dependent $\delta>0$) scaling of the stress. This scaling is further weakened in tall boxes, where a large-scale dynamo action makes the stress less sensitive to the dissipation scale. These results suggest that the geometry and ${\rm Pm}$ dependence of turbulent transport is deeply connected to the ability of the domain to support a large-scale dynamo.

Despite the restricted setup of zero-net-flux, unstratified, ideal MHD, the physical principle behind long-period dynamo cycles is clear: they emerge from coherent beats between nearly degenerate MHD waves. In this paper, we have identified this beating mechanism explicitly for magnetorotational dynamos, but the principle is more general and need not be tied to the MRI. Kelvin--Helmholtz dynamos in shear layers \citep[][]{Tripathi.etal.26} and buoyancy-driven dynamos in sheared flows with zero net rotation \citep[][]{Squire.etal.25} are alternate cases where analogous beats are presumably in action, and we intend to study such non-magnetorotational dynamo pathways in future work. This broader interpretation has direct implications for cyclic magnetic activity across astrophysical systems, from stellar and planetary dynamos to quasi-periodic variability in accretion disks around stars, planets, and black holes. In accreting systems, magnetic cycles could lead to episodic accretion and quasi-periodic flaring and jet launching, which could in principle introduce cyclic features to observed light curves \citep[][]{Alston.etal.16}, e.g., low-frequency quasi-periodic oscillations in AGNs \citep[][]{ONeill.etal.11,Zhou.Lai.24}. In some cases, cyclic dynamo activity could mimic variability otherwise attributed to binaries, e.g., putative binary supermassive black holes in AGN disks \citep[][]{Zhou.Lai.24}. The solar cycle is often attributed to a near-surface \citep[][]{Vasil.etal.24} or near-tachocline MRI-type dynamo \citep[][]{Kagan.Wheeler.14}. It would be interesting to see if our theory can be generalized to the study of stellar and planetary dynamos.

\begin{acknowledgments}
The authors are thankful to Callum Fairbairn, Bindesh Tripathy, James Beattie, Alexander Dittmann, George Wong, Jeremy Goodman, Eliot Quataert, and Sihao Cheng, for stimulating discussions and valuable suggestions. UB is especially grateful to Callum Fairbairn for expert advice regarding \texttt{Athena++} simulations. The authors acknowledge the role of GPT-5.5 Thinking for assistance in symbolic manipulation. This research is supported by the Bezos Member Fund and the Fund for Memberships in Natural Sciences at the Institute for Advanced Study.
\end{acknowledgments}

\appendix

\section{Derivation of the quasilinear emf}\label{App:emf}

Here we compute the quasilinear emf from the linear perturbations in the velocity and magnetic fields. The evolution of the second order perturbations, $\bv_2$ and $\bB_2$, is given by the force and induction equations perturbed up to second order:
\begin{equation}
\begin{split}
\frac{\partial \bB_2}{\partial t}
&= \nabla\times \left(\bv_0\times \bB_2\right)
 + \nabla\times \left(\bv_2\times \bB_0\right)
 + \nabla\times \left(\bv_1\times \bB_1\right),
\\
\frac{\partial \bv_2}{\partial t}
&+ \bv_2\cdot\nabla\bv_0
 + \bv_0\cdot\nabla\bv_2
 + \frac{\nabla\left(\bB_0\cdot\bB_2\right)-\bB_0\cdot\nabla\bB_2}{4\pi\rho_0}
\\
&= -\bv_1\cdot\nabla \bv_1
-\frac{1}{4\pi\rho_0}
\left[
\frac{\nabla\left(\bB_1\cdot\bB_1\right)}{2}
-\bB_1\cdot\nabla\bB_1
\right].
\end{split}
\end{equation}
The linear perturbations given by equations~(\ref{lin_eig}) need to be substituted above to obtain the time evolution of $\bv_2$ and $\bB_2$.

The induction can be computed as follows:
\begin{equation}
\begin{split}
\pmb{\calE}_\bk
&= i\bk\times \pmb{\varepsilon}_\bk
= i\bk\times \int \rmd \bk'\,
\bv_{1\bk - \bk'}(t)\times \bB_{1\bk'}(t)
\\
&= i\int \rmd \bk'\,
\left[
\bk\cdot\bB_{1\bk'}(t)\,\bv_{1\bk-\bk'}(t)
-\bk\cdot\bv_{1\bk-\bk'}(t)\,\bB_{1\bk'}(t)
\right]
\\
&= i\int \rmd \bk'\,
\left(k_R-k_z\frac{k'_R}{k'_z}\right)\nonumber\\
&\times\Big[
B_{1\bk' R}(t)\,\bv_{1\bk-\bk'}(t)
-v_{1\bk' R}(t)\,\bB_{1\bk-\bk'}(t)
\Big],
\end{split}
\end{equation}
where we have used the fact that $\bk'\cdot\bB_{1\bk'} = 0$ and $\bk'\cdot\bv_{1\bk'} = 0$. The azimuthal/toroidal and radial components of the induction are given by
\begin{equation}
\begin{split}
\calE_{\bk\phi}
&= i\int \rmd \bk'
\left(k_R-k_z\frac{k'_R}{k'_z}\right)\nonumber\\
&\times\Big[
B_{1\bk' R}(t)\,v_{1\bk-\bk'\phi}(t)
-v_{1\bk' R}(t)\,B_{1\bk-\bk'\phi}(t)
\Big],
\\
\calE_{\bk R} &= -\frac{k_z}{k_R} \calE_{\bk z}\nonumber\\
&= -i k_z\int \rmd \bk'\,\frac{k'_R}{k'_z}
\Big[
B_{1\bk' R}(t)\,v_{1\bk-\bk'R}(t)
-v_{1\bk' R}(t)\,B_{1\bk-\bk'R}(t)
\Big].
\end{split}
\label{emf_1}
\end{equation}
Plugging in the linear perturbations from equations~(\ref{lin_eig}) in the above equation, and defining $s_{i\bk'} = (c^+_{i\bk'} + c^-_{i\bk'})/2$ and $d_{i\bk'} = (c^+_{i\bk'} - c^-_{i\bk'})/2$, we can rewrite the toroidal induction in the following concise form:
\begin{align}
{\bf\calE}_{\bk\phi}\left(\bB,t\right)
&= \Omega \int \rmd \bk'\,
\left(k_z \frac{k'_R}{k'_z} - k_R\right)\,
\bk'\cdot\bB\;
\delta v_{\bk'R}\,\delta v_{\bk-\bk'R}
\nonumber\\
&\times
\sum_{i,j}
\frac{
\calR^{(\phi)}_{\bk\bk'ij}\left(\bB,t\right)
+i\calI^{(\phi)}_{\bk\bk'ij}\left(\bB,t\right)
}{
\omega^2_{j\,\bk-\bk'} - {\left(\bk'\cdot\bv_\rmA\right)}^2
},
\nonumber\\[1ex]
\calR^{(\phi)}_{\bk\bk'ij}\left(\bB,t\right)
&=
\left(s_{i\bk'}s_{j\bk-\bk'} - d_{i\bk'}d_{j\bk-\bk'}\right)
\nonumber\\
&\times
\left[
1 - \frac{\omega_{j\bk-\bk'}}{2\omega_{i\bk'}}
\left(
\frac{\kappa^2}{2\Omega^2}
-\frac{{\left(\bk'\cdot\bv_\rmA\right)}^2}{\omega^2_{j\bk-\bk'}}
\frac{\rmd\ln\Omega}{\rmd \ln R}
\right)
\right]
\nonumber\\
&\times
\cos\!\left[\left(\omega_{i\bk'}-\omega_{j\bk-\bk'}\right)t\right]
\nonumber\\
&+
\left(s_{i\bk'}s_{j\bk-\bk'} + d_{i\bk'}d_{j\bk-\bk'}\right)
\nonumber\\
&\times
\left[
1 + \frac{\omega_{j\bk-\bk'}}{2\omega_{i\bk'}}
\left(
\frac{\kappa^2}{2\Omega^2}
-\frac{{\left(\bk'\cdot\bv_\rmA\right)}^2}{\omega^2_{j\bk-\bk'}}
\frac{\rmd\ln\Omega}{\rmd \ln R}
\right)
\right]
\nonumber\\
&\times
\cos\!\left[\left(\omega_{i\bk'}+\omega_{j\bk-\bk'}\right)t\right],
\nonumber\\[1ex]
\calI^{(\phi)}_{\bk\bk'ij}\left(\bB,t\right)
&=
\left(s_{i\bk'}d_{j\bk-\bk'} - d_{i\bk'}s_{j\bk-\bk'}\right)
\nonumber\\
&\times
\left[
1 - \frac{\omega_{j\bk-\bk'}}{2\omega_{i\bk'}}
\left(
\frac{\kappa^2}{2\Omega^2}
-\frac{{\left(\bk'\cdot\bv_\rmA\right)}^2}{\omega^2_{j\bk-\bk'}}
\frac{\rmd\ln\Omega}{\rmd \ln R}
\right)
\right]
\nonumber\\
&\times
\sin\!\left[\left(\omega_{i\bk'}-\omega_{j\bk-\bk'}\right)t\right]
\nonumber\\
&-
\left(s_{i\bk'}d_{j\bk-\bk'} + d_{i\bk'}s_{j\bk-\bk'}\right)
\nonumber\\
&\times
\left[
1 + \frac{\omega_{j\bk-\bk'}}{2\omega_{i\bk'}}
\left(
\frac{\kappa^2}{2\Omega^2}
-\frac{{\left(\bk'\cdot\bv_\rmA\right)}^2}{\omega^2_{j\bk-\bk'}}
\frac{\rmd\ln\Omega}{\rmd \ln R}
\right)
\right]
\nonumber\\
&\times
\sin\!\left[\left(\omega_{i\bk'}+\omega_{j\bk-\bk'}\right)t\right].
\label{emf_1_simp}
\end{align}
Similarly, the radial induction can be written as
\begin{align}
{\bf\calE}_{\bk R}\left(\bB,t\right)
&= \frac{k_z}{2} \int \rmd \bk'\,\frac{k'_R}{k'_z}\,
\bk'\cdot\bB\;
\delta v_{\bk'R}\,\delta v_{\bk-\bk'R}
\nonumber\\
&\quad\times
\sum_{i,j}
\left[
\calR^{(R)}_{\bk\bk'ij}\left(\bB,t\right)
+i\calI^{(R)}_{\bk\bk'ij}\left(\bB,t\right)
\right],
\nonumber\\[1ex]
\calR^{(R)}_{\bk\bk'ij}\left(\bB,t\right)
&=
\left(s_{i\bk'}s_{j\bk-\bk'} - d_{i\bk'}d_{j\bk-\bk'}\right)
\nonumber\\
&\quad\times
\left(\frac{1}{\omega_{i\bk'}} - \frac{1}{\omega_{j\bk-\bk'}}\right)
\sin\!\left[\left(\omega_{i\bk'}-\omega_{j\bk-\bk'}\right)t\right]
\nonumber\\
&\quad+
\left(s_{i\bk'}s_{j\bk-\bk'} + d_{i\bk'}d_{j\bk-\bk'}\right)
\nonumber\\
&\quad\times
\left(\frac{1}{\omega_{i\bk'}} + \frac{1}{\omega_{j\bk-\bk'}}\right)
\sin\!\left[\left(\omega_{i\bk'}+\omega_{j\bk-\bk'}\right)t\right],
\nonumber\\[1ex]
\calI^{(R)}_{\bk\bk'ij}\left(\bB,t\right)
&=
-\left(s_{i\bk'}d_{j\bk-\bk'} - d_{i\bk'}s_{j\bk-\bk'}\right)
\nonumber\\
&\quad\times
\left(\frac{1}{\omega_{i\bk'}} - \frac{1}{\omega_{j\bk-\bk'}}\right)
\cos\!\left[\left(\omega_{i\bk'}-\omega_{j\bk-\bk'}\right)t\right]
\nonumber\\
&\quad+
\left(s_{i\bk'}d_{j\bk-\bk'} + d_{i\bk'}s_{j\bk-\bk'}\right)
\nonumber\\
&\quad\times
\left(\frac{1}{\omega_{i\bk'}} + \frac{1}{\omega_{j\bk-\bk'}}\right)
\cos\!\left[\left(\omega_{i\bk'}+\omega_{j\bk-\bk'}\right)t\right].
\label{emf_2_simp}
\end{align}
Since $\pmb{\calE}_\bk = i\bk\times \pmb{\varepsilon}_\bk$, we have
$\calE_{\bk R} = -i k_z \varepsilon_{\bk \phi}$,
$\calE_{\bk\phi} = i\left(k_z\varepsilon_{\bk R} - k_R\varepsilon_{\bk z}\right)$,
and $\calE_{\bk z} = i k_R\varepsilon_{\bk \phi}$.
The emf $\pmb{\varepsilon}_\bk = \left(\varepsilon_{\bk R},\varepsilon_{\bk \phi},\varepsilon_{\bk z}\right)$ can thus be trivially obtained from the above expression for the induction.

\section{Computation of the dynamo cycle period}
\label{App:Deltaomega_large_k}

The cycle period depends on the frequency difference between
the two eigenfrequencies, $\Delta\omega_{\bk'} = \omega_{1\bk'}-\omega_{2\bk'}$, in the MRI stable regime, $\bk'\cdot\bv_\rmA/\Omega\gtrsim 1$ (see section~\ref{sec:cycle_period_analytic_QLT}). Here we derive the large-wavenumber expansion of this frequency difference or beat frequency, which yields a simple expression for the cycle period. Define
\begin{equation}
x\equiv \frac{\bk'\cdot\bv_{\rm A}}{\Omega},\quad
\mu\equiv \frac{k'_z}{k'},\quad
s\equiv \frac{\kappa^2}{\Omega^2}.
\end{equation}
The two branches may be written as
\begin{align}
\frac{\omega_{1\bk',2\bk'}^2}{\Omega^2}
&=
x^2+\frac{\mu^2s}{2}
\left[
1\pm
\sqrt{
1+\frac{16x^2}{\mu^2s^2}
}
\right],
\label{eq:MRI_branches_dimless}
\end{align}
where $\omega_{1\bk'}$ and $\omega_{2\bk'}$ denote the larger and smaller positive frequencies. Expanding these in the limit of large $x$ (MRI stable regime) gives
\begin{align}
\frac{\omega_{1\bk'}}{\Omega}
&=
x+\mu
+\frac{\mu^2}{x}
\left(
\frac{s}{4}-\frac{1}{2}
\right)
+\frac{\mu^3}{x^2}
\left(
\frac{1}{2}-\frac{s}{4}+\frac{s^2}{32}
\right)
\nonumber\\
&\quad
-\frac{\mu^4}{x^3}
\left(
\frac{s^2-6s+10}{16}
\right)
\nonumber\\
&\quad
+\frac{\mu^5}{x^4}
\left(
\frac{7}{8}
-\frac{5s}{8}
+\frac{9s^2}{64}
-\frac{s^3}{128}
-\frac{s^4}{2048}
\right)
+\mathcal O(x^{-5}),
\label{eq:omega_1_large_x}
\\
\frac{\omega_{2\bk'}}{\Omega}
&=
x-\mu
+\frac{\mu^2}{x}
\left(
\frac{s}{4}-\frac{1}{2}
\right)
-\frac{\mu^3}{x^2}
\left(
\frac{1}{2}-\frac{s}{4}+\frac{s^2}{32}
\right)
\nonumber\\
&\quad
-\frac{\mu^4}{x^3}
\left(
\frac{s^2-6s+10}{16}
\right)
\nonumber\\
&\quad
-\frac{\mu^5}{x^4}
\left(
\frac{7}{8}
-\frac{5s}{8}
+\frac{9s^2}{64}
-\frac{s^3}{128}
-\frac{s^4}{2048}
\right)
+\mathcal O(x^{-5}).
\label{eq:omega_2_large_x}
\end{align}
The common $x^{-1}$ and $x^{-3}$ terms cancel in the difference. Thus,
\begin{align}
\Delta\omega_{\bk'}
&\equiv \omega_{1\bk'}-\omega_{2\bk'}
\nonumber\\
&=
2\mu\Omega\,
\Bigg[
1+
\frac{\mu^2}{4x^2}
\left(
2-s+\frac{s^2}{8}
\right)
\nonumber\\
&\quad
+
\frac{\mu^4}{x^4}
\left(
\frac{7}{8}
-\frac{5s}{8}
+\frac{9s^2}{64}
-\frac{s^3}{128}
-\frac{s^4}{2048}
\right)+\mathcal O(x^{-6})
\Bigg].
\label{eq:Deltaomega_large_k_s}
\end{align}
Restoring $s=\kappa^2/\Omega^2$ and
$x=(\bk'\cdot\bv_{\rm A})/\Omega$, this becomes
\begin{align}
\Delta\omega_{\bk'}
&=
\frac{2k'_z}{k'}\Omega\,
\Bigg[
1+
\frac{k_z'^2}{4k'^2}
{\left(
\frac{\bk'\cdot\bv_{\rm A}}{\Omega}
\right)}^{-2}
\left(
2-\frac{\kappa^2}{\Omega^2}
+\frac{\kappa^4}{8\Omega^4}
\right)
\nonumber\\
&\quad
+
\frac{k_z'^4}{k'^4}
{\left(
\frac{\bk'\cdot\bv_{\rm A}}{\Omega}
\right)}^{-4}
\left(
\frac{7}{8}
-\frac{5}{8}\frac{\kappa^2}{\Omega^2}
+\frac{9}{64}\frac{\kappa^4}{\Omega^4}
\right.
\nonumber\\
&\quad\left.
-\frac{1}{128}\frac{\kappa^6}{\Omega^6}
-\frac{1}{2048}\frac{\kappa^8}{\Omega^8}
\right)
+\mathcal O\!\left[
{\left(
\frac{\bk'\cdot\bv_{\rm A}}{\Omega}
\right)}^{-6}
\right]
\Bigg].
\label{eq:Deltaomega_large_k_kappa}
\end{align}

We next write this in terms of the shear parameter
$q\equiv-\rmd\ln\Omega/\rmd\ln R$, for which
$\kappa^2=(4-2q)\Omega^2$. The two correction coefficients become
\begin{equation}
2-\frac{\kappa^2}{\Omega^2}
+\frac{\kappa^4}{8\Omega^4}
=
\frac{q^2}{2},
\end{equation}
and
\begin{align}
&\frac{7}{8}
-\frac{5}{8}\frac{\kappa^2}{\Omega^2}
+\frac{9}{64}\frac{\kappa^4}{\Omega^4}
-\frac{1}{128}\frac{\kappa^6}{\Omega^6}
-\frac{1}{2048}\frac{\kappa^8}{\Omega^8} =
\frac{q^3(16-q)}{128}.
\end{align}
Therefore
\begin{align}
\Delta\omega_{\bk'}
&=
\frac{2k'_z}{k'}\Omega\,
\Bigg[
1+
\frac{q^2}{8}
\frac{k_z'^2}{k'^2}
{\left(
\frac{\bk'\cdot\bv_{\rm A}}{\Omega}
\right)}^{-2}
\nonumber\\
&\quad
+
\frac{q^3(16-q)}{128}
\frac{k_z'^4}{k'^4}
{\left(
\frac{\bk'\cdot\bv_{\rm A}}{\Omega}
\right)}^{-4}
+\mathcal O\!\left[
{\left(
\frac{\bk'\cdot\bv_{\rm A}}{\Omega}
\right)}^{-6}
\right]
\Bigg].
\label{eq:Deltaomega_large_k_q}
\end{align}

We now evaluate the finite-$k'$ correction at $k'=2k'_{\rm c}$ (this is roughly the lowest $k'$ that contributes to the quasi-periodic emf; see section~\ref{sec:cycle_period_analytic_QLT}). The marginal
MRI condition gives
\begin{align}
\left(
\frac{\bk'_{\rm c}\cdot\bv_{\rm A}}{\Omega}
\right)^2
&=
\left(
4-\frac{\kappa^2}{\Omega^2}
\right)
\frac{k_z'^2}{k'^2}
=
2\mu^2 q.
\end{align}
Therefore, at $k'=2k'_{\rm c}$, $x^2=8\mu^2 q$. Substituting this into
equation~(\ref{eq:Deltaomega_large_k_q}) gives
\begin{align}
\Delta\omega_{\rm c}\equiv
\Delta\omega_{\bk'}\big|_{k'=2k'_{\rm c}}&=
2\mu\Omega
\left[
1+\frac{q}{64}
+\frac{q(16-q)}{8192}
\right]
\nonumber\\
&=
2\mu\Omega
\left[
1+\frac{q(144-q)}{8192}
\right].
\label{eq:Deltaomega_c_q}
\end{align}
The asymptotic value is $\Delta\omega_\infty=2\mu\Omega$. Thus, with
$t_{\rm orb}=2\pi/\Omega$, the two oscillation periods of the emf (see section~\ref{sec:cycle_period_analytic_QLT}) are given by
\begin{align}
T_{\Delta1}
&=
\frac{4\pi}
{\left|\Delta\omega_\infty-\Delta\omega_{\rm c}\right|}=
\frac{8192}{q(144-q)}
\frac{k'}{k'_z}\,
t_{\rm orb},
\label{eq:TDelta1_q}
\\
T_{\Delta2}
&=
\frac{4\pi}
{\Delta\omega_\infty+\Delta\omega_{\rm c}}=
\frac{8192}{16384+q(144-q)}
\frac{k'}{k'_z}\,
t_{\rm orb}.
\label{eq:TDelta2_q}
\end{align}

For a mode geometry characterized by $k'_R/k'_z\simeq k_R/k_z= a$, we have
$k'/k'_z\simeq\sqrt{1+a^2}$. More generally, the geometry factor may be
written as
\begin{equation}
\frac{k'}{k'_z} = \mathcal A(a)=
\begin{cases}
\sqrt{1+a^2}, & \bk'\parallel\bk,\\[3pt]
\sqrt{1+a^{-2}}, & \bk'\perp\bk .
\end{cases}
\end{equation}
Therefore
\begin{align}
T_{\Delta1}
&\simeq
\frac{8192}{q(144-q)}
\mathcal A(a)\,
t_{\rm orb},
\\
T_{\Delta2}
&\simeq
\frac{8192}{16384+q(144-q)}
\mathcal A(a)\,
t_{\rm orb}.
\end{align}
For Keplerian shear, $q=3/2$, these become
\begin{align}
T_{\Delta1}
&\simeq
38.3\,\mathcal A(a)\,t_{\rm orb},
\\
T_{\Delta2}
&\simeq
0.494\,\mathcal A(a)\,t_{\rm orb}.
\end{align}

For comparison, the periods may also be computed directly from the unexpanded
branch frequencies. At $k'=2k'_{\rm c}$, define
\begin{equation}
s=4-2q,\quad
\chi(q)=
\left[
1+\frac{128q}{s^2}
\right]^{1/2}.
\end{equation}
The beat frequency at this scale is
\begin{align}
\Delta\omega_{\rm c}^{\rm ex}
&=
\mu\Omega\,
\Bigg[
\left(
8q+\frac{s}{2}
\left[
1+\chi(q)
\right]
\right)^{1/2}
-
\left(
8q+\frac{s}{2}
\left[
1-\chi(q)
\right]
\right)^{1/2}
\Bigg].
\label{eq:Deltaomega_c_exact}
\end{align}
Writing $\Delta\omega_{\rm c}^{\rm ex}=\mu\Omega F(q)$, where
\begin{align}
F(q)
&\equiv
\left(
8q+\frac{s}{2}
\left[
1+\chi(q)
\right]
\right)^{1/2}
-
\left(
8q+\frac{s}{2}
\left[
1-\chi(q)
\right]
\right)^{1/2},
\end{align}
the exact beat periods are
\begin{align}
T_{\Delta1}^{\rm ex}
&=
\frac{2}{|F(q)-2|}
\mathcal A(a)\,
t_{\rm orb},
\\
T_{\Delta2}^{\rm ex}
&=
\frac{2}{F(q)+2}
\mathcal A(a)\,
t_{\rm orb}.
\end{align}
For Keplerian shear, $q=3/2$, the exact values are
\begin{align}
T_{\Delta1}^{\rm ex}
&\simeq
37.6\,\mathcal A(a)\,t_{\rm orb},
\\
T_{\Delta2}^{\rm ex}
&\simeq
0.493\,\mathcal A(a)\,t_{\rm orb}.
\end{align}

\section{Computation of the dynamo coefficients}\label{App:dynamo_coeff}

Since we are interested in the large-scale dynamo, we can perform a Taylor series expansion of the emf in $\bk$. In other words, we can expand the induction $\pmb{\calE}_{\bk}$ as
$\pmb{\calE}_{\bk} = \pmb{\calE}_{\bk}^{(0)} + \pmb{\calE}_{\bk}^{(1)} + \cdots$,
where $\pmb{\calE}_{\bk}^{(i)}$ is $\left(i+1\right)^{(\rm th)}$ order in $\bk$. The corresponding emf is $i^{(\rm th)}$ order in $\bk$.

\subsection{$\alpha$ dynamo}\label{App:alpha}

Taking $k \lesssim k'$ in $\bv_{1\bk - \bk'}$, $\bB_{1\bk - \bk'}$, noting that $\bQ_{1-\bk'} = \bQ^\ast_{1\bk'}$ for $\bQ = \bv$ or $\bB$, $s_{i-\bk'} = s^\ast_{i\bk'}$ and $d_{i-\bk'} = -d^\ast_{i\bk'}$, we obtain the following leading order contribution to the $\phi$ induction:
\begin{align}
{\bf\calE}^{(0)}_{\bk\phi}\left(\bB,t\right)
&=
\bB \cdot \Omega \int \rmd \bk'\,
\left(k_z \frac{k'_R}{k'_z} - k_R\right)\,\bk'\;
{\left|\delta v_{\bk'R}\right|}^2
\nonumber\\
&\times
\sum_{i,j}
\frac{
\calR^{(\phi)}_{\bk'ij}\left(\bB,t\right)
+i\calI^{(\phi)}_{\bk'ij}\left(\bB,t\right)
}{
\omega^2_{j\,\bk'} - {\left(\bk'\cdot\bv_\rmA\right)}^2
},
\nonumber\\[1ex]
\calR^{(\phi)}_{\bk'ij}\left(\bB,t\right)
&=
\left(s_{i\bk'}s^\ast_{j\bk'} - d_{i\bk'}d^\ast_{j\bk'}\right)
\nonumber\\
&\times
\left[
1 - \frac{\omega_{j\bk'}}{2\omega_{i\bk'}}
\left(
\frac{\kappa^2}{2\Omega^2}
-\frac{{\left(\bk'\cdot\bv_\rmA\right)}^2}{\omega^2_{j\bk'}}
\frac{\rmd\ln\Omega}{\rmd \ln R}
\right)
\right]
\nonumber\\
&\times
\cos\!\left[\left(\omega_{i\bk'}-\omega_{j\bk'}\right)t\right]
\nonumber\\
&+
\left(s_{i\bk'}s^\ast_{j\bk'} + d_{i\bk'}d^\ast_{j\bk'}\right)
\nonumber\\
&\times
\left[
1 + \frac{\omega_{j\bk'}}{2\omega_{i\bk'}}
\left(
\frac{\kappa^2}{2\Omega^2}
-\frac{{\left(\bk'\cdot\bv_\rmA\right)}^2}{\omega^2_{j\bk'}}
\frac{\rmd\ln\Omega}{\rmd \ln R}
\right)
\right]
\nonumber\\
&\times
\cos\!\left[\left(\omega_{i\bk'}+\omega_{j\bk'}\right)t\right],
\nonumber\\[1ex]
\calI^{(\phi)}_{\bk'ij}\left(\bB,t\right)
&=
-\left(s_{i\bk'}d^\ast_{j\bk'} + d_{i\bk'}s^\ast_{j\bk'}\right)
\nonumber\\
&\times
\left[
1 - \frac{\omega_{j\bk'}}{2\omega_{i\bk'}}
\left(
\frac{\kappa^2}{2\Omega^2}
-\frac{{\left(\bk'\cdot\bv_\rmA\right)}^2}{\omega^2_{j\bk'}}
\frac{\rmd\ln\Omega}{\rmd \ln R}
\right)
\right]
\nonumber\\
&\times
\sin\!\left[\left(\omega_{i\bk'}-\omega_{j\bk'}\right)t\right]
\nonumber\\
&+
\left(s_{i\bk'}d^\ast_{j\bk'} - d_{i\bk'}s^\ast_{j\bk'}\right)
\nonumber\\
&\times
\left[
1 + \frac{\omega_{j\bk'}}{2\omega_{i\bk'}}
\left(
\frac{\kappa^2}{2\Omega^2}
-\frac{{\left(\bk'\cdot\bv_\rmA\right)}^2}{\omega^2_{j\bk'}}
\frac{\rmd\ln\Omega}{\rmd \ln R}
\right)
\right]
\nonumber\\
&\times
\sin\!\left[\left(\omega_{i\bk'}+\omega_{j\bk'}\right)t\right].
\label{emf_1_alpha}
\end{align}

It is not hard to see that, since ${\left|\delta v_{\bk'R}\right|}^2$ is even in $\bk'$, and so are $s_{i\bk'}s^\ast_{j\bk'}$ or $d_{i\bk'}d^\ast_{j\bk'}$, the integrand for the real part is odd in $\bk'$ and thus the real part is zero. The imaginary part involves the cross term between $s_{i\bk'}$ and $d^\ast_{j\bk'}$, which makes it odd in $\bk'$. The already existing $\bk'$ factor in the integrand then makes it even, and we have a non-zero imaginary part. And, since $\calE_{\bk\phi} = i\left(k_z\varepsilon_{\bk R} - k_R\varepsilon_{\bk z}\right)$, we have $\pmb{\varepsilon}^{(0)} = \pmb{\alpha}\cdot\bB$ with $\pmb{\alpha}$ a tensor, whose components are given by
\begin{equation}
\begin{split}
\alpha_{RR} &= \alpha_{zR}\nonumber\\
&=
\Omega \int \rmd \bk'\,\frac{k'^2_R}{k'_z}\,
{\left|\delta v_{\bk'R}\right|}^2
\sum_{i,j}
\frac{1}{\omega^2_{j\,\bk'} - {\left(\bk'\cdot\bv_\rmA\right)}^2}\,
\calI^{(\phi)}_{\bk'ij}\left(\bB,t\right),
\\
\alpha_{Rz} &= \alpha_{zz}\nonumber\\
&=
\Omega \int \rmd \bk'\,k'_z\,
{\left|\delta v_{\bk'R}\right|}^2
\sum_{i,j}
\frac{1}{\omega^2_{j\,\bk'} - {\left(\bk'\cdot\bv_\rmA\right)}^2}\,
\calI^{(\phi)}_{\bk'ij}\left(\bB,t\right).
\end{split}
\label{alpha_Rz}
\end{equation}
These are not the only non-zero components of $\pmb{\alpha}$. In equation~(\ref{emf_2_simp}) for the radial induction, similar arguments as above yield the real (imaginary) part zero (non-zero). Noting that $\calE_{\bk R} = -i k_z \varepsilon_{\bk\phi}$ and comparing it to equation~(\ref{emf_2_simp}), we can obtain the remaining non-zero components of $\pmb{\alpha}$:
\begin{equation}
\begin{split}
\alpha_{\phi R}
&= -\frac{1}{2}
\int \rmd \bk'\,\frac{k'^2_R}{k'_z}
{\left|\delta v_{\bk'R}\right|}^2
\sum_{i,j}\calI^{(R)}_{\bk\bk'ij}\left(\bB,t\right),
\\
\alpha_{\phi z}
&= -\frac{1}{2}
\int \rmd \bk'\,k'_z\,
{\left|\delta v_{\bk'R}\right|}^2
\sum_{i,j}\calI^{(R)}_{\bk\bk'ij}\left(\bB,t\right),
\end{split}
\label{alpha_phi}
\end{equation}
where $\calI^{(R)}_{\bk\bk'ij}$ is given by
\begin{equation}
\begin{split}
&\calI^{(R)}_{\bk\bk'ij}\left(\bB,t\right)\nonumber\\
&=
\left(s_{i\bk'}d^\ast_{j\bk'} + d_{i\bk'}s^\ast_{j\bk'}\right)
\left(\frac{1}{\omega_{i\bk'}} - \frac{1}{\omega_{j\bk'}}\right)
\cos\!\left[\left(\omega_{i\bk'}-\omega_{j\bk'}\right)t\right]
\\
&\quad-
\left(s_{i\bk'}d^\ast_{j\bk'} - d_{i\bk'}s^\ast_{j\bk'}\right)
\left(\frac{1}{\omega_{i\bk'}} + \frac{1}{\omega_{j\bk'}}\right)
\cos\!\left[\left(\omega_{i\bk'}+\omega_{j\bk'}\right)t\right].
\end{split}
\label{I_R}
\end{equation}

Now let us try to connect $\pmb{\alpha}$ to known quantities. We will show that $\alpha_{ij}$ ($i,j=R$ or $z$) can be approximately expressed in terms of the mean kinetic helicity $\left<\bv_1\cdot\pmb{\omega}_1\right>$ and the mean current helicity $\left<\bJ_1\cdot\bB_1\right>/(4\pi\rho_0)$. Upon substituting the linear eigenfunctions, the mean kinetic helicity can be expressed as
\begin{equation}
\begin{split}
\calH_{\rm kin}
&= i\int\rmd\bk'\,\bv^\ast_{1\bk'}\cdot\left(\bk'\times\bv_{1\bk'}\right)
\nonumber\\
&= -2\int\rmd\bk'\,\frac{k'^2}{k'_z}\,
{\rm Im}\left(v^\ast_{1\bk'\phi} v_{1\bk'R}\right)
\nonumber\\
&=
\Omega\int\rmd\bk'\,\frac{k'^2}{k'_z}
{\left|\delta v_{\bk'R}\right|}^2
\sum_{i,j}
\frac{\omega_{j\bk'}}{\omega^2_{j\bk'}-{\left(\bk'\cdot\bv_\rmA\right)}^2}
\nonumber\\
&\times
\left[
-\frac{\kappa^2}{2\Omega^2}
+\frac{{\left(\bk'\cdot\bv_\rmA\right)}^2}{\omega^2_{j\bk'}}
\frac{\rmd\ln\Omega}{\rmd\ln R}
\right]
\nonumber\\
&\times
\Big[
\left(s_{i\bk'}d^\ast_{j\bk'}+d_{i\bk'}s^\ast_{j\bk'}\right)
\cos\!\left[\left(\omega_{i\bk'}-\omega_{j\bk'}\right)t\right]
\nonumber\\
&+
\left(s_{i\bk'}d^\ast_{j\bk'}-d_{i\bk'}s^\ast_{j\bk'}\right)
\cos\!\left[\left(\omega_{i\bk'}+\omega_{j\bk'}\right)t\right]
\Big].
\end{split}
\end{equation}
Similarly, the mean current helicity can be written as
\begin{equation}
\begin{split}
\calH_{\rm cur}
&=
\frac{i}{4\pi\rho_0}
\int\rmd\bk'\,\bB^\ast_{1\bk'}\cdot\left(\bk'\times\bB_{1\bk'}\right)
\nonumber\\
&=
-\frac{2}{4\pi\rho_0}
\int\rmd\bk'\,\frac{k'^2}{k'_z}\,
{\rm Im}\left(B^\ast_{1\bk'\phi} B_{1\bk'R}\right)
\nonumber\\
&=
-2\Omega\int\rmd\bk'\,\frac{k'^2}{k'_z}
{\left|\delta v_{\bk'R}\right|}^2
{\left(\bk'\cdot\bv_\rmA\right)}^2\nonumber\\
&\times\sum_{i,j}
\frac{1}{\omega_{i\bk'}\left(\omega^2_{j\bk'}-{\left(\bk'\cdot\bv_\rmA\right)}^2\right)}
\nonumber\\
&\times
\Big[
\left(s_{i\bk'}d^\ast_{j\bk'}+d_{i\bk'}s^\ast_{j\bk'}\right)
\cos\!\left[\left(\omega_{i\bk'}-\omega_{j\bk'}\right)t\right]
\nonumber\\
&-
\left(s_{i\bk'}d^\ast_{j\bk'}-d_{i\bk'}s^\ast_{j\bk'}\right)
\cos\!\left[\left(\omega_{i\bk'}+\omega_{j\bk'}\right)t\right]
\Big].
\end{split}
\end{equation}
Now, the difference between these two helicities is given by
\begin{align}
\calH_{\rm cur}-\calH_{\rm kin}
&=
2\Omega \int\rmd\bk'\,\frac{k'^2}{k'_z}
|\delta v_{\bk'R}|^2\nonumber\\
&\times\sum_{i,j}
\frac{\omega_{i\bk'}}{\omega^2_{j\bk'}-{\left(\bk'\cdot\bv_\rmA\right)}^2}
\calG_{\bk'ij}^{(\phi)}(\bB,t),
\nonumber\\
\calG_{\bk'ij}^{(\phi)}(\bB,t)
&=
-\left(s_{i\bk'}d^\ast_{j\bk'} + d_{i\bk'}s^\ast_{j\bk'}\right)
\nonumber\\
&\times
\left[
\frac{{\left(\bk'\cdot\bv_\rmA\right)}^2}{\omega^2_{i\bk'}}
-
\frac{\omega_{j\bk'}}{2\omega_{i\bk'}}
\left(
\frac{\kappa^2}{2\Omega^2}
-\frac{{\left(\bk'\cdot\bv_\rmA\right)}^2}{\omega^2_{j\bk'}}
\frac{\rmd\ln\Omega}{\rmd \ln R}
\right)
\right]
\nonumber\\
&\times
\cos\!\left[\left(\omega_{i\bk'}-\omega_{j\bk'}\right)t\right]
\nonumber\\
&+
\left(s_{i\bk'}d^\ast_{j\bk'} - d_{i\bk'}s^\ast_{j\bk'}\right)
\nonumber\\
&\times
\left[
\frac{{\left(\bk'\cdot\bv_\rmA\right)}^2}{\omega^2_{i\bk'}}
+
\frac{\omega_{j\bk'}}{2\omega_{i\bk'}}
\left(
\frac{\kappa^2}{2\Omega^2}
-\frac{{\left(\bk'\cdot\bv_\rmA\right)}^2}{\omega^2_{j\bk'}}
\frac{\rmd\ln\Omega}{\rmd \ln R}
\right)
\right]
\nonumber\\
&\times
\cos\!\left[\left(\omega_{i\bk'}+\omega_{j\bk'}\right)t\right].
\end{align}
Noting that $\omega_{i\bk'}\approx \bk'\cdot\bv_\rmA$ at large enough $\bk'$ and $\omega_{1\bk'}<\omega_{2\bk'}$ for all $\bk'$, it is not difficult to see by comparing the above equation to $\pmb{\alpha}$ in equations~(\ref{alpha_Rz}) and $\calI^{(\phi)}_{\bk'}$ in equations~(\ref{emf_1_alpha}), that
\begin{align}
\alpha_{RR}+\alpha_{Rz} 
&=\alpha_{RR}+\alpha_{zz}
=
\alpha_{zR}+\alpha_{Rz}
=\alpha_{zR}+\alpha_{zz}\nonumber\\
&\sim
\int \rmd t\,\left(\calH_{\rm cur}-\calH_{\rm kin}\right).
\end{align}
Similarly, it is not hard to see that $\alpha_{\phi R}$ and $\alpha_{\phi z}$ are related to but not quite equal to the radial part of the mean cross helicity $\left<\bv_1\cdot\bB_1\right>$, $\int \rmd \bk'\, {\rm Re}\left(v_{1\bk'R}^\ast B_{1\bk'R}\right)$.

\subsection{$\beta$ dynamo}\label{App:beta}

Let us now obtain the next order contribution, $\pmb{\calE}^{(1)}_{\bk}$. The toroidal induction at this order is given by
\begin{align}
{\bf\calE}^{(1)}_{\bk\phi}
&\approx
\Omega \int \rmd \bk'\,
\left(k_z \frac{k'_R}{k'_z} - k_R\right)\,
\bk'\cdot\bB
\sum_{n=1}^3 \calJ^{(n)}_{\bk\bk'\phi}\left(\bB,t\right),
\nonumber\\
\calJ^{(1)}_{\bk\bk'\phi}\left(\bB,t\right)
&=
-\frac{k_l}{2}
\frac{\partial}{\partial k'_l}
{\left|\delta v_{\bk'R}\right|}^2
\sum_{i,j}
\frac{
\calR^{(\phi)}_{\bk\bk'ij}\left(\bB,t\right)
}{
\omega^{2}_{j\bk'} - {\left(\bk'\cdot\bv_\rmA\right)}^2
},
\nonumber\\
\calJ^{(2)}_{\bk\bk'\phi}\left(\bB,t\right)
&=
-k_l\,{\left|\delta v_{\bk'R}\right|}^2
\sum_{i,j}
\frac{
\partial \omega^2_{j\bk'} / \partial k'_l
}{
{\left(\omega^2_{j\bk'} - {\left(\bk'\cdot\bv_\rmA\right)}^2\right)}^2
}
\calR^{(\phi)}_{\bk\bk'ij}\left(\bB,t\right),
\nonumber\\
\calJ^{(3)}_{\bk\bk'\phi}\left(\bB,t\right)
&=
\frac{k_l}{2}\,{\left|\delta v_{\bk'R}\right|}^2
\sum_{i,j}
\frac{1}{\omega^{2}_{j\bk'} - {\left(\bk'\cdot\bv_\rmA\right)}^2}\,
\frac{1}{\omega_{i\bk'}}
\frac{\partial \omega_{j\bk'}}{\partial k'_l}
\nonumber\\
&\quad\times
\left(
\frac{\kappa^2}{2\Omega^2}
+\frac{{\left(\bk'\cdot\bv_\rmA\right)}^2}{\omega^{\ast 2}_{j\bk'}}
\frac{\rmd\ln\Omega}{\rmd \ln R}
\right)
\calS^{(\phi)}_{\bk\bk'ij}\left(\bB,t\right).
\label{emf_5_app}
\end{align}
where
\begin{align}
\calR^{(\phi)}_{\bk\bk'ij}\left(\bB,t\right)
&=
\left(s_{i\bk'}s_{j\bk'}^\ast + d_{i\bk'}d_{j\bk'}^\ast\right)
\nonumber\\
&\quad\times
\left[
1 - \frac{\omega_{j\bk'}}{2\omega_{i\bk'}}
\left(
\frac{\kappa^2}{2\Omega^2}
-\frac{{\left(\bk'\cdot\bv_\rmA\right)}^2}{\omega^2_{j\bk'}}
\frac{\rmd\ln\Omega}{\rmd \ln R}
\right)
\right]
\nonumber\\
&\quad\times
\cos\!\Bigg[
\Bigg(
\omega_{i\bk'}-\omega_{j\bk'}
+\frac{\partial \omega_{j\bk'}}{\partial k'_l} k_l
\Bigg)t
\Bigg]
\nonumber\\
&\quad+
\left(s_{i\bk'}s_{j\bk'}^\ast - d_{i\bk'}d_{j\bk'}^\ast\right)
\nonumber\\
&\quad\times
\left[
1 + \frac{\omega_{j\bk'}}{2\omega_{i\bk'}}
\left(
\frac{\kappa^2}{2\Omega^2}
-\frac{{\left(\bk'\cdot\bv_\rmA\right)}^2}{\omega^2_{j\bk'}}
\frac{\rmd\ln\Omega}{\rmd \ln R}
\right)
\right]
\nonumber\\
&\quad\times
\cos\!\Bigg[
\Bigg(
\omega_{i\bk'}+\omega_{j\bk'}
-\frac{\partial \omega_{j\bk'}}{\partial k'_l} k_l
\Bigg)t
\Bigg],
\nonumber\\
\calS^{(\phi)}_{\bk\bk'ij}\left(\bB,t\right)
&=
\left(s_{i\bk'}s^\ast_{j\bk'} + d_{i\bk'}d^\ast_{j\bk'}\right)
\nonumber\\
&\quad\times
\cos\!\Bigg[
\Bigg(
\omega_{i\bk'}-\omega_{j\bk'}
+\frac{\partial \omega_{j\bk'}}{\partial k'_l} k_l
\Bigg)t
\Bigg]
\nonumber\\
&\quad-
\left(s_{i\bk'}s^\ast_{j\bk'} - d_{i\bk'}d^\ast_{j\bk'}\right)
\nonumber\\
&\quad\times
\cos\!\Bigg[
\Bigg(
\omega_{i\bk'}+\omega_{j\bk'}
-\frac{\partial \omega_{j\bk'}}{\partial k'_l} k_l
\Bigg)t
\Bigg].
\label{Rkij_app}
\end{align}
Note that the superscript in $\pmb{\calE}^{(1)}_{\bk}$ stands for first order in $\bk$, while that in $\calJ^{(n)}_{\bk\bk'\phi}$ indicates the index for each term in $\calE^{(1)}_{\bk}$. We have expanded $\omega_{j\bk-\bk'}\approx \omega_{j\bk'} - k_l\,\partial\omega_{j\bk'}/\partial k'_l$. From the expression for the frequencies in equation~(\ref{MRI_freqs}), we find that
\begin{align}
\frac{1}{2}\frac{\partial \omega^2_{\bk'}}{\partial k'_R}
&=
\left(\bk'\cdot\bv_\rmA\right) v_{\rmA R}
\nonumber\\
&\quad
-\frac{k'_R k'^2_z}{2 k'^4}\,\kappa^2
\left(
1 \pm \sqrt{
1 + \frac{16\,k'^2}{k'^2_z}
\frac{\Omega^2 {\left(\bk'\cdot\bv_\rmA\right)}^2}{\kappa^4}}
\right)
\nonumber\\
&\quad\pm
\frac{4\Omega^2}{\kappa^2}
\frac{
\bk'\cdot\bv_\rmA
\left(
\dfrac{k'_R}{k'^2}\,\bk'\cdot\bv_\rmA + v_{\rmA R}
\right)
}{
\sqrt{
1 + \dfrac{16\,k'^2}{k'^2_z}
\dfrac{\Omega^2 {\left(\bk'\cdot\bv_\rmA\right)}^2}{\kappa^4}}
},
\nonumber\\[1ex]
\frac{1}{2}\frac{\partial \omega^2_{\bk'}}{\partial k'_z}
&=
\left(\bk'\cdot\bv_\rmA\right) v_{\rmA z}
\nonumber\\
&\quad
+\frac{k'_z k'^2_R}{2 k'^4}\,\kappa^2
\left(
1 \pm \sqrt{
1 + \frac{16\,k'^2}{k'^2_z}
\frac{\Omega^2 {\left(\bk'\cdot\bv_\rmA\right)}^2}{\kappa^4}}
\right)
\nonumber\\
&\quad\pm
\frac{4\Omega^2}{\kappa^2}
\frac{
\bk'\cdot\bv_\rmA
\left(
-\dfrac{k'^2_R}{k'_z k'^2}\,\bk'\cdot\bv_\rmA + v_{\rmA z}
\right)
}{
\sqrt{
1 + \dfrac{16\,k'^2}{k'^2_z}
\dfrac{\Omega^2 {\left(\bk'\cdot\bv_\rmA\right)}^2}{\kappa^4}}
}.
\end{align}
Noting that $\bk\cdot\bv_\rmA = 0$, we have that
\begin{equation}
\begin{split}
\frac{k_l}{2}\frac{\partial \omega^2_{\bk'}}{\partial k'_l}
&=
\left(k_z k'_R - k_R k'_z\right)
\frac{k'_R k'_z}{k'^4}\nonumber\\
&\times\Bigg[
\frac{\kappa^2}{2}
\left(
1 \pm \sqrt{
1 + \frac{16\,k'^2}{k'^2_z}
\frac{\Omega^2 {\left(\bk'\cdot\bv_\rmA\right)}^2}{\kappa^4}}
\right)
\\
&
\mp
\frac{4\Omega^2}{\kappa^2}
\frac{k'^2}{k'^2_z}
\frac{{\left(\bk'\cdot\bv_\rmA\right)}^2}{
\sqrt{
1 + \dfrac{16\,k'^2}{k'^2_z}
\dfrac{\Omega^2 {\left(\bk'\cdot\bv_\rmA\right)}^2}{\kappa^4}}
}
\Bigg].
\end{split}
\end{equation}
Plugging this in the third and fourth of equations~(\ref{emf_5_app}) and assuming that we have a generic form for unstratified MRI turbulence, i.e., ${\left|\delta v_{\bk'R}\right|}^2 = \calN\, {\left|k'_\perp\right|}^{-\delta}$, with $k'_\perp = \left|\bk'\times \hat{\bB}\right|
= \left(k'_z B_R - k'_R B_z\right)/\sqrt{B^2_R + B^2_z}$ and $\delta > 0$, we can approximately compute the integral in equation~(\ref{emf_5_app}) to obtain the following form for the toroidal induction:
\begin{align}
{\bf\calE}_{\bk\phi}
&\approx
\Omega\;
\frac{{\left(k_z B_R - k_R B_z\right)}^2}{B_z}\,
\calN_\phi
\sum_{n=1}^3 \calL^{(n)}_{\bk\phi}\left(\bB,t\right),
\nonumber\\
\calL^{(1)}_{\bk\phi}\left(\bB,t\right)
&=
\frac{\delta}{2}\,
{\rm sgn}\left(k_z B_R - k_R B_z\right)
\int_{k'_\rmc}^\infty \rmd k'\, {k'}^{1-\delta}
\nonumber\\
&\times
\sum_{i,j}
\frac{
\calR^{(\phi)}_{\bk k'ij}\left(\bB,t\right)
}{
\omega^{2}_{jk'} - k'^2 v^2_\rmA
},
\nonumber\\
\calL^{(2)}_{\bk\phi}\left(\bB,t\right)
&=
-\frac{2 B_R B_z}{B^2_R + B^2_z}
\int_{k'_\rmc}^\infty \rmd k'\, {k'}^{1-\delta}
\nonumber\\
&\times
\sum_{i,j}
\frac{
\calK_{k'j}\left(\bB\right)\,
\calR^{(\phi)}_{\bk k'ij}\left(\bB,t\right)
}{
{\left(\omega^{2}_{j k'} - k'^2 v^2_\rmA\right)}^2
},
\nonumber\\
\calL^{(3)}_{\bk\phi}\left(\bB,t\right)
&=
\frac{B_R B_z}{B^2_R + B^2_z}
\int_{k'_\rmc}^\infty \rmd k'\, {k'}^{1-\delta}
\nonumber\\
&\times
\sum_{i,j}
\frac{
1
}{
\omega_{i k'} \omega_{j k'}
\left(\omega^{2}_{jk'} - k'^2 v^2_\rmA\right)
}
\nonumber\\
&\times
\left(
\frac{\kappa^2}{2\Omega^2}
+ \frac{k'^2 v^2_\rmA}{\omega^{\ast 2}_{jk'}}
\frac{\rmd\ln\Omega}{\rmd \ln R}
\right)
\calK_{k'j}\left(\bB\right)
\calS^{(\phi)}_{\bk\bk'ij}\left(\bB,t\right).
\label{emf_6_app}
\end{align}
In reducing the dimensionality of the induction integral from $2$ to $1$, we have made use of the fact that the dominant contribution to the integral comes from $\left|k'_\perp\right| \to 0$, i.e., $\bk'$ nearly parallel to $\bB$, due to the ${\left|k'_\perp\right|}^{-\delta}$ dependence of the fluctuation power spectrum. The key simplification is that the $k_l\, \partial/\partial k'_l$ factor yields a $(k_R B_z - k_z B_R) = -i J_{\bk\phi}$ factor everywhere it appears. The factors arising from integrating over $\phi_{\bk'} = \cos^{-1}\left(\hat{\bk'}\cdot\hat{\bB}\right)$ have been absorbed in the normalization $\calN$, which we now call $\calN_\phi$. This contains information about the level of fluctuations. The integrand diverges at $k'\to k'_\rmc$ since $\omega_{2\bk'}\to 0$ in this limit. This can be mitigated by assuming a broadening of the marginally stable region due to higher order wave-wave coupling, which would typically lead to a width of $\sim k'_\rmc$ around $k' = k'_\rmc$. Here we restrict the integration bound to the stable/oscillatory region, $k'>k'_\rmc$, since the dynamo cycles arise from the interference of the oscillatory modes. $\calR^{(\phi)}_{\bk k'ij}\left(\bB,t\right)$ is given by equation~(\ref{Rkij_app}) but with $\bk' = k' \left(B_R \hat{R} + B_z \hat{z}\right) / \sqrt{B^2_R + B ^2_z}$, and $\calK_{k'j}$ is given by
\begin{equation}
\begin{split}
\calK_{k'j}\left(\bB\right)
&=
\frac{\kappa^2}{2}
\left[
1 + s_j \sqrt{
1 + 16\left(1 + \frac{B^2_R}{B^2_z}\right)
\frac{\Omega^2 k'^2 v^2_\rmA}{\kappa^4}}
\right]
\\
&\quad
- s_j \frac{4\Omega^2}{\kappa^2}
\left(1 + \frac{B^2_R}{B^2_z}\right)
\frac{k'^2 v^2_\rmA}{
\sqrt{
1 + 16\left(1 + \dfrac{B^2_R}{B^2_z}\right)
\dfrac{\Omega^2 k'^2 v^2_\rmA}{\kappa^4}}
},
\end{split}
\label{Kj}
\end{equation}
with $s_j = 1(-1)$ for $j=1(2)$. The $\omega_{j\bk'}$ in $\calR^{(\phi)}_{\bk k'ij}$ is to be evaluated at $\bk' = k' \left(B_R \hat{R} + B_z \hat{z}\right) / \sqrt{B^2_R + B ^2_z}$.

\begin{figure*}
\centering
\includegraphics[width=0.6\textwidth]{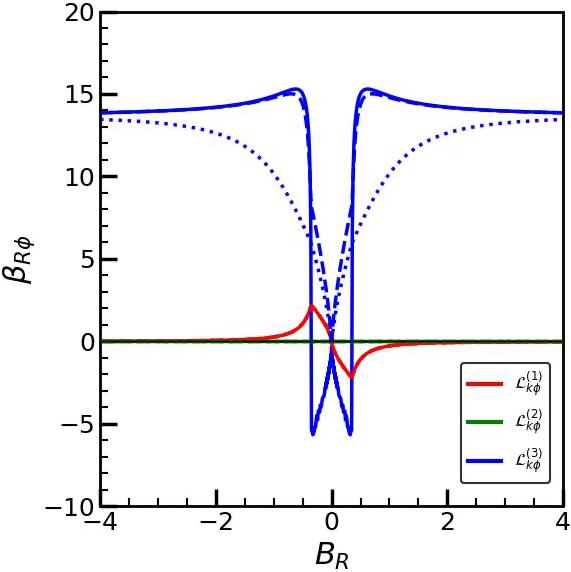}
\caption{The coefficient $\beta_{R \phi}$ responsible for growing the toroidal emf from the azimuthal shear current through the emf $\varepsilon_R = \beta_{R \phi} J_\phi$. The different contributions are shown by different colors. $\calL_{k\phi}^{(2)}$ is identically zero, $\calL_{k\phi}^{(3)}\sim v_{1\bk \phi} B_{1\bk R}$ is the dominant contributor at large $B_R$, and $\calL_{k\phi}^{(1)}\sim v_{1\bk R} B_{1\bk \phi}$ dominates at small $B_R$. The solid, dashed and dotted lines represent broadening values $\epsilon_\omega = 10^{-2}, 10^{-1}$ and $1$ (see the text after equation~[\ref{beta_phiR_app}]).}
\label{fig:beta}
\end{figure*}

Since the toroidal induction $\calE_{\bk\phi}$ is equal to $i\left(k_z \varepsilon_{\bk R} - k_R \varepsilon_{\bk z}\right)$, with $\pmb{\varepsilon}_{\bk}$ the Fourier transform of the emf, and the azimuthal current $J_{\bk\phi}$ is equal to $i(k_z B_R - k_R B_z)$, it is clear from equation~(\ref{emf_6_app}) that, to leading order, the emf can be recast into the form
\begin{equation}
\pmb{\varepsilon} \approx \pmb{\beta} \cdot \bJ,
\end{equation}
with $\pmb{\beta}$ a tensor, given by
\begin{equation}
\beta_{R\phi}
= - \Omega \,\frac{B_R}{B_z}\,
\calN_\phi
\sum_{n=1}^3 \calL^{(n)}_\phi\left(\bB,t\right),
\label{beta_phiR_app}
\end{equation}
where $\calL_{\phi}^{(n)} = \lim_{\bk\to 0}\calL_{\bk\phi}^{(n)}$. Let us now inspect the sign of the mean part of $\calE_{\bk\phi}$. It is not hard to see that the $j=1$ and $2$ terms in $\calL^{(2)}_{\bk\phi}$ exactly cancel, so $\calL^{(2)}_{\bk\phi}$ is identically zero. The $\calL^{(1)}_{\bk\phi}$ is odd in $B_R$ while $\calL^{(3)}_{\bk\phi}$ is even. The $\calL^{(3)}_{\bk\phi}$ term is typically much larger due to the $\omega_{2\bk'}$ factor in the denominator, which diverges at marginal stability $\left(\bk'\to\bk'_\rmc\right)$. To avoid this divergence, we regularize the $1/\omega_{2\bk'}$ divergence by replacing it with $\omega_{2\bk'}/(\omega_{2\bk'}^2 + \epsilon_\omega^2)$. We plot the different terms as a function of $B_R$ (assuming $k_R = 1$, $k_z = 2$, $\delta = 8/3$) for $\epsilon_\omega = 10^{-2}, 10^{-1}$ and $1$ in Fig.~\ref{fig:beta}. Note that $\calL^{(3)}_{\bk\phi}$ is of order $\langle v_{1\bk\phi} B_{1\bk R}\rangle$ while $\calL^{(1)}_{\bk\phi}$ is of order $\langle v_{1\bk R} B_{1\bk \phi}\rangle$. All in all, $\beta_{R \phi}$ is nearly always positive, except at small $B_R$ for small broadening. Since the third term almost always dominates, a very good approximation for $\beta_{R \phi}$ is
\begin{align}
\beta_{R \phi}
&\approx
-\calN_\phi \Omega\,
\frac{1}{1 + \dfrac{B^2_z}{B^2_R}}
\int_{k'_\rmc}^\infty \rmd k'\,{k'}^{1-\delta}
\nonumber\\
&\times
\sum_{i,j}
\frac{
1
}{
\omega_{i k'} \omega_{j k'}
\left(\omega^{2}_{jk'} - k'^2 v^2_\rmA\right)
}
\nonumber\\
&\times
\left(
\frac{\kappa^2}{2\Omega^2}
+ \frac{k'^2 v^2_\rmA}{\omega^{\ast 2}_{jk'}}
\frac{\rmd\ln\Omega}{\rmd \ln R}
\right)
\calK_{k'j}\left(\bB\right)
\calS^{(\phi)}_{\bk\bk'ij}\left(\bB,t\right),
\nonumber\\
\calS^{(\phi)}_{\bk\bk'ij}\left(\bB,t\right)
&=
\left(s_{i\bk'}s^\ast_{j\bk'} + d_{i\bk'}d^\ast_{j\bk'}\right)
\nonumber\\
&\times
\cos\!\left[
\left(
\omega_{i\bk'}-\omega_{j\bk'}
+\frac{\partial \omega_{j\bk'}}{\partial k'_l} k_l
\right)t
\right]
\nonumber\\
&-
\left(s_{i\bk'}s^\ast_{j\bk'} - d_{i\bk'}d^\ast_{j\bk'}\right)
\nonumber\\
&\times
\cos\!\left[
\left(
\omega_{i\bk'}+\omega_{j\bk'}
-\frac{\partial \omega_{j\bk'}}{\partial k'_l} k_l
\right)t
\right].
\end{align}
The integrand is dominated by $k'$ near marginal stability, since $\omega_{2\bk'}\to 0$ in this limit and boosts the integrand. From equation~(\ref{Kj}), it is evident that the first term in $\calK_{k'j}$ dominates at $k'\gtrsim k'_\rmc$. And since the dominant contribution to the integrand comes from $j=2$, $\calK_{k'j}$ is negative for most of $k'$. From equation~(\ref{Rkij_app}), we find that $\calS^{(\phi)}_{\bk\bk'ij}\left(\bB,t\right) \sim \cos^2\left(2\omega_{2\bk'}t\right)$. Putting this altogether, we find that $\beta_{R \phi}$ is usually positive, as also seen from Fig.~\ref{fig:beta}.

We can rewrite the radial induction
\begin{align}
{\bf\calE}_{\bk R}\left(\bB,t\right)
&=
\frac{k_z}{2} \int \rmd \bk'\,\frac{k'_R}{k'_z}\,\bk'\cdot\bB
\sum_{n=1}^2 \calJ^{(n)}_{\bk\bk'R}(\bB,t),
\nonumber\\
\calJ^{(1)}_{\bk\bk'R}(\bB,t)
&=
-k_l\frac{\partial}{\partial k'_l}{\left|\delta v_{\bk'R}\right|}^2
\sum_{i,j} \calR^{(R)}_{\bk\bk'ij}(\bB,t),
\nonumber\\
\calJ^{(2)}_{\bk\bk'R}(\bB,t)
&=
{\left|\delta v_{\bk'R}\right|}^2
\sum_{i,j}
\frac{k_l}{\omega^2_{j\bk'}}
\frac{\partial \omega_{j\bk'}}{\partial k'_l}
\calS^{(R)}_{\bk\bk'ij}\left(\bB,t\right),
\nonumber\\
\calR^{(R)}_{\bk\bk'ij}\left(\bB,t\right)
&=
\left(s_{i\bk'}s^\ast_{j\bk'} + d_{i\bk'}d^\ast_{j\bk'}\right)
\nonumber\\
&\times
\left(\frac{1}{\omega_{i\bk'}} - \frac{1}{\omega_{j\bk'}}\right)
\sin\!\left[
\left(
\omega_{i\bk'}-\omega_{j\bk'}
+\frac{\partial \omega_{j\bk'}}{\partial k'_l} k_l
\right)t
\right]
\nonumber\\
&+
\left(s_{i\bk'}s^\ast_{j\bk'} - d_{i\bk'}d^\ast_{j\bk'}\right)
\nonumber\\
&\times
\left(\frac{1}{\omega_{i\bk'}} + \frac{1}{\omega_{j\bk'}}\right)
\sin\!\left[
\left(
\omega_{i\bk'}+\omega_{j\bk'}
-\frac{\partial \omega_{j\bk'}}{\partial k'_l} k_l
\right)t
\right],
\nonumber\\
\calS^{(R)}_{\bk\bk'ij}\left(\bB,t\right)
&=
-\left(s_{i\bk'}s^\ast_{j\bk'} + d_{i\bk'}d^\ast_{j\bk'}\right)
\nonumber\\
&\times
\sin\!\left[
\left(
\omega_{i\bk'}-\omega_{j\bk'}
+\frac{\partial \omega_{j\bk'}}{\partial k'_l} k_l
\right)t
\right]
\nonumber\\
&+
\left(s_{i\bk'}s^\ast_{j\bk'} - d_{i\bk'}d^\ast_{j\bk'}\right)
\nonumber\\
&\times
\sin\!\left[
\left(
\omega_{i\bk'}+\omega_{j\bk'}
-\frac{\partial \omega_{j\bk'}}{\partial k'_l} k_l
\right)t
\right].
\end{align}

Again, making use of the fact that the dominant contribution to the integral comes from $\left|k'_\perp\right| \to 0$, i.e., $\bk'$ nearly parallel to $\bB$, we can rewrite the above as
\begin{equation}
\begin{split}
{\bf\calE}_{\bk R}\left(\bB,t\right)
&=
\frac{k_z}{2}\frac{B_R}{B_z}
\left(k_z B_R - k_R B_z\right)
\calN_R
\sum_{n=1}^2\calL^{(n)}_{\bk R}(\bB,t),
\\
\calL^{(1)}_{\bk R}\left(\bB,t\right)
&=
{\delta}\,
{\rm sgn}\left(k_z B_R - k_R B_z\right)
\int_{k'_\rmc}^\infty \rmd k'\,{k'}^{1-\delta}
\sum_{i,j}
\calR^{(R)}_{\bk k'ij}\left(\bB,t\right),
\\
\calL^{(2)}_{\bk R}\left(\bB,t\right)
&=
\frac{B_R B_z}{B^2_R + B^2_z}
\int_{k'_\rmc}^\infty \rmd k'\,{k'}^{1-\delta}
\sum_{i,j}
\frac{\calK_{k'j}\left(\bB\right)}{\omega^3_{j k'}}
\calS^{(R)}_{\bk k'ij}\left(\bB,t\right).
\end{split}
\end{equation}
where constant factors arising from the angle integral have been absorbed into the normalization $\calN$, which we now call $\calN_R$. Since the radial induction $\calE_{\bk R}$ is $-i k_z \varepsilon_{\bk \phi}$, and $J_{\bk\phi}$ is equal to $i(k_z B_R - k_R B_z)$, we find that $\varepsilon_{\bk R} = \beta_{\phi\phi} J_{\bk\phi}$ with $\beta_{\phi\phi}$ given by
\begin{equation}
\beta_{\phi\phi}
=
\frac{B_R}{B_z}\frac{\calN_R}{2}
\sum_{n=1}^2 \calL^{(n)}_R\left(\bB,t\right),
\end{equation}
where $\calL_{R}^{(n)} = \lim_{\bk\to 0}\calL_{\bk R}^{(n)}$. Similar arguments as in the toroidal case imply that the dominant contribution to the integrand comes from $k'$ near marginal stability, and therefore the second term almost always dominates over the first. Hence, to very good approximation, we can write
\begin{equation}
\begin{split}
\beta_{\phi \phi}
&\approx
\frac{\calN_R}{2}\,
\frac{1}{1 + \dfrac{B^2_z}{B^2_R}}
\int_{k'_\rmc}^\infty \rmd k'\,{k'}^{1-\delta}
\sum_{i,j}
\frac{\calK_{k'j}\left(\bB\right)}{\omega^3_{j k'}}
\calS^{(R)}_{\bk k'ij}\left(\bB,t\right).
\end{split}
\end{equation}
Since $j=2$ dominates the integrand and $\calK_{k'j}$ is negative in that case, the mean part of $\beta_{\phi\phi}$ turns out to be negative, as opposed to $\beta_{R \phi}$. Thus we see that the only non-zero components of $\pmb{\beta}$ are $\beta_{R \phi}$, $\beta_{\phi z} = (B_z/B_R)\,\beta_{R \phi} = -(k_R/k_z)\,\beta_{R \phi}$ and $\beta_{\phi\phi}$. While $\beta_{R \phi}$ grows the toroidal field, $\beta_{\phi\phi}$ drains the radial field, acting as a turbulent diffusion. Note that $\beta_{R \phi}$ generates the radial emf and the toroidal field, and is therefore different from the $\beta_{\phi R}$ of \citet[][]{Squire.Bhattacharjee.16} that generates the toroidal emf and the radial field. The $\beta_{R\phi} J_\phi$ dependence of the emf here arises from the axisymmetric modes, while the $\beta_{\phi R} J_R$ dependence would require the non-axisymmetric modes since $J_\phi \sim -i k_z B_\phi$.

\subsection{Magnetic helicity conservation and saturation}\label{App:helicity_constraint}

The magnetic helicity flux places additional constraints on the dynamo coefficients. The helicity constraint is best understood by separating the large- and small-scale magnetic helicities. For a mean-field split, $\bB_{\rm tot}=\bB+\delta\bB$ and the vector potential $\bA_{\rm tot}=\bA+\delta\bA$, the corresponding helicity equations are schematically
\begin{align}
&\partial_t(\bA\cdot\bB)
=
2\pmb{\varepsilon}\cdot\bB
-\nabla\cdot\pmb{\calF}_{\rm LS}
-2\eta\,\bJ\cdot\bB,\nonumber\\
&\partial_t\langle\delta\bA\cdot\delta\bB\rangle
=
-2\pmb{\varepsilon}\cdot\bB-\nabla\cdot\pmb{\calF}_{\rm SS}
-2\eta\,\langle\delta\bJ\cdot\delta\bB\rangle,
\end{align}
where $\eta$ is the resistivity, $\delta\bJ = (c/4\pi)\nabla\times \delta\bB$ is the current fluctuation, and $\pmb{\calF}_{\rm LS}$ and $\pmb{\calF}_{\rm SS}$ are the large-scale (LS) and small-scale (SS) helicity fluxes. Thus, if the small-scale helicity has reached a statistically steady state and resistive dissipation is negligible, the parallel emf must be balanced by a small-scale helicity flux,
\[
\pmb{\varepsilon}\cdot\bB
=
-\frac12\nabla\cdot\pmb{\calF}_{\rm SS}.
\]
This imposes the constraint
\[
\bB\cdot\pmb{\alpha}\cdot\bB=0
\]
on the zeroth-order $\alpha$ contribution to the emf, since it does not involve any derivatives of $\bB$ and thus cannot be written as a total divergence. For the MRI $\pmb{\alpha}$ tensor derived above, this condition is
\[
(B_R+B_z)(\alpha_{RR}B_R+\alpha_{Rz}B_z)
+
B_\phi(\alpha_{\phi R}B_R+\alpha_{\phi z}B_z)=0,
\]
where we used $\alpha_{RR}=\alpha_{zR}$ and $\alpha_{Rz}=\alpha_{zz}$. Therefore the tensor $\pmb{\alpha}$ need not vanish; rather, its contraction with the mean field must vanish. Since $B_\phi$ is typically much larger than $B_R$ and $B_z$ in the MRI dynamo, even small $\alpha_{\phi R}$ and $\alpha_{\phi z}$ components, as in the unstratified case, can allow this helicity constraint to be satisfied. 

If $\pmb{\alpha}$ is non-zero, then the helicity constraint provides the relative saturated ratio of the toroidal and poloidal components:

\begin{align}
&\frac{B_\phi}{B_R + B_z} = -\frac{\alpha_{RR}B_R+\alpha_{Rz}B_z}{\alpha_{\phi R}B_R+\alpha_{\phi z}B_z}.
\end{align}
This toroidal-to-poloidal ratio is controlled by the relative amplitudes of
$\calI^{(\phi)}_{ij}/[\omega_j^2-(\bk'\cdot\bv_{\rm A})^2]$ and
$\calI^{(R)}_{ij}$. Near marginal stability, $\omega_{2\bk'}\to0$.
If the slow--slow helicity combination
$H^+_{22}\equiv s_{2\bk'}d^\ast_{2\bk'}-d_{2\bk'}s^\ast_{2\bk'}$
survives, the $22$ term is the most singular and gives
\[
\left|\frac{B_\phi}{B_R+B_z}\right|_{22}
\approx
\frac{1}{4}\left(4-\frac{\kappa^2}{\Omega^2}\right)
\frac{\Omega}{\omega_{\rm br}} ,
\]
where $\omega_{\rm br} = \omega_{2\bk'}((1+\epsilon)k'_\rmc)$ is the broadened version of $\omega_{2\bk'}(k'_\rmc)=0$. Writing
$\Delta k'=\epsilon k'_\rmc$, we take
$\omega_{\rm br}^2\approx
(\partial\omega_{2\bk'}^2/\partial k')_{k'_\rmc}\Delta k'$. The width
$\epsilon$ may be estimated by equating the linear and quadratic terms in
the expansion of $\omega_{2\bk'}^2$ about $k'_\rmc$,
$(\partial\omega_{2\bk'}^2/\partial k')_{k'_\rmc}\Delta k'
\approx
(\partial^2\omega_{2\bk'}^2/\partial k'^2)_{k'_\rmc}(\Delta k')^2/2$,
which gives
\[
\epsilon
=
\frac{2\left(8-\kappa^2/\Omega^2\right)^2}
{\left(8-\kappa^2/\Omega^2\right)^2+64} \approx 1.
\]
Using
\[
\frac{\bk'_\rmc\cdot\bv_{\rm A}}{\Omega}
=
\left(4-\frac{\kappa^2}{\Omega^2}\right)^{1/2}
\frac{k'_z}{k'},
\quad
\frac{k'}{k'_z}=\sqrt{1+a^2},
\]
the surviving-$H^+_{22}$ estimate becomes
\begin{align}
\left|\frac{B_\phi}{B_R+B_z}\right|_{22}
&\approx
\frac{\sqrt{1+a^2}}{4}
\left[
\frac{8-\kappa^2/\Omega^2}{2\epsilon}
\right]^{1/2}\nonumber\\
&=\frac{\sqrt{1+a^2}}{8}
\left[
\frac{\left(8-\kappa^2/\Omega^2\right)^2+64}
{8-\kappa^2/\Omega^2}
\right]^{1/2}\nonumber\\
&\approx 0.47\sqrt{1+a^2},\quad \kappa=\Omega.
\end{align}
Since $B_R\approx B_z$ at saturation (as seen from the Athena++ simulations), we have

\begin{align}
\left|\frac{B_\phi}{B_R}\right| \approx \left|\frac{B_\phi}{B_z}\right| \approx \sqrt{1+a^2}.
\end{align}

If instead $H^+_{22}$ vanishes, the leading
singular pieces are the cross-branch terms. Defining
$H_{12}^2\equiv |H^-_{12}|^2+|H^+_{12}|^2$ and similarly for $H_{21}$,
one finds
\begin{align}
\left|\frac{B_\phi}{B_R+B_z}\right|_{12/21}
&\approx
\frac{2\sqrt{1+a^2}}
{\left(8-\kappa^2/\Omega^2\right)^{1/2}}\nonumber\\
&\times\left[
\frac{
\left[\left(4-\kappa^2/\Omega^2\right)/4\right]^2H_{12}^2
+H_{21}^2
}{
H_{12}^2+H_{21}^2
}
\right]^{1/2}.
\end{align}
Thus both cases give a toroidal-to-poloidal ratio of order
$\sqrt{1+a^2}$, with an order-unity prefactor set by
$\kappa/\Omega$, the marginal-mode broadening, and the relative
cross-branch helicity amplitudes. Even if the $\pmb{\alpha}$ term in the emf is typically small compared to the $\pmb{\beta}$ term for unstratified MRI, a non-zero $\pmb{\alpha}$ is required for the saturation of the toroidal-to-poloidal ratio via helicity conservation. It should be noted, though, that above estimation is only an order-of-magnitude, approximate one.

The next order term in the emf, which is the shear current term $\pmb{\beta}\cdot\bJ$, is proportional to $J_{\bk\phi} = i(k_z B_R - k_R B_z)$, and can therefore be recast as a total divergence form (in the Fourier space). The same can be said of the higher order terms which typically scale as powers of $J_{\bk\phi}$. Therefore, the quasilinear emf satisfies the conservation of magnetic helicity.





\bibliographystyle{apsrev4-2}
\bibliography{references}

\end{document}